\documentclass[11pt,english,british]{article}
\usepackage[T1]{fontenc}
\usepackage[latin9]{inputenc}
\usepackage{geometry}
\usepackage{color}
\usepackage{babel}
\usepackage{verbatim}
\usepackage{float}
\usepackage{amsmath}
\usepackage{amssymb}
\usepackage{graphicx}
\usepackage{esint}
\usepackage[unicode=true,pdfusetitle,
 bookmarks=true,bookmarksnumbered=false,bookmarksopen=false,
 breaklinks=false,pdfborder={0 0 0},pdfborderstyle={},backref=false,colorlinks=true]
 {hyperref}
\hypersetup{
 pdfborderstyle={},urlcolor=blue,citecolor=red,linkcolor=blue}
\usepackage{breakurl}

\makeatletter
\numberwithin{equation}{section}
\newenvironment{lyxcode}
{\par\begin{list}{}{
\setlength{\rightmargin}{\leftmargin}
\setlength{\listparindent}{0pt}
\raggedright
\setlength{\itemsep}{0pt}
\setlength{\parsep}{0pt}
\normalfont\ttfamily}%
 \item[]}
{\end{list}}

\usepackage{babel}

\@ifundefined{showcaptionsetup}{}{%
 \PassOptionsToPackage{caption=false}{subfig}}
\usepackage{subfig}
\makeatother

\begin{document}

\title{Overlap singularity and time evolution in integrable quantum field
theory}

\author{D. X. Horv{á}th$^{1,2}$, M. Kormos$^{1,2}$ and G. Tak{á}cs$^{1,2}$~\\
 $^{1}${\small{}{}BME \textquotedbl{}Momentum\textquotedbl{} Statistical
Field Theory Research Group}\\
 {\small{}{}1111 Budapest, Budafoki {ú}t 8, Hungary}\\
 $^{2}${\small{}{}Department of Theoretical Physics, }\\
 {\small{}{} Budapest University of Technology and Economics}\\
 {\small{}{}1111 Budapest, Budafoki {ú}t 8, Hungary} }

\date{6th July 2018}
\maketitle
\begin{abstract}
We study homogeneous quenches in integrable quantum field theory where
the initial state contains zero-momentum particles. We demonstrate
that the two-particle pair amplitude necessarily has a singularity
at the two-particle threshold. Albeit the explicit discussion is carried
out for special (integrable) initial states, we argue that the singularity
is inevitably present and is a generic feature of homogeneous quenches
involving the creation of zero momentum particles. We also identify
the singularity in quenches in the Ising model across the quantum
critical point, and compute it perturbatively in phase quenches in
the quantum sine\textendash Gordon model which are potentially relevant
to experiments. We then construct the explicit time dependence of
one-point functions using a linked cluster expansion regulated by
a finite volume parameter. We find that the secular contribution normally
linear in time is modified by a $t\ln t$ term. We additionally encounter
a novel type of secular contribution which is shown to be related
to parametric resonance. It is an interesting open question to resum
the new contributions and to establish their consequences directly
observable in experiments or numerical simulations.

\tableofcontents{} 
\end{abstract}

\section{Introduction}

Understanding the out-of-equilibrium dynamics of isolated quantum
many-body systems is one of the most challenging problems in contemporary
physics. Due to the direct insight into quantum statistical physics
provided by the experimental realisability of closed quantum systems,
significant progress has been made both on the experimental and theoretical
side in the study of non-equilibrium behaviour. Using cold atomic
gases it has become possible to engineer and manipulate isolated quantum
systems \cite{NewtonCradle,ExperimentalNoThermalization1,ExperimentalNoThermalization3,GGEExperimental,ColdAtomSchm1,ColdAtomSchm2,Nagerl,Fukuhara,Kaufman},
and recent studies have led to a series of interesting discoveries
such as the experimental observation of the lack of thermalisation
in quantum integrable systems \cite{NewtonCradle,ExperimentalNoThermalization1,ExperimentalNoThermalization3,ExperimentalNoThermalization2}.

A paradigmatic framework for studying non-equilibrium dynamics is
provided by quantum quenches \cite{CardyCalabrese} which correspond
to a sudden change of some parameters of a system prepared in an equilibrium
state, typically in its ground state. For a long time, the focus of
the theoretical investigations was the description of the late time
asymptotic steady state. To explain the stationary state of integrable
quantum systems, the concept of the generalised Gibbs ensemble (GGE)
was proposed \cite{GGEProposal} and later experimentally confirmed
\cite{GGEExperimental}. However, specifying the complete set of conserved
charges for the GGE in interacting integrable models proved to be
a non-trivial problem \cite{CauxNoGGE,PozsgayNoGGE,Pozsi2,Goldstein,EsslerMussardo,ProsenCGGE,IlievskiQuasiLocal}
.

Beyond the steady state it is also of interest to describe the actual
time evolution and identify universal features of the non-equilibrium
dynamics. Theoretical description of the out-of-equilibrium time evolution
is much more difficult and less understood than the characterisation
of the steady state, and analytical results have mainly been obtained
in systems that can be mapped to free particles \cite{Cazalilla,Silva,SotiriadisCalabreseCardy,FiorettoMussardo,D=00003D0000F3raZar=00003D0000E1nd,CalabreseEsslerFagotti1,CalabreseEsslerFagotti2,CalabreseEsslerFagotti3,EsslerEvangelisti,ColluraSotiriadis,HeylPolkovnikov,BucciantiniKormosCalabrese,KormosColluraHardCore,SpyrosCalabrese,SpyrosMemory}
and in conformal field theory \cite{CardyCalabrese} and in a few
cases in interacting integrable systems \cite{Andrei1,Andrei2,Andrei3,Andrei4,NardisPiroliCaux,PiroliPozsgayVernier1,PiroliPozsgayVernier2}.

Quantum field theory (QFT) provides an effective and universal description
of quantum systems near their critical point. Therefore small quenches
in the vicinity of the critical point are expected to be described
by a non-equilibrium QFT, capturing universal physics even out of
equilibrium. Quenches in quantum field theories, on the other hand,
are interesting also in their own right being relevant to high energy
physics, and for certain experiments as well \cite{Schmiedmayer,SchmiedmayerPhase}.

In suitably small quantum field theory quenches, the semi-classical
approach \cite{Igloi1,Igloi2,Evangelesti,SineGSemiClassical} and
approaches based on form factor expansions \cite{SchurichtEssler,BertiniSineG,SchurichtCubero,DelfinoOscOlder,DelfinoOscNewer}
lead to analytical predictions for the time evolution of certain observables.
Whereas the perturbative approach in \cite{DelfinoOscOlder,DelfinoOscNewer}
can be applied to any quench in which the pre-quench Hamiltonian is
integrable, analytical results have only been obtained by perturbation
theory up to first order in the quench amplitude. The method developed
in \cite{SineGSemiClassical,SchurichtEssler,BertiniSineG,SchurichtCubero}
can be applied whenever the post-quench Hamiltonian is integrable
and the post-quench particle density is suitably small. This latter
approach also requires to consider specific, so-called integrable
quenches, for which the initial state can be cast in a squeezed coherent
form in the post-quench basis as

\begin{equation}
|\Omega\rangle=\mathcal{N}\exp\left(\int\frac{d\vartheta}{4\pi}K_{ab}(\vartheta)Z_{a}^{\dagger}(-\vartheta)Z_{b}^{\dagger}(\vartheta)\right)|0\rangle\;,\label{IntegrableQuench}
\end{equation}
written in terms of the post-quench Faddeev-Zamolodchikov creation
operators $Z_{a}^{\dagger}(\vartheta)$ for particle species $a$,
the post-quench vacuum $|0\rangle$ and the pair-amplitude $K_{ab}$,
which we will often refer to simply as the $K$ function. It must
be stressed that, for models with one particle species as well as
for the repulsive regime of the sine\textendash Gordon model, the
smallness of the quench essentially guarantees the structure of (\ref{IntegrableQuench}),
whereas for a generic quench protocol in a generic integrable QFT
it is not known and probably not true \cite{DelfinoOscNewer,E8Quench}
that the initial state obeys this particular structure in the post-quench
basis. For interacting integrable field theories, the determination
of the $K$ function for a specific quench protocol assumed to be
integrable is a difficult and unsolved problem, with some progress
in the numerical determination \cite{E8Quench,sGOverlaps} and analytic
approximations based on form factors \cite{SotiriadisTakacsMussardo,InitalStateIntEqHierarchcy}.

Nevertheless, integrable quenches are known to exist for lattice spin
systems \cite{PiroliPozsiVernier}, and are an ideal starting point
for analytic considerations in field theory due to their close resemblance
to the integrable boundary states introduced by Ghoshal and Zamolodchikov
\cite{GhoshalZamo}. In fact, following the case of boundary states
it is possible to extend (\ref{IntegrableQuench}) with zero-momentum
particles yielding 
\begin{lyxcode}
\begin{equation}
|\Omega\rangle=\mathcal{N}\exp\left(\frac{g_{a}}{2}Z_{a}^{\dagger}(0)+\int\frac{d\vartheta}{4\pi}K_{ab}(\vartheta)Z_{a}^{\dagger}(-\vartheta)Z_{b}^{\dagger}(\vartheta)\right)|0\rangle\;,\label{IntegrableQuench1pt}
\end{equation}
\end{lyxcode}
where $g_{a}$ is called the one-particle coupling which corresponds
to the quench breaking (particle number) parity. Such an initial state
is also motivated by experiments \cite{SchmiedmayerPhase} which show
the presence of oscillations with a frequency corresponding to the
energy of the zero-momentum particle (i.e. the particle mass).

For an initial state (\ref{IntegrableQuench1pt}), time evolution
of the vertex operator $e^{i\beta\phi/2}$ in the attractive regime
of the sine\textendash Gordon model was studied in \cite{SchurichtCubero}.
In the homogeneous (translationally invariant) quenches considered
here, the presence of zero-momentum solitons or antisolitons is excluded
if the initial state is annihilated by the topological charge which
is a typical situation including e.g quenches by local operators starting
from the ground state of the model with a different coupling. For
stationary breathers $B_{n}$, however, the one-particle couplings
$g_{B_{n}}$ can appear in (\ref{IntegrableQuench1pt}). The conclusion
of \cite{SchurichtCubero} was that one-particle oscillations of time
dependence $e^{-imt}$ show an exponential decay with several relaxation
rate contributions. In particular, with only one breather species
in the model the contribution of the first breather to the relaxation
rate of one-particle oscillations is

\begin{equation}
\tau_{B_{1}}^{-1}=\frac{m_{B_{1}}}{\pi}\int_{0}^{\infty}d\vartheta\left(1-S_{B_{1}B_{1}}(\vartheta)\right)\left|K_{B_{1}B_{1}}(\vartheta)\right|^{2}\sinh\vartheta\;.\label{TauBSchuricht}
\end{equation}

For boundary field theories, however, it is known that the existence
of a one-particle coupling implies a first order pole in the corresponding
$K$ function at the origin,

\begin{equation}
K_{aa}(\vartheta)\propto-i\frac{g_{a}^{2}}{2}\frac{1}{\vartheta}\ ;\label{KSing}
\end{equation}
the correct relation of the residue to the one-particle coupling was
first found in \cite{gDorey} and later proven in \cite{gPallaBajnokTakacs2}.
In this paper we show that this relation extends to the case of integrable
quenches corresponding to an initial state of the form (\ref{IntegrableQuench1pt})
as well. As an immediate consequence, the integral in (\ref{TauBSchuricht})
becomes divergent since $S_{B_{1}B_{1}}(0)=-1$ . Even though much
of the derivation in \cite{SchurichtCubero} remains valid, the singular
expressions clearly need to be corrected.

In this paper we first demonstrate the presence of the singularity
(\ref{KSing}) whenever a zero-momentum particle is present in the
initial state. Starting from this observation, we follow a linked
cluster expansion similar to the one performed in \cite{SchurichtEssler,BertiniSineG,SchurichtCubero}
but now supported by a finite volume regularisation to treat the singularities
and compute the time dependence of one-point functions $\langle\mathcal{O}(t)\rangle$
up to terms containing five particles focusing on the one-particle
oscillations $e^{-imt}$. As a result of these calculations, we show
that the linearly time dependent secular contribution in \cite{SchurichtCubero}
is modified by a $mt\ln mt$ term and a novel type of secular term
is encountered which has analogies with the phenomenon of parametric
resonance.

The detailed structure of our paper is the following. In Section \ref{sec:SingularK},
first a general argument is presented for the existence of the singularity
(\ref{KSing}), and then we consider two particular quench protocols
that illustrate our arguments. In the Ising model, quenches from the
ferromagnetic to the paramagnetic phase are discussed and the singular
structure of the $K$ function is explicitly demonstrated in the continuum
limit. Then we consider a quench protocol in the sine\textendash Gordon
model which consists of shifting the field by a constant. Computing
the quench overlaps using an expansion in the quench magnitude, we
demonstrate explicitly that the pair amplitude for the first breather
possesses the anticipated singularity up to leading order. In Section
\ref{sec:TimeDep} we perform a linked cluster expansion for the time-dependent
one-point function, using a finite volume regularisation which was
first introduced in \cite{FinVolFF1and2}. This allows us to refine
the argument for the existence of the singularity (\ref{KSing}),
and also the explicit construction of the contributions for $\langle\mathcal{O}(t)\rangle$
up to four particle terms, from which the terms corresponding to one-particle
oscillations are extracted. We then consider the five-particle terms,
but only present the leading order emergent secular contributions.
In Section \ref{subsec:Interpretation} we collect and present our
formulas describing the time evolution, address the question of resummation
in the linked cluster expansion and generalise our results to cases
involving more than one particle species. Finally we discuss a class
of secular terms linked to a mechanism analogous to parametric resonance,
and conclude in Section \ref{sec:Conclusions}. Due to the large amount
of tedious calculations involved, most of their details are relegated
to appendices. In Appendix \ref{sub:Appendix-A-FinVol}, the finite
volume formalism is reviewed, while Appendices \ref{sec:IsingAppendix}
and \ref{subsec:ExponentialQuenches} are devoted to the quenches
in the Ising and sine\textendash Gordon models. The technical details
of the linked cluster expansion and the calculation of the time evolution
can be found in Appendices \ref{sec:Some-useful-relations}, \ref{sub:Eucl1point},
\ref{sec:D23Res}, \ref{sec:CInt} and \ref{sec:EvaluatingKernel}.
To confirm the validity of our calculations they were numerically
cross-checked at several points; details of these checks are presented
in Appendix \ref{sec:Numerics}.

\section{Integrable quenches with one-particle coupling and the singularity
of $K$\label{sec:SingularK}}

In this section we show that in integrable quenches with non-zero
one-particle coupling $g$ the $K$ function necessarily possesses
a first order pole of the form (\ref{KSing}) at the origin. For simplicity,
we focus on models with one particle species but the argument carries
over to systems with several particle species. First we present a
general argument based on an analogy with the one-point functions
of bulk operators in the presence of boundaries discussed in \cite{OnePointFunctions}.
The core of the argument is based on the cancellation of divergent
terms in the expectation value which is evaluated using finite volume
regulators. We then proceed to a concrete example of the quench in
the Ising field theory crossing the phase boundary, and discuss an
interesting quench in the sine\textendash Gordon model, for which
the one-particle coupling and the singular part of the $K$ function
can be calculated to lowest order in the quench parameter and the
presence of the singularity in $K$ and the relation of its residue
to $g$ can be verified explicitly.

\subsection{Boundary one-point functions}

In the following we briefly review the boundary problem discussed
in \cite{OnePointFunctions}. Let us consider an integrable field
theory with a single massive particle constrained on a finite line
$x\in[0,\mathcal{R}]$ with integrable boundary conditions $\alpha$
and $\beta$ at the two ends; for simplicity we consider the case
when the left/right boundary conditions are identical i.e. $\alpha=\beta=B$.
The vacuum expectation value 
\begin{equation}
\langle\mathcal{O}(x)\rangle^{B}\label{objective}
\end{equation}
taken with respect to the ground state of the finite volume Hamiltonian
$H_{\mathcal{R}}^{B}$ can be rewritten using an Euclidean rotation
\begin{equation}
\langle\mathcal{O}(x)\rangle^{B}=\frac{\langle B|\ e^{-Hx}\mathcal{O}(0)\ e^{-H(\mathcal{R}-x)}\:|B\rangle}{\langle B|e^{-H\mathcal{R}}|B\rangle}\;,\label{ezkellene}
\end{equation}
where the coordinate $x$ plays the role of the Euclidean time variable,
$H$ is the infinite volume Hamiltonian in the crossed channel and
$|B\rangle$ is the boundary state corresponding to the boundary condition
$B$. When the boundary state contains zero-momentum particles associated
with a non-zero coupling of a single particle state to the boundary
in the original channel, $|B\rangle$ can be expanded in the asymptotic
multi-particle basis \cite{GhoshalZamo} as

\begin{equation}
|B\rangle=\mathcal{N}\exp\left(\bar{g}_{B}Z^{\dagger}(0)+\int\frac{d\vartheta}{4\pi}K_{B}(\vartheta)Z^{\dagger}(-\vartheta)Z^{\dagger}(\vartheta)\right)|0\rangle\;,\label{Bstate}
\end{equation}
where for simplicity we assumed that the model has only one species,
\[
K_{B}(\vartheta)=R_{B}(i\pi/2-\vartheta)
\]
in terms of the boundary reflection factor $R_{B}(\vartheta)$, and
$Z^{\dagger}(\vartheta)$ are the Faddeev\textendash Zamolodchikov
creation operators satisfying the commutation relations 
\[
Z^{\dagger}(\vartheta_{1})Z^{\dagger}(\vartheta_{2})=S(\vartheta_{1}-\vartheta_{2})Z^{\dagger}(\vartheta_{2})Z^{\dagger}(\vartheta_{1})\;.
\]
$\bar{g}_{B}$ is the one-particle coupling to the boundary and the
amplitudes $K_{B}(\vartheta)$ satisfy the boundary crossing-unitary
equation \cite{GhoshalZamo} 
\[
K_{B}(\vartheta)=S(2\vartheta)K_{B}(-\vartheta)\;,
\]
which serves as a consistency relation of (\ref{Bstate}) and also
implies $K_{B}(\vartheta)^{*}=K_{B}(-\vartheta)$. The single particle
coupling implies a pole for the reflection factor 
\[
R_{B}(\vartheta)\sim\frac{i}{2}\frac{g_{B}^{2}}{\theta-i\pi/2}\;,
\]
which then yields a pole at the origin

\begin{equation}
K_{B}(\vartheta)\sim-\frac{i}{2}\frac{g_{B}^{2}}{\vartheta}\;,\label{eq:K_BSing}
\end{equation}
for the boundary $K$ function. Whereas it was argued that $\bar{g}_{B}=g_{B}$
in \cite{GhoshalZamo}, it was later demonstrated in \cite{gDorey,gPallaBajnokTakacs}
that the proper normalisation is $\bar{g}_{B}=g_{B}/2$, with a general
proof given in \cite{gPallaBajnokTakacs2}.

The relation between the residue of $K$ and the boundary one-particle
coupling is crucial for the consistency of a number of theoretical
constructs, such as the boundary form factor bootstrap considered
in \cite{bff}. For the one-point function, the approach of \cite{OnePointFunctions}
is to put the theory in a finite volume $L$ in the crossed channel
(with periodic boundary conditions) and consider the limit 
\begin{equation}
\langle\mathcal{O}(x)\rangle^{B}=\lim_{L\to\infty}\langle\mathcal{O}(x)\rangle_{L}^{B}=\lim_{L\to\infty}\frac{\langle B_{L}|\ e^{-H_{L}x}\mathcal{O}(0)\ e^{-H_{L}(\mathcal{R}-x)}\ |B_{L}\rangle}{\langle B_{L}|e^{-H_{L}\mathcal{R}}|B_{L}\rangle}\;,\label{ezkell}
\end{equation}
where $H_{L}$ is the finite volume Hamiltonian with periodic boundary
conditions and $|B_{L}\rangle$ represents the boundary state in finite
volume. The finite volume is introduced here as a regulator for disconnected
contributions arising from the matrix elements of the operator $\mathcal{O}$
which appear once $|B_{L}\rangle$ is expanded in terms of the corresponding
finite volume multi-particle eigenstates of $H_{L}$. The disconnected
terms lead to positive powers of the dimensionless volume variable
$mL$ which only cancel if the singularity of $K_{B}$ is exactly
as in (\ref{eq:K_BSing}). Therefore the singularity of $K_{B}$ and
the relation of its residue to the one-particle coupling are consistency
criteria for the existence of well-defined one-point functions in
the infinite volume theory.

The time evolution of the expectation value of a local operator $\mathcal{O}$
after a quantum quench with initial state 
\[
|\varOmega\rangle=\mathcal{N}\exp\left(\frac{g}{2}Z^{\dagger}(0)+\int\frac{d\vartheta}{4\pi}K(\vartheta)Z^{\dagger}(-\vartheta)Z^{\dagger}(\vartheta)\right)|0\rangle\;,
\]
and post-quench Hamiltonian $H$ is given by the expression 
\[
\langle\mathcal{O}(t)\rangle=\frac{\langle\varOmega|\ e^{itH}\mathcal{O}(0)\ e^{-itH}|\varOmega\rangle}{\langle\varOmega|\varOmega\rangle}\;,
\]
which is just a real time analogue of the boundary expectation value.
The only difference is that the $K$ function appearing in $|\varOmega\rangle$
is not related to any reflection factor; in fact, Ghoshal\textendash Zamolodchikov
boundary states are not normalisable, while for quench initial states
the factor $\mathcal{N}$ is chosen to ensure $\langle\varOmega|\varOmega\rangle=1$.
However, it is clear from the calculations performed in \cite{OnePointFunctions}
that the condition for the cancellation of singularities in unaffected
by these details, therefore a well-defined expectation value $\langle\mathcal{O}(t)\rangle$
after an integrable quench with one-particle coupling $g/2$, the
$K$ function must have a first order pole at the origin with a residue
equal to $-ig^{2}/2$.

In the next subsections we discuss two examples where this relation
can be verified explicitly, while in Section \ref{sec:TimeDep} we
consider the real time evolution and show that this condition must
hold for consistency.

\subsection{Quench in the Ising field theory from the ferromagnetic to the paramagnetic
phase\label{subsec:Quench-in-the}}

In this subsection we study the Ising field theory and quenches across
the two phases of the model, as in the Ising model only quenches from
the ferromagnetic to the paramagnetic (PM) phase can account for a
zero-momentum particle in the initial state. In our approach, we perform
the continuum limit of various quantities obtained in the lattice
model and in particular, we show that the pair-amplitude possesses
a pole with the residue expected from the above considerations.

Quenches in the lattice model were discussed in \cite{CalabreseEsslerFagotti1,CalabreseEsslerFagotti2,CalabreseEsslerFagotti3}
and for particular quenches within the ferromagnetic phase, calculations
also in the continuum model \cite{SchurichtEssler} were performed
and numerically checked \cite{Tibi}. Although for the $FM\rightarrow PM$
quench such calculations in the QFT were not carried out, the scaling
limit of the analogous quantities in the lattice model make perfect
sense, which we regard as the characteristics of the QFT quench problem.
Throughout the subsection, we only discuss the most important features
and formulas, and for a more detailed treatment of the topic, Appendix
\ref{sec:IsingAppendix} is recommended to consult.

We first recall, that the transverse field quantum Ising model (TQIM)
is defined by the Hamiltonian

\begin{equation}
H=-J\sum_{i=1}^{N}\left(\sigma_{i}^{x}\sigma_{i+1}^{x}+h\sigma_{i}^{z}\right)\;,\label{IsingChainHamiltonianSec2.2}
\end{equation}
where $\sigma_{i}^{\alpha}$ denotes the Pauli matrices at site $i$,
$J>0$, $h$ is the transverse field and the boundary conditions are
assumed to be periodic. By applying the Jordan\textendash Wigner transformation,
the Hamiltonian (\ref{IsingChainHamiltonianSec2.2}) can be mapped
to spinless Majorana fermions with dispersion relation \cite{IsingDiag1,IsingDiag2}

\begin{equation}
\varepsilon_{h}(k)=2J\sqrt{1+h^{2}-2h\cos k}\;,\label{IsingDispRelSec2.2}
\end{equation}
and with an energy gap $\Delta=2J|1-h|$. The model possesses a quantum
critical point at $h=1$ separating the paramagnetic or disordered
phase for $h>1$ and the ferromagnetic, ordered phase for $h<1$.
In the disordered phase, the expectation value of $\sigma_{i}^{x}$,
i.e. that of the magnetisation operator vanishes, while in the ferromagnetic
phase its value is non-zero. The Hilbert space of the model consists
of two sectors with respect to fermion number parity. In the Neveu\textendash Schwarz
and Ramond sectors states with even and odd number of fermions are
present, respectively, resulting in the quantisation condition for
the wave numbers

\begin{equation}
\begin{split}k_{n}= & \frac{2\pi}{N}\left(n+\frac{1}{2}\right)\quad\text{Neveu\textendash Schwarz}\\
p_{n}= & \frac{2\pi}{N}n\qquad\qquad\quad\text{Ramond}\;,
\end{split}
\end{equation}
where $n$ is a positive integer.

Performing a quench in the transverse field $h$, the pre- and post-quench
excitations can be related via a Bogoliubov transformation if the
initial state is the pre-quench vacuum. As a consequence, the squeezed-coherent
form of the initial state in the post quench basis (\ref{IntegrableQuench})
is guaranteed. Focusing on quenching from the ground state of the
FM phase to the PM phase, ( $h_{0}\rightarrow h$ with $h_{0}<1$
and $h>1$), one can write \cite{CalabreseEsslerFagotti1,CalabreseEsslerFagotti2,CalabreseEsslerFagotti3}

\begin{equation}
\begin{split}\frac{|0,h_{0}\rangle_{NS}^{FM}\pm|0,h_{0}\rangle_{R}^{FM}}{\sqrt{2}}= & \frac{1}{\sqrt{2}N_{NS}}\exp\left(i\sum_{k\subset NS}K(k)a_{-k}^{\dagger}a_{k}^{\dagger}\right)|0,h\rangle_{NS}^{PM}\\
\pm & \frac{1}{\sqrt{2}N_{R}}\exp\left(i\sum_{p\subset R\setminus\{0\}}K(p)a_{-p}^{\dagger}a_{p}^{\dagger}\right)a_{0}^{\dagger}|0,h\rangle_{R}^{PM}\;,
\end{split}
\label{IsingFMInitStateSec2.2}
\end{equation}
where $N_{NS}$ and $N_{R}$ are normalisation constants

\begin{equation}
\begin{split}N_{R}^{2}= & \prod_{n\subset\mathbb{N}^{+}}\left(1+\left|K\left(\frac{2\pi}{L}n\right)\right|^{2}\right)\;,\\
N_{NS}^{2}= & \prod_{m\subset\mathbb{N}+\frac{1}{2}}\left(1+\left|K\left(\frac{2\pi}{L}m\right)\right|^{2}\right)\;,
\end{split}
\end{equation}
$a_{k}$ and $a_{k}^{\dagger}$ are fermionic operators and $K$ can
be found in Appendix \ref{sec:IsingAppendix} in (\ref{IsingKSq}).

In the scaling limit of the TQIM, $J$ is sent to infinity together
with $h\rightarrow1$ such that the gap associated with the fermion
mass remains finite

\begin{equation}
M=2J|1-h|\;.
\end{equation}
In addition, the lattice spacing is sent to zero as $a=\frac{1}{2J}$.
In particular, for the $FM\rightarrow PM$ quench, we send $\delta h\rightarrow0$
in the following way:

\begin{equation}
\begin{split}h=1+\delta h\;,\quad & h_{0}=1-\frac{M_{0}}{M}\delta h\;,\\
J=\frac{M}{2\delta h}\;,\quad & a=\frac{\delta h}{M}\;,
\end{split}
\end{equation}
which ensures that the dispersion relation in the post- and pre-quench
model is $\varepsilon_{h}(pa)\rightarrow\sqrt{M^{2}+p^{2}}$ and $\varepsilon_{h_{0}}(pa)\rightarrow\sqrt{M_{0}^{2}+p^{2}}$
respectively, i.e. the mass in the PM and FM phase is $M$ and $M_{0}$.
Upon the substitution $k=pa$, the continuum limit of the square of
the lattice $K$ function (\ref{IsingKSq}) in (\ref{IsingFMInitStateSec2.2})
is

\begin{equation}
|K(p)|^{2}=\frac{\sqrt{p^{2}+M^{2}}\sqrt{p^{2}+M_{0}^{2}}-p^{2}+MM_{0}}{\sqrt{p^{2}+M^{2}}\sqrt{p^{2}+M_{0}^{2}}+p^{2}-MM_{0}}\;.\label{IsingKSqQFTSec2.2}
\end{equation}

From (\ref{IsingKSqQFTSec2.2}) it is easily seen that this function
has a $1/p^{2}$ singularity at the origin with the coefficient 
\[
\frac{4M_{0}^{2}M^{2}}{(M+M_{0})^{2}}\;.
\]
This singularity corresponds to the presence of a zero-momentum particle
in the Ramond contribution to the initial state (\ref{IsingFMInitStateSec2.2}).
As $p=M\sinh\vartheta$, the coefficient of the singularity of $|K(p)|^{2}$
equals $\frac{g^{4}}{4}M^{2}$ , therefore $K(\vartheta)$ can be
written in the form (\ref{KSing}) with 
\begin{equation}
g=2\sqrt{\frac{M_{0}}{M+M_{0}}}\;.\label{eq:Ising_gM0}
\end{equation}
Now we show that the one-particle coupling expressed with $M$ and
$M_{0}$ in (\ref{IsingFMInitStateSec2.2}) equals $g$. To calculate
the latter, we put the theory into finite volume, where (cf. also
Section \ref{sec:TimeDep}) the finite volume expansion of the integrable
initial state reads

\begin{equation}
|\Omega\rangle_{L}=\mathcal{G}(L)\left(|0\rangle_{L}+\frac{g}{2}\sqrt{ML}|\{0\}\rangle_{L}+\sum_{I}K(\vartheta)N_{2}(\vartheta,L)|\{-I,I\}\rangle_{L}.+...\right){\color{black}{\color{red}\:{\color{black},}}}\label{eq:FinVolState1partTerm}
\end{equation}
where the $I$ denote quantum numbers labelling the finite volume
states and $N_{2}$ can be found in (\ref{N1N2}). Then from (\ref{IsingFMInitStateSec2.2})

\begin{equation}
\sqrt{ML}\frac{g}{2}=\frac{N_{NS}}{N_{R}}\:,
\end{equation}
must hold. It is convenient to calculate the logarithm of their ratio:

\begin{equation}
\begin{split}\ln\frac{N_{NS}^{2}}{N_{R}^{2}}= & \ln\prod_{n\subset\mathbb{N}^{+}}\frac{1+\left|K\left(\frac{2\pi}{L}(n-1/2)\right)\right|^{2}}{1+\left|K\left(\frac{2\pi}{L}(n)\right)\right|^{2}}\end{split}
\;,\label{CsunyaLog-1}
\end{equation}
which in Appendix \ref{sec:IsingAppendix} is shown to be equal to
\[
\ln\frac{MM_{0}L}{(M+M_{0})}
\]
when $L\rightarrow\infty$, so (\ref{eq:Ising_gM0}) indeed holds.

\subsection{Phase quenches in the sine\textendash Gordon model \label{subsec:PhaseQuench}}

Consider the sine\textendash Gordon model defined by the action

\begin{equation}
\mathcal{A}=\int d^{2}x\left(\frac{1}{2}\partial_{\mu}\Phi\partial^{\mu}\Phi+\frac{\mu^{2}}{\beta^{2}}\cos\beta\Phi\right)\:,\label{eq:SineGAction-1}
\end{equation}
in a finite volume $L$ with quasi-periodic boundary conditions 
\[
\Phi(t,x+L)=\Phi(t,x)+\frac{2\pi}{\beta}n\quad n\in\mathbb{Z}\;.
\]
The quench protocol consists of abruptly shifting the sine\textendash Gordon
field $\Phi\rightarrow\Phi+\delta/\beta$ at $t=0$, i.e. changing
the phase of the cosine potential regarding the pre-quench vacuum
as initial state for the post-quench evolution. The peculiarity of
this protocol is that, as shown in Appendix \ref{subsec:ExponentialQuenches},
it is possible to relate the pre- and post-quench ground states by
a unitary transformation

\begin{equation}
|\Omega\rangle_{L}=\exp\left(i\frac{\delta}{\beta}\Pi_{0}\right)|0\rangle_{L}\:,\label{eq:Omegainitial}
\end{equation}
where 
\[
\Pi_{0}=\int_{0}^{L}dx\Pi(t,x)=i\int_{0}^{L}[H,\Phi(t,x)]dx
\]
is the zero mode of the conjugate momentum field $\Pi=\dot{\Phi}$.
This allows one to derive a form factor expansion for the overlaps
with an arbitrary state $|\chi\rangle$

\begin{equation}
\vphantom{O}_{L}\langle\chi|\Omega\rangle_{L}=\vphantom{O}_{L}\langle\chi|\exp\left(i\frac{\delta}{\beta}\Pi_{0}\right)|0\rangle_{L}\,,\label{eq:OverlapDef}
\end{equation}
by expanding the exponential into power series, 
\begin{equation}
\begin{split}\vphantom{O}_{L}\langle\chi|\Omega\rangle_{L}=\sum_{l=0}^{\infty}\frac{(-1)^{l}}{l!}\mbox{\ensuremath{\left({\displaystyle \frac{\delta}{\beta}}\right)}}^{l}\sum_{\alpha_{1}}\ldots\sum_{\alpha_{l-1}}\int\left(\prod_{i}dx_{i}\right)\vphantom{O}_{L}\langle\chi|e^{i\hat{P}x_{1}}\Phi(0)e^{-i\hat{P}x_{1}}|\alpha_{1}\rangle_{L}(E_{\chi}-E_{\alpha_{1}})\times\\
\vphantom{O}_{L}\langle\alpha_{1}|e^{i\hat{P}x_{2}}\Phi(0)e^{-i\hat{P}x_{2}}|\alpha_{2}\rangle_{L}(E_{\alpha_{1}}-E_{\alpha_{2}})...\vphantom{O}_{L}\langle\alpha_{l-1}|e^{i\hat{P}x_{l}}\Phi(0)e^{-i\hat{P}x_{l}}|0\rangle_{L}(E_{\alpha_{l-1}}-E_{0}),
\end{split}
\end{equation}
where $\hat{P}$ is the momentum operator and the $\alpha_{i}$ index
$l-1$ complete sets of asymptotic multi-particle eigenstates. As
the initial state is the ground state of a translational invariant
Hamiltonian, the overlaps are non-zero only for states $|\chi\rangle$
of total spatial momentum zero. Due to the integrals over $x_{i},$
this also restricts the intermediate states $|\alpha_{i}\rangle_{L}$
to have zero total momentum so we can write 
\begin{align}
\vphantom{O}_{L}\langle\chi|\Omega\rangle_{L}= & \sum_{l=0}^{\infty}(-1)^{l}\mbox{\ensuremath{\left({\displaystyle \frac{\delta}{\beta}}\right)}}^{l}\frac{L^{l}}{l!}\tilde{\sum_{\alpha_{1}}}\ldots\tilde{\sum_{\alpha_{l-1}}}\vphantom{O}_{L}\langle\chi|\Phi(0)|\alpha_{1}\rangle_{L}(E_{\chi}-E_{\alpha_{1}})\times\label{eq:OverlapFFExpansion}\\
 & \vphantom{O}_{L}\langle\alpha_{1}|\Phi(0)|\alpha_{2}\rangle_{L}(E_{\alpha_{1}}-E_{\alpha_{2}})...\vphantom{O}_{L}\langle\alpha_{l-1}|\Phi(0)|0\rangle_{L}(E_{\alpha_{l-1}}-E_{0})\;,
\end{align}
where the tildes over the sums mean that only zero momentum states
are included.

We now compute the overlap for the first breather to first order in
$\delta/\beta$. Matching the finite volume expression of the initial
state (\ref{eq:Omegainitial}) with the general case (\ref{eq:FinVolState1partTerm}),
and using (\ref{N1N2}) and the finite volume form factors (\ref{eq:FinVolFF}),
we obtain  
\begin{equation}
\begin{split}\sqrt{m_{1}L}\,\frac{g}{2}= & _{L}\langle\{0\}|\Omega\rangle_{L}\\
= & -m_{1}L{\displaystyle \frac{\delta}{\beta}}{}_{L}\langle\{0\}|\Phi(0)|0\rangle_{L}\\
= & -\mbox{\ensuremath{\left({\displaystyle \frac{\delta}{\beta}}\right)}}m_{1}L\frac{F_{B_{1}}^{*}}{\sqrt{m_{1}L}}\;,
\end{split}
\end{equation}
from which 
\begin{equation}
\frac{g}{2}=-\mbox{\ensuremath{{\displaystyle \frac{\delta}{\beta}}}}F_{B_{1}}^{*}\;,\label{eq:gper2}
\end{equation}
where $F_{B_{1}}$ is the infinite volume one-breather form factor
of $\Phi$ (\ref{eq:FB1}).

Since the form factors of $\Phi$ with even number of breathers $B_{1}$
vanishes, the lowest non-trivial order for{} the pair amplitude $K$
is $\left(\delta/\beta\right)^{2}$ : 
\begin{equation}
\begin{split}N_{2}(\vartheta,L)K(\vartheta)= & _{L}\langle\{I,-I\}|\Omega\rangle_{L}\\
= & \mbox{\ensuremath{\left({\displaystyle \frac{\delta}{\beta}}\right)}}^{2}\frac{L^{2}}{2}\tilde{\sum_{\alpha_{1}}}{}_{L}\langle\{I,-I\}|\Phi(0)|\alpha_{1}\rangle_{L}(2m_{1}\cosh\vartheta-E_{\alpha_{1}}){}_{L}\langle\alpha_{1}|\Phi(0)|0\rangle_{L}E_{\alpha_{1}}\,,
\end{split}
\label{eq:Kexpansion}
\end{equation}
where $\vartheta$ is related to the quantum number $I$ via the Bethe
quantisation condition 
\begin{equation}
m_{1}L\sinh\vartheta+\delta_{B_{1}B_{1}}(2\vartheta)=2\pi I\,,\label{BetheQuantizationPhaseQuench}
\end{equation}
$N_{2}(\theta,L)$ is a finite volume normalisation factor{} (\ref{N1N2})
$\delta_{B_{1}B_{1}}$ and its derivative $\varphi_{B_{1}B_{1}}(\vartheta)$
is defined by $S_{B_{1}B_{1}}(\vartheta)=-e^{i\delta_{B_{1}B_{1}}(\vartheta)}$,
and $\theta_{1},...,\theta_{n}$ are the particle rapidities in the
state $|\alpha_{1}\rangle_{L}$ determined by finite volume quantisation
relations analogous to (\ref{BetheQuantizationPhaseQuench}). Using
the expression for finite volume form factors (\ref{eq:FinVolFF})
it can be written as 
\begin{equation}
\begin{split} & \mbox{\ensuremath{\left({\displaystyle \frac{\delta}{\beta}}\right)}}^{2}\frac{L^{2}}{2}\sum_{n=1}^{\infty}\sum_{\{\beta\}_{n}}\frac{F_{B_{1}B_{1}B_{i_{1}}...B_{i_{n}}}(i\pi+\vartheta,i\pi-\vartheta,\theta_{1},...\theta_{n})}{\sqrt{\left(m_{1}L\cosh\vartheta\right)^{2}+\left(m_{1}L\cosh\vartheta\right)\,\varphi_{B_{1}B_{1}}(\vartheta)}\rho_{n}(\theta_{1},...\theta_{n})}\times\\
 & \left(2m_{1}\cosh\vartheta-\sum_{i=1}^{n}m_{i}\cosh\theta_{i}\right)F_{B_{i_{1}}...B_{i_{n}}}^{*}(\theta_{1},...\theta_{n})\left(\sum_{i=1}^{n}m_{i}\cosh\theta_{i}\right)\,.
\end{split}
\end{equation}
In Appendix \ref{subsec:ExponentialQuenches} it is shown that in
the limits $L\rightarrow\infty$ and $\vartheta\rightarrow0$ only
the $n=1$ term survives where the single particle in the intermediate
state is also a $B_{1}$, from which 
\begin{equation}
K(\vartheta)\sim\mbox{\ensuremath{\left({\displaystyle \frac{\delta}{\beta}}\right)}}^{2}\frac{F_{B_{1}B_{1}B_{1}}(i\pi+\vartheta,i\pi-\vartheta,0)(2\cosh\vartheta-1)F_{B_{1}}^{\text{*}}}{2\cosh\vartheta}\,.
\end{equation}
Using the form factor kinematical singularity equation (\ref{eq:kinematicalaxiom})
one can extract that for small $\vartheta$

\begin{equation}
F_{B_{1}B_{1}B_{1}}(i\pi+\vartheta,i\pi-\vartheta,0)\sim-\frac{4i}{\mathbf{\vartheta}}F_{B_{1}}\,,
\end{equation}
so 
\begin{equation}
\begin{split}K(\vartheta)\sim & -2i\left({\displaystyle \frac{\delta}{\beta}}F_{B_{1}}\right)^{2}\frac{1}{\vartheta}+O(\vartheta^{0})\\
= & -i\frac{g^{2}}{2}\frac{1}{\vartheta}+O(\vartheta^{0})\,,
\end{split}
\end{equation}
where we used (\ref{eq:gper2}) which establishes the relation (\ref{eq:K_BSing})
between the one-particle coupling of the first breather and the singularity
of its pair amplitude for this particular quench in the sine\textendash Gordon
model.

We remark that for this particular quench protocol we do not know
whether it leads to an integrable initial state of the generalised
squeezed form (\ref{IntegrableQuench1pt}). However, we recall that
the proof of the analogous relation in boundary quantum field theory
presented in \cite{gPallaBajnokTakacs2} does not depend on integrability
either (in fact, it works in general $D+1$ space-time dimensions).

Note also that the above argument straightforwardly generalises to
a much larger class of quench protocols, the ``exponential quenches''
when the initial state is related to the post-quench ground state
via

\begin{equation}
|\Omega\rangle_{L}=\exp\left(i\lambda\int dx\Psi(x)\right)|0\rangle_{L}\:,
\end{equation}
where $\Psi(x)$ is a local field which breaks particle number parity.

\section{Linked cluster expansion in finite volume \label{sec:TimeDep}}

To describe the time evolution of expectation values of local operator,
we follow the approach developed in \cite{SchurichtEssler,BertiniSineG,SchurichtCubero}
and apply a linked cluster expansion, combined with the finite volume
regularisation scheme used in \cite{OnePointFunctions,FiniteTCorr}.
The latter is based on the finite volume form factor formalism developed
in \cite{FinVolFF1and2}; the ingredients necessary for our calculations
are described in Appendix \ref{sub:Appendix-A-FinVol}.

For a quench starting from an initial state written in terms of post-quench
multi-particle states as in (\ref{IntegrableQuench1pt}), a natural
approach to compute the one-point function of a local operator $\mathcal{O}$
is to decompose into contributions from states with different number
of particles, which results in an expansion in terms of form factors
of the local operator. However, in infinite volume form factors possess
pole singularities due to (\ref{eq:kinematicalaxiom}), and for quenches
with one-particle coupling the $K$ functions also possess singularities.
As a result, the contributions are ill-defined and need to be regularised
which can be done by putting the theory in finite volume, where due
to the quantisation of the particle momenta neither the kinematic
singularities of the form factors nor the singularities of the pair-amplitude
contribute. The finite volume $L$ can be considered as a physical
regulator and the expectation value is then written as\textit{ 
\begin{equation}
\begin{split}\langle\mathcal{O}(t)\rangle= & \frac{\langle\varOmega|\ e^{itH}\mathcal{O}(0)\ e^{-itH}|\varOmega\rangle}{\langle\varOmega|\varOmega\rangle}\\
= & \lim_{L\rightarrow\infty}\frac{_{L}\langle\varOmega|\ e^{itH}\mathcal{O}(0)\ e^{-itH}|\varOmega\rangle_{L}}{_{L}\langle\varOmega|\varOmega\rangle_{L}}\;,
\end{split}
\label{tervszerint}
\end{equation}
}which is first evaluated for finite $L$ where one can verify the
cancellation of singular terms explicitly and then take the limit
$L\rightarrow\infty$.

To perform the calculation, one needs an expression for the initial
state in finite volume which was derived in \cite{OnePointFunctions}:

\textit{
\begin{multline}
|\varOmega\rangle_{L}=\mathcal{G}(L)\Big(|0\rangle_{L}+\frac{g}{2}\sqrt{mL}|\{0\}\rangle_{L}+\sum_{I}K(\vartheta)N_{2}(\vartheta,L)|\{-I,I\}\rangle_{L}\\
+\sum_{I}\frac{g}{2}K(\vartheta)N_{3}(\vartheta,L)|\{-I,I,0\}\rangle_{L}+\frac{1}{2}\sum_{I\neq J}K(\vartheta_{1})K(\vartheta_{2})N_{4}(\vartheta_{1},\vartheta_{2},L)|\{-I,I,-J,J\}\rangle_{L}\Big)+\ldots\;,\label{FinVolInitState}
\end{multline}
}where $I,J$ are Bethe quantum numbers and we used the notations
introduced in Appendix \ref{sub:Appendix-A-FinVol}. To simplify notations,
one can write the following shorthand for\textit{ 
\[
|\varOmega\rangle_{L}=\mathcal{G}(L)\sum_{n=0}^{\infty}|\varOmega\rangle^{(n)}\;,
\]
}where $|\varOmega\rangle^{(0)}=|0\rangle_{L}\;,|\varOmega\rangle^{(1)}=\frac{g}{2}N_{1}(L)|\{0\}\rangle_{L}\;,$
etc. denote the contributions with a fixed number of particles.

To ensure the convergence of the expansion for high energies one can
introduce a regulator parameter $R$ and consider

\textit{
\begin{equation}
\langle\mathcal{O}(t,R)\rangle=\frac{\langle\varOmega|\ e^{\left(-\frac{R}{2}+it\right)H}\mathcal{O}(0)\ e^{\left(-\frac{R}{2}-it\right)H}|\varOmega\rangle}{\langle\varOmega|e^{-RH}|\varOmega\rangle}\;,\label{tervszerint-2}
\end{equation}
}where $R>0$. When recasting the quantum number sums in terms of
contour integrals, $R$ ensures that the integrals themselves are
convergent and therefore allows appropriate manipulations of the contours.
At the end of the calculation the parameter $R$ is sent to zero.

Following the procedure introduced in \cite{EsslerKonik}, one can
separate contributions indexed by particle number as follows\textit{
\[
C_{kl}\,=\,^{(k)}\langle\varOmega|e^{\left(-\frac{R}{2}+it\right)H}\mathcal{O}(0)\ e^{\left(-\frac{R}{2}-it\right)H}|\varOmega\rangle^{(l)}\;,
\]
}and for a proper normalisation of the state one must also divide
by the ``partition function'' 
\[
Z=\sum_{n}Z_{n}=\sum_{n}{}_{L}^{(n)}\langle\varOmega|e^{-RH}|\varOmega\rangle_{L}^{(n)}\,.
\]
In particular for $Z$, the first few terms are 
\begin{equation}
Z_{0}=1\;,\qquad\qquad Z_{1}=\frac{g^{2}}{4}mLe^{-mR}\;,
\end{equation}
and 
\begin{equation}
Z_{2}=\sum_{I}K^{*}(\vartheta)K(\vartheta)N_{2}(\vartheta,L)^{2}e^{-2mR\cosh\vartheta}\;.
\end{equation}
Let us turn to the issue of the expansion parameter. Whereas in our
calculations $R$ is eventually sent to zero at the end, for $\langle\mathcal{O}(t,R)\rangle$
is expected to be well-defined for any finite $R$. Therefore let
us first treat $R$ as a large positive quantity, and introduce the
parameters $z=e^{-m(R/2+it)}$ and $\bar{z}=e^{-m(R/2-it)}$. Then
the order of $C_{kl}$ is $\bar{z}^{k}z^{l}$ , and that of $Z_{n}$
is $(z\bar{z})^{n}$. The inverse of the partition function, thus
can be expanded in powers of $z\bar{z}$ as 
\[
Z^{-1}=\sum_{n}\bar{Z}_{n}\;,
\]
where the first few terms read 
\[
\bar{Z}_{0}=1\;,\qquad\bar{Z}_{1}=-Z_{1}\;,\qquad\bar{Z}_{2}=Z_{1}^{2}-Z_{2}\;.
\]
Putting these ingredients together, in a finite volume $L$ we obtain
\textit{ 
\begin{equation}
\langle\mathcal{O}(t,R)\rangle_{L}=\frac{1}{Z}\sum C_{kl}=\sum\tilde{D}_{kl}\;,\label{complete}
\end{equation}
}\textit{\emph{where analogously to }}to \cite{OnePointFunctions}
and \cite{EsslerKonik}, $\tilde{D}_{nm}$ is introduced as 
\begin{equation}
\tilde{D}_{kl}=\sum_{j}C_{k-j,l-j}\bar{Z}_{j}\;,\label{dnm}
\end{equation}
with the first few terms having the form 
\begin{align*}
\tilde{D}_{1l} & =C_{1l}-Z_{1}C_{0,l-1}\;, & l=1,2,\dots\;,\\
\tilde{D}_{2l} & =C_{2l}-Z_{1}C_{1,l-1}+(Z_{1}^{2}-Z_{2})C_{0,l-2}\;, & l=2,3,\dots\;.
\end{align*}
Since the $\tilde{D}_{kl}$ are of order $z^{k}\bar{z}^{l}$, they
must separately be well-defined as $L\to\infty$: 
\begin{equation}
D_{kl}=\lim_{L\to\infty}\tilde{D}_{kl}\;,\label{dnm-uj}
\end{equation}
and so the infinite volume limit can be written as 
\[
\langle\mathcal{O}(t,R)\rangle=\sum_{k,l}D_{kl}\;.
\]
For convenience and later use, we also introduce the quantities

\begin{equation}
\tilde{G}_{n}=\sum_{l=0}^{n}\tilde{D}_{n-l,l}\;,\label{eq:Gndef}
\end{equation}
whose infinite volume limit is denoted by $G_{n}$.

The expressions individual $C_{kl}$ contain finite volume form factors,
which in general are given by \cite{FinVolFF1and2} 
\begin{equation}
\begin{split}_{L}\langle\{I_{1},\dots,I_{k}|O|\{J_{1},\dots,J_{l}\}\rangle_{L}=\frac{F_{k+l}^{O}(\vartheta_{1}+i\pi,\dots,\vartheta_{k}+i\pi,\vartheta'_{1},\dots,\vartheta'_{l})}{\sqrt{\rho_{k}(\vartheta_{1},\dots,\vartheta_{k})\rho_{l}(\vartheta'_{1},\dots,\vartheta'_{l})}}+\mathcal{O}(e^{-\mu L})\;,\end{split}
\label{fftcsalap}
\end{equation}
where it is understood that the rapidities $\{\vartheta_{1},\dots,\vartheta_{k}\}$
and $\{\vartheta'_{1},\dots,\vartheta'_{l}\}$ are solutions to the
corresponding Bethe\textendash Yang equations with quantum numbers
$\{I_{n}\},\{J_{n}\}$. Formula (\ref{fftcsalap}) is valid whenever
there are no coinciding rapidities; otherwise a more complicated formula
taking into account disconnected contributions is necessary. In this
paper we are interested in contributions to one-particle oscillations,
for which coinciding rapidities cannot occur, the numbers of particles
in the two multi-particle states differ by an odd number which excludes
the two possible cases with disconnected terms (cf. \cite{FinVolFF1and2}).

Note that the equality (\ref{fftcsalap}) is valid up to a suitably
chosen phase factor which can be changed by redefining the phases
of the finite volume eigenstates $|\{I_{1},\dots,I_{n}\}\rangle_{L}$.
This includes also the fact that the ordering of the particles is
not determined by first principles and any exchange leads to an $S$-matrix
factor according to (\ref{eq:exchangeaxiom}). It is clear that all
such ambiguities cancel in the expectation value (\ref{tervszerint-2});
however, for a practical calculation one must fix the phases of the
multi-particle contributions to the matrix elements consistently.
Here we make use of the consistent prescription introduced in \cite{OnePointFunctions}:
any time the amplitudes $K(\vartheta_{i})$ and $K^{*}(\vartheta_{i})$
appear with some $\vartheta_{i}$, the explicit order of the rapidities
substituted into the relevant form factor is given by $(-\vartheta_{i},\vartheta_{i})$
and $(\vartheta_{i}+i\pi,-\vartheta_{i}+i\pi)$, respectively. Exchanging
any two pairs of rapidities does not make any difference, therefore
the phase of the form factors is completely fixed by the above rule.
Note that the presence of zero-momentum particles does not produce
any additional ambiguities.

\subsection{The singularity of K revisited}

Now we are ready to complete our arguments why the pair-amplitude
must be singular in integrable quenches with one-particle coupling.
Recall that in the boundary problem \cite{OnePointFunctions} the
ordering of the terms was performed according to powers of $e^{-m(\mathcal{R}-x)}$
and $e^{-mx}$, which resulted in the same expressions for the boundary
version of $\tilde{D}_{kl}$ that we obtained in (\ref{dnm}) with
the expansion parameters $z=e^{-m(R/2+it)}$ and $\bar{z}=e^{-m(R/2-it)}$.
In the boundary problem, the existence of the infinite volume limit
(\ref{dnm-uj}) and eventually $\langle\mathcal{O}(x)\rangle^{B}$
requires the presence of the singularity 
\[
K_{B}(\vartheta)\sim-\frac{i}{2}\frac{g_{B}^{2}}{\vartheta}\;,
\]
therefore 
\[
K(\vartheta)\sim-\frac{i}{2}\frac{g^{2}}{\vartheta}
\]
must hold for the quench problem as well. The easiest way to see that
is to consider the one-particle contribution $-Z_{1}$ which behaves
as $mL$. To make $C_{12}-Z_{1}C_{01}$ finite in the infinite volume
limit, $C_{12}$ must have a similar volume-dependence, which is ensured
by the singularity of $K$ involved in $C_{12}$.

At the end of the calculations the regulator $R$ is sent to zero.
The contribution $G_{n}$ (\ref{eq:Gndef}) is of order $|z|^{n}=(e^{-mR/2})^{n}$
and just as in \cite{OnePointFunctions} it turns out that the coefficient
of the largest power of the $mL$ term is always of order $g^{n}$
. Focussing on the behaviour of the singular term, the small parameter
of the linked-cluster calculation can be identified with $g$. As
the singularity of the $K$ is of order $g^{2}$ one can formally
treat $K$ as a term of order $g^{2}$. This way of counting the orders
results in the same classification of contributions $\tilde{D}_{kl}$
(\ref{eq:Gndef}) that we obtained by considering $z$ and $\bar{z}$
as the expansion parameters.

This counting of orders is clearest for a perturbative quench corresponding
to changing the Hamiltonian as 
\[
\delta H=\lambda\int dx\Psi(x)\;,
\]
where $\Psi$ is a purely odd operator (i.e. whose form factors with
an even number of particles in the pre-quench system is zero). Using
perturbation theory to compute the overlaps following \cite{DelfinoOscOlder,DelfinoOscNewer}
one obtains that the one-particle coupling is of order $\lambda$,
while the pair amplitude $K$ is always of order $\lambda^{2}.$

When the perturbing operator has even matrix elements as well, the
pair-amplitude can also possess a $\lambda$ order term. Whereas this
term is not singular due to the regular behaviour of $F(i\pi+\vartheta,i\pi-\vartheta)$
at $\vartheta=0$, at order $\lambda^{2}$ a singular term similar
to the one found in Section \ref{sec:SingularK} is always present.
Therefore the presence of the zero-rapidity pole singularity of $K$
is generic.

\subsection{Contributions up to $4^{th}$ order: analytic continuation of the
boundary result \label{subsec:TimeDepG4}}

In this section we present all the terms up to fourth order using
the Euclidean quantities computed in \cite{OnePointFunctions} and
continuing them to real time. The Euclidean one-point function 
\begin{equation}
\langle\mathcal{O}(x)\rangle^{B}=\frac{\langle B|\ e^{-Hx}\mathcal{O}(0)\ e^{-H(R-x)}\ |B\rangle}{\langle B|e^{-HR}|B\rangle}=\sum_{k,l}D_{kl}\label{ezkell-1}
\end{equation}
was computed up to contributions $D_{kl}$ with $k+l\le4$ which are
collected in Appendix \ref{sub:Eucl1point}.

For the analytic continuation, we apply the $R\rightarrow0$ and $x\rightarrow-it$
substitutions together with $K_{B}\rightarrow K$ and $g_{B}\rightarrow g$
which give \begin{subequations} \label{TermsUptog4} 
\begin{equation}
\begin{split}G_{0}:\quad & \langle0|\mathcal{O}|0\rangle\quad,\\
G_{1}:\quad & g\,\Re eF_{1}^{\mathcal{O}}e^{-imt},\\
G_{2}:\quad & \frac{g^{2}}{4}F_{2}^{\mathcal{O}}(i\pi,0)+\Re e\int_{-\infty}^{\infty}\frac{d\theta}{2\pi}K(\theta)F_{2}^{\mathcal{O}}(-\vartheta,\vartheta)e^{-imt2\cosh\vartheta}\;,\\
G_{3}:\quad & \frac{g}{2}\Re e\int_{-\infty}^{\infty}\frac{d\vartheta}{2\pi}K(\vartheta)F_{3}^{\mathcal{O}}(-\vartheta,\vartheta,0)e^{-imt\left(2\cosh\vartheta+1\right)}\;,\\
 & +\frac{g}{2}\Re e\int_{-\infty}^{\infty}\frac{d\vartheta}{2\pi}\left\{ K(\vartheta)F_{3}^{\mathcal{O}}(i\pi,-\vartheta,\vartheta)e^{-imt\left(2\cosh\vartheta-1\right)}-2g^{2}\frac{\cosh\vartheta}{\sinh^{2}\vartheta}F_{1}^{\mathcal{O}}e^{-imt}\right\} \\
 & +2g^{3}\varphi(0)\Re eF_{1}^{\mathcal{O}}e^{-imt}\;,
\end{split}
\label{TermsUptog4_1}
\end{equation}
\begin{equation}
\begin{split}G_{4}:\quad & \frac{1}{4}\Re e\int_{-\infty}^{\infty}\frac{d\vartheta_{1}}{2\pi}\frac{d\vartheta_{2}}{2\pi}K(\vartheta_{1})K(\vartheta_{2})F_{4}^{\mathcal{O}}(-\vartheta_{1},\vartheta_{1},-\vartheta_{2},\vartheta_{2})e^{-imt\left(2\cosh\vartheta_{1}+2\cosh\vartheta_{2}\right)}\\
 & +\frac{g^{2}}{4}\Re e\int_{-\infty}^{\infty}\frac{d\vartheta}{2\pi}K(\vartheta)F_{4}^{\mathcal{O}}(-\vartheta+i\pi,\vartheta+i\pi,i\pi,0)e^{-imt2\cosh\vartheta}\\
 & +\frac{1}{4}\int_{-\infty}^{\infty}\frac{d\vartheta_{1}}{2\pi}\frac{d\vartheta_{2}}{2\pi}K(\vartheta_{1})K(\vartheta_{2})F_{4}^{\mathcal{O}}(i\pi-\vartheta_{1}i\pi+\vartheta_{1},-\vartheta_{2},\vartheta_{2})e^{imt2\left(\cosh\vartheta_{1}-\cosh\vartheta_{2}\right)}\\
 & +F_{2}^{\mathcal{O}}(i\pi,0)\int_{-\infty}^{\infty}\frac{d\vartheta}{2\pi}\left\{ |K(\vartheta)|^{2}-\frac{g^{4}\cosh\vartheta}{4\sinh^{2}\vartheta}\right\} \\
 & +\frac{g^{4}}{8}F_{2}^{\mathcal{O}}(i\pi,0)\varphi(0)\quad,
\end{split}
\label{TermsUptog4_2}
\end{equation}
where $F_{2,s}^{\mathcal{O}}=F_{2}^{\mathcal{O}}(i\pi,0)$ and 
\begin{equation}
\varphi(\vartheta)=-i\frac{\partial\log S(\vartheta)}{\partial\vartheta}\;.
\end{equation}
\end{subequations}

Note that these integrals remain well-defined even when there is a
pole in the amplitude $K(\vartheta)$ at $\vartheta=0$ because the
form factors possess a zero $\vartheta_{i}=0$ as a consequence of
the exchange axiom (\ref{eq:exchangeaxiom}) and the general property
$S(0)=-1$. Concerning the large rapidity behaviour, normalisability
of the initial state requires that $K$ tends to zero fast enough
for large $\vartheta$, again ensuring the existence of the integrals.

We now turn to analysing the time dependence originating from (\ref{TermsUptog4}).
Due to the oscillatory integrands it is convenient to apply the stationary
phase approximation (SPA) briefly discussed in Appendix \ref{subsec:SPA}.
Both SPA and direct analysis leads to the following type of terms
expected from (\ref{TermsUptog4}): 
\begin{equation}
e^{-inmt}t^{\alpha}\:,
\end{equation}
where $\alpha$ is either integer or half integer and $n$ is an integer.
For terms $D_{kl}$ the lower bound of the oscillation frequency is
always $nm=(k-l)m$.

As the main objective of this paper is to study the time dependence
of one-particle oscillations, we concentrate here on terms with $n=1$
but we will also briefly comment on the time dependence of the non-oscillatory
part of $\langle\mathcal{O}\rangle$. The non-oscillatory parts include
the static contributions 
\begin{equation}
\begin{split}G_{0}:\quad & \langle0|\mathcal{O}|0\rangle\;,\\
G_{2}:\quad & \frac{g^{2}}{4}F_{2}^{\mathcal{O}}(i\pi,0)\;,\\
G_{4}:\quad & F_{2}^{\mathcal{O}}(i\pi,0)\int_{-\infty}^{\infty}\frac{d\vartheta}{2\pi}\left\{ |K(\vartheta)|^{2}-\frac{g^{4}\cosh\vartheta}{4\sinh^{2}\vartheta}\right\} \\
+ & \frac{g^{4}}{8}F_{2}^{\mathcal{O}}(i\pi,0)\varphi(0)\;,
\end{split}
\end{equation}
whereas for the only time dependent integral, 
\begin{equation}
\frac{1}{4}\int_{-\infty}^{\infty}\frac{d\vartheta_{1}}{2\pi}\frac{d\vartheta_{2}}{2\pi}K(\vartheta_{1})K(\vartheta_{2})F_{4}^{\mathcal{O}}(i\pi-\vartheta_{1}i\pi+\vartheta_{1},-\vartheta_{2},\vartheta_{2})e^{imt2\left(\cosh\vartheta_{1}-\cosh\vartheta_{2}\right)}\;,
\end{equation}
the SPA (\ref{eq:StacPhase}) can be applied, yielding 
\begin{equation}
\ \frac{\mathcal{C}}{16\pi mt}\;,\qquad1\ll mt\:,
\end{equation}
where 
\begin{equation}
\mathcal{C}=\lim_{\vartheta_{2}\rightarrow0}\lim_{\vartheta_{1}\rightarrow0}K(\vartheta_{1})K(\vartheta_{2})F_{4}^{\mathcal{O}}(i\pi-\vartheta_{1}i\pi+\vartheta_{1},-\vartheta_{2},\vartheta_{2})\label{eq:Cexpr}
\end{equation}
which is finite.

For terms with one-particle oscillation, one finds 
\begin{equation}
\begin{split}G_{1}:\quad & g\Re e\,F_{1}^{\mathcal{O}}e^{-imt}\,,\\
G_{3}:\quad & 2g^{3}\varphi(0)\Re e\,F_{1}^{\mathcal{O}}e^{-imt}\\
+ & \frac{g}{2}\Re e\,\int_{-\infty}^{\infty}\frac{d\vartheta}{2\pi}\left\{ K(\vartheta)F_{3}^{\mathcal{O}}(i\pi,-\vartheta,\vartheta)e^{-imt\left(2\cosh\vartheta-1\right)}-2g^{2}\frac{\cosh\vartheta}{\sinh^{2}\vartheta}F_{1}^{\mathcal{O}}e^{-imt}\right\} \,.
\end{split}
\end{equation}
For the last term, SPA cannot be applied directly, therefore we shift
the contour off the real axis where (as shown in \cite{OnePointFunctions})
the contribution from the term 
\[
\frac{\cosh\vartheta}{\sinh^{2}\vartheta}
\]
vanishes and reintroduce the regulator $R$ . We rewrite the resulting
expression using (\ref{DistCoshSinh2}) after which the SPA (\ref{eq:StacPhase})
can be applied, and finally perform the $R\rightarrow0$ limit. The
result is 
\begin{equation}
\frac{g^{3}F_{1}^{\mathcal{O}}\left(\varphi^{2}(0)-2/3\right)}{2\sqrt{4\pi mt}}\Re e\,e^{-imt}e^{-i\pi/4}+g^{3}\sqrt{\frac{mt}{\pi}}\Re e\,F_{1}^{\mathcal{O}}e^{-imt}\frac{-\sqrt{2}-\sqrt{2}i}{2}\,,\qquad mt\gg1\,.\label{TimeDepg3}
\end{equation}
For a more detailed derivation, the interested reader is referred
to in Appendix \ref{sec:D12App}. While the first term has the standard
$\sim1/\sqrt{t}$ time dependence, the second one behaves as $\sim\sqrt{t}$
for large time. We return to this peculiar finding in Sec. \ref{sub:paramres}.

\subsection{Leading order time dependence from $G_{5}$ }

The contributions involving five particles are $D_{05}$, $D_{14}$,
$D_{23}$ and their complex conjugates $D_{50}$,$D_{41}$, $D_{32}$
. Based on (\ref{dnm}), the expressions to evaluate are

\begin{equation}
\begin{split}D_{05}= & \lim_{L\rightarrow\infty}C_{05}\\
D_{14}= & \lim_{L\rightarrow\infty}C_{14}-Z_{1}C_{03}\\
D_{23}= & \lim_{L\rightarrow\infty}C_{23}-Z_{1}C_{12}-(Z_{2}-Z_{1}^{2})C_{01}\;.
\end{split}
\end{equation}
The calculation of these expressions is very tedious; even in Appendices
\ref{sec:D23Res}, \ref{sec:CInt} and \ref{sec:EvaluatingKernel}
where the details of the calculations are presented, we focused only
on leading secular contributions to the one-particle oscillations,
i.e. terms of the form $e^{-imt}t^{\alpha}$ with the highest power
$\alpha$. As one-particle oscillations originate exclusively from
$D_{kl}$ with $|k-l|=1$, we can focus on $D_{23}$ and its conjugate
$D_{32}$.

The leading order secular terms have two origins: a residue contribution
from encircling the poles of the form factors $F_{5}(i\pi+\vartheta_{1},i\pi-\vartheta_{1},-\vartheta_{2},\vartheta_{2},0)$
when $\vartheta_{1}\approx\vartheta_{2}$ and a contribution from
these poles when a contour integral is performed with an integration
contour just above the real axis.

Concerning the former term, the explicit expression reads

\begin{equation}
D_{23}^{Res}(t)=\frac{g}{2}F_{1}^{\mathcal{O}}e^{-imt}\left(imt\right)\int_{-\infty}^{\infty}\frac{d\vartheta}{2\pi}|K(\vartheta)|^{2}\Im\text{m}S(\vartheta)\sinh\vartheta\;,\label{D23Res}
\end{equation}
whose derivation can be found in Appendix \ref{sec:D23Res}. Unlike
time dependent terms discussed so far, deriving (\ref{D23Res}) involves
no SPA, hence it is valid also for small times. Note that the coefficient
of the oscillatory factor $e^{-imt}$ is purely imaginary and linear
in time.

For the other term denoted by $D_{23}^{CInt}(t)$, the explicit formula
reads

\begin{equation}
\begin{split}D_{23}^{CInt}(t)= & \frac{g}{2}F_{1}^{\mathcal{O}}e^{-imt}\left(-imt\right)\int_{-\infty}^{\infty}\frac{d\vartheta}{2\pi}|K(\vartheta)|^{2}\tanh(\vartheta_{1})\left(Ker_{a}(\vartheta,t)\right)^{'}\\
+ & \frac{g}{2}F_{1}^{\mathcal{O}}e^{-imt}\left(-imt\right)\int_{-\infty}^{\infty}\frac{d\vartheta}{2\pi}|K(\vartheta)|^{2}Ker(\vartheta,t)\\
+ & \frac{g}{2}F_{1}e^{-imt}\int_{-\infty}^{\infty}\frac{d\vartheta}{2\pi}\frac{1}{4\sinh\vartheta}\frac{d}{d\vartheta}\left\{ \vphantom{\frac{OOO}{OOO}}|K(\vartheta)|^{2}\Re e\,Ker(\vartheta,t,0)\times\right.\\
 & \left.\qquad\times\Re e\,\left(\frac{F_{5}^{\varepsilon}(\vartheta)}{F_{1}\Omega(\vartheta)}+\frac{K'(\vartheta)}{K(\vartheta)}-\frac{F_{5}^{\varepsilon}(-\vartheta)}{F_{1}\Omega(\vartheta)}-\frac{K'(-\vartheta)}{K(-\vartheta)}\right)\tanh\vartheta\right\} \\
 & +\mathcal{O}(\sqrt{t})\quad,
\end{split}
\label{D23CInt}
\end{equation}
where

\begin{equation}
\begin{split}Ker^{a}(\vartheta_{1},t)= & \lim_{R\rightarrow0}e^{imt}\Omega(\vartheta_{1})\int_{-\infty}^{\infty}\frac{d\vartheta_{2}}{2\pi}\frac{\left[h(\vartheta_{1}|\vartheta_{2},\{0\})_{R}\left(\sinh\vartheta_{2}-\frac{\sinh\vartheta_{1}}{\cosh\vartheta_{2}-\vartheta_{1}}\right)\right]}{\sinh\vartheta_{2}-\vartheta_{1}}\;,\end{split}
\label{KerA}
\end{equation}
and

\begin{equation}
Ker(\vartheta_{1},t)=e^{imt}\Omega(\vartheta_{1})\int_{-\infty}^{\infty}\frac{d\vartheta_{2}}{2\pi}\frac{\left[h(\vartheta_{1}|\vartheta_{2},\{0\})-h(\vartheta_{1}|\vartheta_{1},\{0\})\right]\sinh\vartheta_{1}}{\sinh(\vartheta_{2}-\vartheta_{1})\cosh(\vartheta_{2}-\vartheta_{1})}\label{KerB}
\end{equation}
with 
\begin{align}
\Omega(\vartheta)= & \left(1-S(-\vartheta)\right)\left(1-S(\vartheta)\right)\;,\\
h(\vartheta_{1}|\vartheta_{2},\{0\})_{R} & =e^{imt(2\cosh\vartheta_{1}-2\cosh\vartheta_{2}-1)}e^{-mR/2(2\cosh\vartheta_{1}+2\cosh\vartheta_{2}+1)}\;,\\
h(\vartheta_{1}|\vartheta_{2},\{0\}) & =h(\vartheta_{1}|\vartheta_{2},\{0\})_{R=0}\;.
\end{align}
Their derivation can be found in Appendices \ref{sec:CInt} and \ref{sec:EvaluatingKernel}.
It is shown in Appendix \ref{sec:EvaluatingKernel} that in the large
time limit $mt\gg1$ the integral kernels behave as

\begin{equation}
\begin{split}Ker_{stac}^{a}(\vartheta,t)= & e^{imt}\Omega(\vartheta)\frac{1}{\sqrt{4\pi mt}}\frac{e^{2imt(\cosh\vartheta-1)}e^{-i\pi/4}}{\cosh\vartheta}\end{split}
,\quad1\ll mt\:,\label{KerAExplicit}
\end{equation}
and

\begin{equation}
\begin{split}Ker_{stac}(\vartheta,t)= & \frac{\sqrt{2(\cosh\vartheta-1)}}{\cosh\vartheta}\Omega(\vartheta)\\
\times & \left\{ \frac{1}{2}\left(F_{S}\left(\sqrt{\frac{4mt(\cosh(\vartheta)-1)}{\pi}}\right)-F_{C}\left(\sqrt{\frac{4mt(\cosh(\vartheta)-1)}{\pi}}\right)\right)\right.\\
- & \left.\frac{1}{2}i\left(F_{C}\left(\sqrt{\frac{4mt(\cosh(\vartheta)-1)}{\pi}}\right)+F_{S}\left(\sqrt{\frac{4mt(\cosh(\vartheta)-1)}{\pi}}\right)\right)\right.\\
+ & \left.i\left(\frac{1}{2}-\frac{\cosh\vartheta\sqrt{\sinh^{2}\vartheta}}{2\sqrt{2(\cosh(\vartheta)-1)}}\right)\right\} ,\quad1\ll mt\:,
\end{split}
\label{KerBExplicit}
\end{equation}
where $F_{S}$ and $F_{C}$ are the Fresnel sine and cosine functions,
respectively. This leads to the final result 
\begin{equation}
D_{23}(t)=\frac{g}{2}F_{1}^{\mathcal{O}}e^{-imt}mt\left(\frac{g^{4}}{4}\left(-\frac{\log(mt)}{\pi}\right)+\gamma_{1}+i\gamma_{2}\right)+\mathcal{O}(e^{-imt}\sqrt{t})\label{D23(t)}
\end{equation}
with

\begin{equation}
\begin{split}\gamma_{1}= & \mathcal{K}+\frac{g^{4}}{4}\frac{3}{\pi}\;,\\
\gamma_{2}= & \frac{g^{4}}{8}+\int_{-\infty}^{\infty}\frac{d\vartheta}{2\pi}|K(\vartheta)|^{2}\Im m\,S(\vartheta)\sinh\vartheta\;,
\end{split}
\label{Gammas}
\end{equation}
where

\begin{equation}
\begin{split}\mathcal{K}=\lim_{t\rightarrow\infty} & \left\{ -\frac{1}{2}\int_{-\infty}^{\infty}\frac{d\vartheta}{2\pi}\Omega(\vartheta)|K(\vartheta)|^{2}\left[\frac{\sqrt{2(\cosh\vartheta-1)}}{\cosh\vartheta}\times\right.\right.\\
 & \left.\left(F_{C}\left(\sqrt{\frac{4mt(\cosh(\vartheta)-1)}{\pi}}\right)+F_{S}\left(\sqrt{\frac{4mt(\cosh(\vartheta)-1)}{\pi}}\right)-1\right)+\sqrt{\sinh^{2}\vartheta}\vphantom{F_{C}\left(\sqrt{\frac{4t(\cosh(\vartheta)-1)}{\pi}}\right)}\right]\\
 & \left.+\frac{g^{4}}{4}\left(\frac{\log(mt)}{\pi}\right)\right\} \;.
\end{split}
\label{CaligraphicK}
\end{equation}

\section{Discussion of the results \label{subsec:Interpretation}}

Before discussing the time evolution of the one-point function, we
collect the leading order time-dependent contributions for the non-oscillatory
and one-particle-oscillatory part of $\langle\mathcal{O}\rangle$
from each term $D_{kl}$ we considered in the previous section. In
the long-time limit $1\ll mt$, these are

\begin{equation}
\begin{split}G_{0}:\quad & \langle0|\mathcal{O}|0\rangle\,,\\
G_{2}:\quad & \frac{g^{2}}{4}F_{2}^{\mathcal{O}}(i\pi,0)\,,\\
G_{4}:\quad & F_{2}^{\mathcal{O}}(i\pi,0)\int_{-\infty}^{\infty}\frac{d\vartheta}{2\pi}\left\{ |K(\vartheta)|^{2}-\frac{g^{4}\cosh\vartheta}{4\sinh^{2}\vartheta}\right\} +\frac{g^{4}}{8}F_{2}^{\mathcal{O}}(i\pi,0)\varphi(0)\\
+ & \frac{\mathcal{C}}{16\pi mt}\;,
\end{split}
\end{equation}
and

\begin{equation}
{\begin{split}{\normalcolor G_{1}:\quad} & {\normalcolor g\Re e\,F_{1}^{\mathcal{O}}e^{-imt}\;,}\\
{\normalcolor G_{3}:\quad} & {\normalcolor 2g^{3}\varphi(0)\Re e\,F_{1}^{\mathcal{O}}e^{-imt}}\\
{\normalcolor +} & {\normalcolor g^{3}\sqrt{\frac{mt}{\pi}}\Re e\,F_{1}^{\mathcal{O}}e^{-imt}\frac{-\sqrt{2}-\sqrt{2}i}{2}}\\
{\normalcolor +} & {\normalcolor \mathcal{O}(1/\sqrt{t})\;,}\\
{\normalcolor G_{5}:\quad} & {\normalcolor g\Re e\,F_{1}^{\mathcal{O}}e^{-imt}mt\left(\frac{g^{4}}{4}\left(-\frac{\log(mt)}{\pi}\right)+\gamma_{1}+i\gamma_{2}\right)}\\
{\normalcolor +} & {\normalcolor \mathcal{O}(\sqrt{t})}\;,
\end{split}
}
\end{equation}
where $\mathcal{C}$ is given in (\ref{eq:Cexpr}) and $\gamma_{1,2}$
in (\ref{Gammas}).

In the preceding subsections we found that the long time asymptotics
of the leading order contributions to the one-particle oscillations
contain, besides the original oscillation $e^{-imt}$ from $G_{1}$,
two new types of terms: one with time dependence $\sqrt{t}e^{-imt}$
from $G_{3}$ and terms of time dependence $te^{-imt}$ and $t\ln te^{-imt}$
from $G_{5}$.

Since these are secular terms growing for large $t,$ it is necessary
to sum up higher order contributions coming from $G_{2n+1}$. Computing
terms $G_{7}$ and higher is extremely tedious and has not yet been
performed, therefore in this work we can only present a limited discussion
of their resummation based on insights gained from earlier works \cite{SchurichtEssler,BertiniSineG,SchurichtCubero}.

In addition to these works, one could also try to compare with the
$FM\rightarrow PM$ quench in the Ising model considered in Subsection
(\ref{subsec:Quench-in-the}); however, we show that this is unfortunately
not possible. In Appendix \ref{sec:IsingAppendix} it is shown that
in the continuum limit the time evolution of the magnetisation is
\begin{equation}
\langle\sigma^{x}(x,t)\rangle=\bar{\sigma}\left(\frac{1}{2}\right)^{\frac{1}{4}}\left(\frac{M_{0}}{M}\right)^{\frac{1}{8}}\left[\cos\left(\sqrt{M^{2}+MM_{0}}t+\alpha'\right)+...\right]e^{-t/\tau}\;,\label{IsingSigmaTimeDep}
\end{equation}
where the relaxation time is given by 
\begin{equation}
\tau^{-1}=\frac{1}{\pi}\left\{ \sqrt{M^{2}+M\,M_{0}}\ln\left(\frac{\sqrt{M^{2}+M\,M_{0}}+M}{\sqrt{M^{2}+M\,M_{0}}-M}\right)-\frac{1}{2}\sqrt{M^{2}-M_{0}^{2}}\ln\left(\frac{M+\sqrt{M^{2}-M_{0}^{2}}}{M-\sqrt{M^{2}-M_{0}^{2}}}\right)\right\} \;\label{IsingZeta}
\end{equation}
for $M>M_{0}$ and

\begin{align}
\tau^{-1} & =\frac{1}{\pi}\left\{ \sqrt{M^{2}+M\,M_{0}}\ln\left(\frac{\sqrt{M^{2}+M\,M_{0}}+M}{\sqrt{M^{2}+M\,M_{0}}-M}\right)-\sqrt{M_{0}^{2}-M^{2}}\left[\tan^{-1}\left(\frac{M}{\sqrt{M_{0}^{2}-M^{2}}}\right)-\frac{\pi}{2}\right]\right\} \label{IsingZeta1}
\end{align}
for $M<M_{0}$.{} At first sight, since the $K$ function is known
for this particular quench, the expansion of (\ref{IsingSigmaTimeDep})
can be matched with the form factor expansion evaluated in our work.
However, note that (\ref{IsingSigmaTimeDep}) and (\ref{IsingZeta})
are non-analytic functions of the pre-quench mass $M_{0}$ around
the origin, and therefore are also non-analytic in $g$ around $g=0$
due to the relation (\ref{eq:Ising_gM0}). This is not surprising
since a quench across a quantum critical point is expected to be a
large quench with possibly non-analytic behaviour, and therefore the
form factor expansion is not expected to be valid at all. This situation
is in marked contrast with the phase quenches in the sine\textendash Gordon
model considered in Subsection (\ref{subsec:PhaseQuench}) where the
shift $\delta/\beta$ can play the role of a small parameter.

\subsection{Connection with previous results and discussion of possible resummation
of $G_{4n+1}$}

First we consider terms $G_{4n+1}$. Contributions to one-particle
oscillations originate from the pieces $D_{2n,2n+1}$. In \cite{SchurichtCubero}
these terms served as the sole source of secular terms which were
shown to sum up to an exponential function to order $t^{2}$, i.e.
the result including the two leading order corrections had the form
\[
\Re e\,\frac{g}{2}F_{1}e^{-imt}\left(1-\frac{t}{\tau}+\frac{1}{2}\frac{t^{2}}{\tau^{2}}+\dots\right)
\]
which is the expansion of 
\[
\Re e\,\frac{g}{2}F_{1}e^{-imt}e^{-t/\tau}\;,
\]
where 
\begin{equation}
\tau^{-1}=\frac{m}{\pi}\int_{0}^{\infty}d\vartheta\left(1-S(\vartheta)\right)|K(\vartheta)|^{2}\sinh\vartheta\;.\label{eq:tauschuricht}
\end{equation}
The real part of the above integral is the relaxation time, while
the imaginary part is a frequency shift. It is easy to see that sending
$g\rightarrow0$ in our expressions (\ref{D23(t)}), which is equivalent
to removing the singularity of $K$, reproduces the result for $\tau$
above as the integrand is non-singular and

\begin{equation}
\lim_{x\rightarrow\infty}F_{C}(x)=\lim_{x\rightarrow\infty}F_{S}=\frac{1}{2}\;.
\end{equation}
The origin of (\ref{eq:tauschuricht}) is essentially the kinematic
singularity of the form factors. In our calculation we have an additional
ingredient, namely the singularity of $K$ which gives some new contributions
according to (\ref{D23(t)}). Assuming an exponentiation similar to
that observed in \cite{SchurichtEssler,BertiniSineG,SchurichtCubero},
the leading order time dependence from $G_{4n+1}$ is

\begin{equation}
D_{2n,2n+1}(t)=\frac{g}{2}F_{1}e^{-imt}\frac{\left(mt\right)^{n}}{n!}\left(-\frac{g^{4}}{4}\frac{\log(mt)}{\pi}+\gamma_{1}+i\gamma_{2}\right)^{n}+...\;,
\end{equation}
which can be resummed into

\begin{equation}
\frac{g}{2}F_{1}e^{-imt}\exp\left[mt\left(-\frac{g^{4}}{4}\frac{\log(mt)}{\pi}+\gamma_{1}+i\gamma_{2}\right)\right]\;.
\end{equation}
Therefore, besides the frequency shift, the relaxation for late time
is naively expected to be super-exponential with a dependence of the
form $e^{-t\ln t}$. This also means, that under the assumption of
exponentialisation, $\tau^{-1}$ in \eqref{eq:tauschuricht} is to
be replaced with $m\left(\frac{g^{4}}{4}\frac{\log(mt)}{\pi}-\gamma_{1}-i\gamma_{2}\right)$
in the exponential function. However, this cannot be concluded safely
without computing at least $D_{45}(t)$ and checking whether one obtains
the correct combinatorial coefficients for the terms involving higher
powers of $-\frac{g^{4}}{4}\frac{\log(mt)}{\pi}mt$ which is the condition
for exponentialisation as established in \cite{SchurichtEssler,BertiniSineG,SchurichtCubero}.
Based on analogy with \cite{SchurichtCubero}, one can argue that
terms containing $\gamma_{1,2}$ exponentialise and their resummation
leads to a relaxation rate and a frequency shift; however, the fate
of the logarithmic term cannot be decided without examining higher
order contributions, whose straightforward evaluation is extremely
complicated. Assuming that the relaxation is of the usual exponentially
decaying form (i.e. the logarithmic part does not exponentialise)
also means that it is not clear at the moment what part of the real
terms (involving $\gamma_{1}$) exponentialises and determines the
relaxation rate.

\subsection{Multiple species}

It is easy to generalise our results to the case of multiple species
following the reasoning in \cite{SchurichtCubero} which studied relaxation
in the attractive regime of the sine\textendash Gordon model for operators
semi-local to soliton excitations (and consequently local with respect
to the soliton-antisoliton bound states, i.e. the breathers $B_{n}$
). In the following we write down the result for semi-local operators

\begin{equation}
{\color{red}\begin{split}{\normalcolor \langle\mathcal{O}(t)\rangle=} & {\normalcolor \left(\langle0|\mathcal{O}|0\rangle+\sum_{j}\frac{g_{j}^{2}}{4}F_{jj}^{\mathcal{O}}(i\pi,0)\right)\left[1-\frac{t}{\tau_{s}}+...\right]}\\
{\normalcolor +} & {\normalcolor \sum_{j}g_{j}\Re e\,\left\{ F_{j}^{\mathcal{O}}e^{-im_{j}t}\left[1-\frac{t}{\tau_{sj}}-\sum_{k}\frac{t}{\tau_{jk}}+...\right]\right\} }\\
{\normalcolor +} & {\normalcolor \sum_{j\neq k}\frac{g_{j}g_{k}}{2}\Re e\,\left\{ F_{jk}^{\mathcal{O}}(i\pi,0)e^{i(m_{j}-m_{k})t}\left[1-\frac{t}{\tau_{jks}}-\sum_{l}\frac{t}{\tau_{jkl}}+...\right]\right\} \;,}
\end{split}
}\label{TimeDepMoreSpecies}
\end{equation}
where

\begin{equation}
\begin{split}\tau_{s}^{-1}= & \frac{2m_{s}}{\pi}\int_{0}^{\infty}d\vartheta|K_{s\bar{s}}(\vartheta)|^{2}\sinh\vartheta\;,\\
\tau_{sj}^{-1}= & \frac{m_{s}}{\pi}\int_{0}^{\infty}d\vartheta\left(1+S_{sj}(\vartheta)\right)|K_{s\bar{s}}(\vartheta)|^{2}\sinh\vartheta\;,\\
\tau_{jk}^{-1}= & \frac{m_{j}}{\pi}\int_{0}^{\infty}d\vartheta\left(1-S_{jk}(\vartheta)\right)|K_{j}(\vartheta)|^{2}\sinh\vartheta\;,\qquad j\neq k\;,\\
\tau_{jks}^{-1}= & \frac{m_{s}}{\pi}\int_{0}^{\infty}d\vartheta\left(1+S_{sj}(\vartheta)S_{sk}(-\vartheta)\right)|K_{s\bar{s}}(\vartheta)|^{2}\sinh\vartheta\;,\\
\tau_{jkl}^{-1}= & \frac{m_{l}}{\pi}\int_{0}^{\infty}d\vartheta\left(1-S_{kl}(\vartheta)S_{jl}(-\vartheta)\right)|K_{l}(\vartheta)|^{2}\sinh\vartheta\;,\qquad j\neq l,k\neq l\;,
\end{split}
\label{Taus}
\end{equation}
and

\begin{equation}
\begin{split}\tau_{jj}^{-1}{\color{red}{\normalcolor (t)}}= & -m_{j}\left(-\frac{g_{j}^{4}}{4}\frac{\log(m_{j}t)}{\pi}+\gamma_{1}^{(j)}+i\gamma_{2}^{(j)}\right)\;,\\
\\
\tau_{lkl}^{-1}{\color{red}{\normalcolor (t)}}=\tau_{kll}^{-1}{\color{red}{\normalcolor (t)}}= & -m_{l}\left(-\frac{g_{l}^{4}}{4}\frac{\log(m_{l}t)}{\pi}+\gamma_{1}^{(llk)}+i\gamma_{2}^{(llk)}\right)\;,\\
\\
\end{split}
\label{NewGammas}
\end{equation}
where $j,k,l,o$ index breather excitations, and $\gamma_{1,2}^{(n)}$
are obtained by substituting $S$ in (\ref{Gammas}) with the $B_{n}-B_{n}$
scattering amplitude $S_{nn}$ and $g$ with the $B_{n}$ one-particle
coupling $g_{n}$. The expressions for $\gamma_{1}^{(llk)}$ and $\gamma_{2}^{(llk)}$
are obtained from (\ref{Gammas}), by replacing $g$ with $g_{j}$
and also

\begin{equation}
\Omega(\vartheta)=\left(1-S(\vartheta)\right)\left(1-S(-\vartheta)\right)\rightarrow\left(1-S_{lk}(\vartheta)S_{ll}(-\vartheta)\right)\left(1-S_{ll}(\vartheta)S_{lk}(-\vartheta)\right)
\end{equation}
in (\ref{KerA}) and (\ref{KerB}); in addition, in the residue term
$D_{23}^{Res}$ (\ref{D23Res}) it is necessary to replace

\begin{equation}
\Im m\,S(\vartheta)\rightarrow\frac{1}{2i}\left(S_{ll}(\vartheta)S_{lk}(-\vartheta)-S_{ll}(-\vartheta)S_{lk}(\vartheta)\right)\;.
\end{equation}
Since solitons can have no one-particle coupling in the quench due
to topological charge conservation, the amplitude $K_{s\bar{s}}$
is regular at the origin.

Note that (\ref{Taus}) are identical to the results in \cite{SchurichtCubero}.
It is easy to understand this from the fact the expressions given
in \cite{SchurichtCubero} for $\tau_{jk}^{-1}$ with $j\neq k$ and
for $\tau_{jkl}^{-1}$ with $j\neq l,k\neq l$, are regular even if
the $K_{j}$ are singular. The only terms to be revisited are $\tau_{jj}^{-1}$
and $\tau_{lkl}^{-1}$ where the naive application of (\ref{Taus})
results in divergent contributions and must be modified using our
calculations performed above.

Results for theories with multiple species with fully diagonal scattering
can be obtained by omitting the solitonic contributions from the above
results and replacing the breathers with the actual particle spectrum.

\subsection{Parametric resonance \label{sub:paramres}}

Finally, we turn to contribution (\ref{TimeDepg3}) from $G_{3}$
\begin{equation}
g^{3}\sqrt{\frac{mt}{\pi}}\Re e\,F_{1}^{\mathcal{O}}e^{-imt}\frac{-\sqrt{2}-\sqrt{2}i}{2}\qquad mt\gg1\:.\label{TimeDepg3-1}
\end{equation}
The origin of this term is the integral

\begin{equation}
\frac{g}{2}\Re e\int_{-\infty}^{\infty}\frac{d\vartheta}{2\pi}\left\{ K(\vartheta)F_{3}^{\mathcal{O}}(i\pi,-\vartheta,\vartheta)e^{-imt\left(2\cosh\vartheta-1\right)}-2g^{2}\frac{\cosh\vartheta}{\sinh^{2}\vartheta}F_{1}^{\mathcal{O}}e^{-imt}\right\} \;,
\end{equation}
which is derived from a finite volume contribution of the form

\begin{equation}
\sum_{I>0}\langle\{0\}|\mathcal{O}|\{-I,I\}\rangle_{L}N_{2}(\vartheta,L)K(\vartheta)e^{-imt\left(2\cosh\vartheta-1\right)}\;,\label{ParamResMatrix}
\end{equation}
where $\vartheta$ is determined by the quantisation rule 
\[
mL\sinh\vartheta+\delta(2\vartheta)=2\pi I\;.
\]
Let us recall the phenomenon of parametric resonance, for which the
simplest example is the Mathieu equation describing a system with
only one degree of freedom,

\begin{equation}
\frac{d^{2}x}{dt^{2}}+\left[\omega_{0}^{2}-2q\cos\omega_{p}t\right]x=0\;.\label{Mathieu}
\end{equation}
This equation has a region of instability in which the solution of
the equation oscillates with an exponentially growing amplitude. This
region is when the ratio $\omega_{p}/\omega_{0}$ is sufficiently
close to 2, where the width of the region of proportional to $q$.

It is clear that the term (\ref{ParamResMatrix}) couples the one-particle
mode with the two-particle modes and satisfies the condition for resonance
on the threshold; however, the interplay between the integration over
rapidity and the singularity of the form factor $F_{3}^{\mathcal{O}}(i\pi,-\vartheta,\vartheta)$
results in a growth of $\sqrt{t}$ instead of being exponential. This
cannot be the final story: since a quench pumps a finite energy density
in the system, the oscillations cannot grow without bound and therefore
higher terms $G_{4n+1}$ must modify this behaviour to keep the amplitude
bounded. Since at present we do not have control over these higher
order terms, we cannot predict the eventual fate of this class of
contributions.

A heuristic analogy can be drawn by noting that in the Mathieu equation
the driving oscillator is external, while in the quench the two-particle
modes are also dynamical and the total energy stored in the system
is conserved. A closer analogy for this dynamics is provided by the
differential equation describing coupled modes

\[
\frac{d^{2}x}{dt^{2}}+\left[\omega_{1}^{2}+qy\right]x=0\qquad\frac{d^{2}y}{dt^{2}}+\omega_{2}^{2}y+\frac{q}{2}x^{2}=0\,.
\]

\begin{figure}
\begin{centering}
\includegraphics[width=0.6\columnwidth]{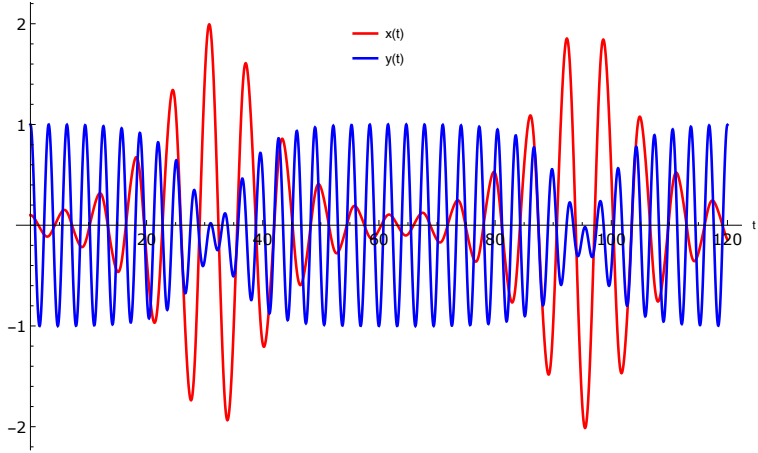} 
\par\end{centering}
\caption{\label{fig:An-example-of} An example of a parametric resonance in
a closed system with two modes. The frequencies are chosen $\omega_{1}=1$
and $\omega_{2}=2$, corresponding to parametric resonance, while
the mode coupling is $q=0.5$. Initial conditions for the particular
motion shown are $x(0)=0.1$, $y(0)=1$ and $\dot{x}(0)=\dot{y}(0)=0$.}
\end{figure}

The mode $x$ is parity odd and is the analogue of the one-particle
mode, while the mode $y$ is parity even and is the analogue of the
two-particle mode. Depending on the choice of the parameters, this
system has solutions in which the energy stored in the modes shows
an oscillatory behaviour in time. With the choice $\omega_{2}=2\omega_{1}$
the condition of parametric resonance is satisfied, but in contrast
to the Mathieu equation, the system has a total energy 
\[
H=\frac{1}{2}\left(\dot{x}^{2}+\omega_{1}^{2}x^{2}\right)+\frac{1}{2}\left(\dot{y}^{2}+\omega_{2}^{2}y^{2}\right)+\frac{q}{2}yx^{2}
\]
which is conserved since the driving and driven modes now form a closed
system. An example of a resonance solution is shown in Fig. \ref{fig:An-example-of};
notice the long plateau in the amplitude of the driving mode $y$
showing the non-linearity of this system. This is in contrast to linearly
coupled oscillators where the energy transfer would itself have a
harmonic dependence on time.

It is an interesting question whether in the full quench situation
such a non-trivial behaviour can be observed in the time evolution
for some choice of the parameters. Since the $\sqrt{t}$ term is of
order $g^{2}$ and the next secular contribution is of order $g^{4}$,
it is quite possible that for small enough quenches there is some
intermediate time window in which the parametric resonance dominates.
To study this requires the detailed analysis and resummation of higher
order terms which is left for further investigation.

\section{Conclusions \label{sec:Conclusions}}

In this work we studied integrable quenches with zero-momentum one-particle
states in the initial state and the subsequent time evolution of one-point
functions. We found that (similarly to the case of integrable boundary
conditions) in the presence of a non-vanishing one-particle coupling
the two-particle amplitude $K$ must have a singularity at the origin
$\vartheta=0$, 
\begin{equation}
K(\vartheta)\propto-i\frac{g^{2}}{2}\frac{1}{\vartheta}\;,\label{eq:pole_residue}
\end{equation}
where $g$ is the one-particle coupling determining the overlap of
the post-quench zero-momentum one-particle state with the initial
state. We gave a general argument based on the relation of the quench
time evolution to a boundary field theory problem discussed in \cite{OnePointFunctions}.
In addition, we presented two explicit examples, the quenches from
the paramagnetic to the ferromagnetic phase in the Ising QFT and phase
quenches in the sine\textendash Gordon model, in which the relation
between the one-particle coupling and the residue of $K$ at zero
rapidity could be established using explicit calculations and perturbation
theory. We also pointed out that the sine\textendash Gordon phase
quench is a particular member of a more general class of ``exponential
quenches'', where the same argument applies.

A general proof establishing the relation (\ref{eq:pole_residue})
also follows from our explicit computation of the time evolution using
finite volume as a regulator, which shows that the infinite volume
limit only exists when (\ref{eq:pole_residue}) holds. We stress that
for the results in \cite{SchurichtCubero} the presence of a pole
in $K$ is problematic as the expressions for the relaxation times
are not well-defined due to divergence of the integrals involved in
the formulas.

We then proceeded with the explicit computation of the time evolution
of one-point functions focusing on the one-particle oscillations with
time dependence $e^{-imt}$ . We used a modification of the linked
cluster expansion introduced in \cite{SchurichtEssler,BertiniSineG,SchurichtCubero},
where instead of rapidity space point splitting, we applied a finite
volume regulator first proposed in \cite{FinVolFF1and2}. The advantage
of this regulator is that it uses a physical parameter, i.e. the system
size $L$, and since the thermodynamic limit must be well-defined,
the computed one-point function must have a finite limit as $L\rightarrow\infty$.
The cancellation of terms containing positive powers of the volume
(resulting from kinematical singularities) provides an important consistency
check for the computation.

We found two important secular contributions to the one-particle oscillations.
The first one takes the following form at leading order 
\begin{equation}
\Re e\,gF_{1}^{\mathcal{O}}e^{-imt}mt\left(-\frac{g^{4}}{4}\frac{\log(mt)}{\pi}+\gamma_{1}+i\gamma_{2}\right)\;.
\end{equation}
The terms $\gamma_{1}$ and $\gamma_{2}$ are directly analogous to
the contributions found in \cite{SchurichtCubero}, with two essential
differences. First, the integrals expressing the coefficients $\gamma_{1,2}$
are now entirely well-defined. If there was no pole in $K(\vartheta)$
(which we argued to be impossible for $g\neq0$), these terms would
reduce to the expressions given in \cite{SchurichtCubero}. The second
difference is the presence of the logarithmic time-dependence for
$g\neq0$. In analogy to \cite{SchurichtCubero} it is expected that
either higher order terms $G_{4n+1}$ get resummed to an exponential
function $e^{-t/\tau}$ with $\tau^{-1}\rightarrow m\left(\frac{g^{4}}{4}\frac{\log(mt)}{\pi}-\gamma_{1}-i\gamma_{2}\right)$,
or alternatively, only terms containing $\gamma_{1,2}$ are resummed
into a relaxation rate and a frequency shift of the one-particle oscillations.
We stress that the fate of the logarithmic term is unclear at present,
which also implies that the part of $\gamma_{1}$ entering the resummation
is not defined until this issue is dealt with, which requires further
investigation.

Our results can be easily extended to local operators in theories
with more than one species with diagonal scattering, and also to operators
semi-local with respect to solitons in the sine\textendash Gordon
model.

The other class of secular contributions to one-particle oscillations
is a novel one having the form

\begin{equation}
\Re e\,gF_{1}^{\mathcal{O}}e^{-imt}g^{2}\sqrt{\frac{mt}{\pi}}\frac{-\sqrt{2}-\sqrt{2}i}{2}\;,
\end{equation}
and its leading order is $g^{3}$. The origin of this secular term
is a physical effect analogous to parametric resonance, and is caused
by the effective coupling between one- and two-particle modes. This
coupling is established by the corresponding form factor of the operator
and the $K$ function resulting in a singularity at the threshold
of the two-particle continuum. At this point the ratio between the
frequencies is exactly two satisfying the condition of parametric
resonance. Note that the singularity of the $K$ function is an essential
ingredient as it is the origin of enhancement in the effective coupling
between the modes.

Unfortunately, it is rather difficult to calculate the next secular
contribution from $G_{7}$ and even having an expression for the next
term, it is not guaranteed that they can be resummed in an effective
manner to extract the long time behaviour. Therefore it is presently
unclear how this phenomenon influences the fate of the one-particle
oscillations. However, it is clear that despite the initial growth
indicated by the presence of $\sqrt{t}$ , the amplitude must eventually
saturate as only finite energy density is injected in the system during
the quench.

In most of our discussions we tacitly assumed that the quenches are
integrable, so that the initial state has the generalised squeezed
state form (\ref{D23Res}). However, we wish to note that the arguments
establishing the singularity of the pair amplitude are valid even
for non-integrable quenches whenever a one-particle coupling is present.
A generic initial state can always be written as \cite{InitalStateIntEqHierarchcy}
\[
|\Omega\rangle=\exp\left\{ \sum_{n=1}^{\infty}\frac{1}{n!}\int\frac{d\theta_{1}}{2\pi}\dots\int\frac{d\theta_{n}}{2\pi}K_{n}(\theta_{1}\dots\theta_{n})Z^{\dagger}(\theta_{1})\dots Z^{\dagger}(\theta_{n})\right\} |0\rangle.
\]
Integrable initial states of the form (\ref{eq:pole_residue}) satisfy
$K_{n}=0$ for $n>2$, but the main conclusions of this paper are
unaffected by the presence of the higher $K_{n}$, which would only
result in additional higher order corrections to the frequency shift
and relaxation terms.

Finally, we wish to comment on the relevance of the phase quenches
in the sine\textendash Gordon model introduced in Subsection \ref{subsec:PhaseQuench}.
Being both experimentally realisable and analytically tractable, the
sine\textendash Gordon theory has attracted a lot of attention. The
sine\textendash Gordon model emerges as an effective description in
cold atom experiments involving tunnel coupled quasi-one-dimensional
condensates, where the sine\textendash Gordon field corresponds to
the relative phase of the condensate \cite{Schmiedmayer,SchmiedmayerPhase}.
Therefore, phase quenches can be an ideal protocol to compare experimental
and theoretical results. This is especially true for moderate sized
quenches, where the small post-quench density of excitations makes
the form factor series valid and it is possible to extract analytic
results about the post-quench time evolution.

\subsection*{Acknowledgements}

The authors are grateful to D. Schuricht for useful discussions. This
research was partially supported by the BME-Nanotechnology FIKP grant
of EMMI (BME FIKP-NAT), and also by the National Research Development
and Innovation Office (NKFIH) under K-2016 grant no. 119204, OTKA
grant no. SNN118028 and the Quantum Technology National Excellence
Program (Project No. 2017-1.2.1- NKP-2017- 00001). M.K. was also supported
by a ``Pr{é}mium\textquotedblright{} postdoctoral grant of the
Hungarian Academy of Sciences and D.X.H. acknowledges support from
the {Ú}NKP-17-3-I New National Excellence Program of the Ministry
of Human Capacities.

{\LARGE{}{}$\qquad\qquad\qquad\qquad\hspace{1em}\hspace{1em}\hspace{1em}$}\includegraphics[scale=0.2]{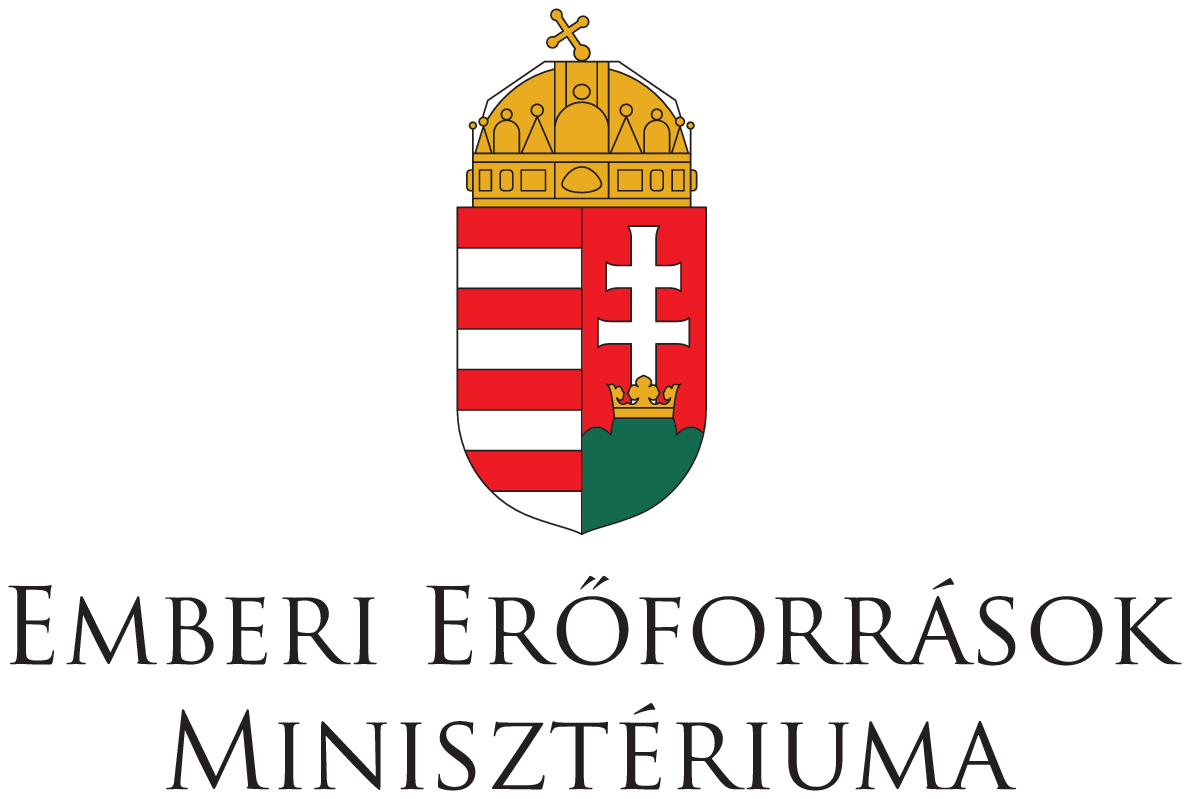}

\appendix

\section{Finite volume formalism\label{sub:Appendix-A-FinVol}}

\subsection{Multi-particle states in finite volume \label{subsec:Multi-particle-states-in}}

Excited states of a massive integrable quantum field theory with one
particle species in a large, but finite volume can be described as
scattering states consisting of $n$ particles with rapidities $\vartheta_{n}$
given by the solution of the Bethe\textendash Yang equations 
\begin{equation}
Q_{k}=mL\sinh\vartheta_{k}+\sum_{j\ne k}\delta(\vartheta_{k}-\vartheta_{j})=2\pi I_{k}\;,\quad\quad k=1,\dots,n\;.\label{eq:BY}
\end{equation}
Using the fact that the effective statistics is fermionic, i.e. $S(0)=-1$,
the two-particle phase shift function is defined as 
\[
S(\vartheta)=-e^{i\delta(\vartheta)}\;,\qquad\qquad\delta(-\vartheta)=-\delta(\vartheta)\;,
\]
and the following prescription is obtained for the quantum numbers
of the particles: 
\[
\begin{split}I_{k}\in\mathbb{Z}\qquad\text{for odd }n\;,\qquad\qquad I_{k}\in\mathbb{Z}+\frac{1}{2}\qquad\text{for even }n\;.\end{split}
\]
The state corresponding to quantum numbers $\{I_{1},\dots,I_{n}\}$
is denoted as 
\[
|\{I_{1}\dots I_{n}\}\rangle_{L}\;,
\]
and it is independent (up to a possible phase ambiguity) of the ordering
of the $I_{k}$. They are normalised so that their scalar products
are 
\[
_{L}\langle\{I_{1}\dots I_{n}\}|\{I'_{1}\dots I'_{m}\}\rangle_{L}=\delta_{nm}\delta_{I_{1},I'_{1}}\dots\delta_{I_{n},I'_{n}}\;,
\]
with the quantum numbers ordered by convention as $I_{1}<\dots<I_{n}$
and $I_{1}'<\dots<I_{m}'$ . The total energy and momentum can be
expressed as 
\[
E=\sum_{i=1}^{n}m\cosh\vartheta_{i}+\mathcal{O}(e^{-\mu L})\;,\qquad\qquad P=\sum_{i=1}^{n}m\sinh\vartheta_{i}+\mathcal{O}(e^{-\mu L})
\]
up to exponential corrections governed by some mass scale $\mu$.
For a systematic treatment of exponential corrections to excitation
energies see \cite{L=00003D0000FCscher,KlassenM} and also \cite{BajnokJanik,Hatsuda}.

The rapidity space density of states in the $n$-particle sector of
the theory is given by the Jacobian 
\begin{equation}
\rho_{n}(\vartheta_{1},\dots,\vartheta_{n})=\det J_{kl}\;,\qquad J_{kl}=\frac{\partial Q_{k}}{\partial\vartheta_{l}}\;.
\end{equation}

\subsection{The initial state in finite volume}

The integrable initial state (\ref{IntegrableQuench}) can be written
in finite volume as \cite{OnePointFunctions}\textit{ 
\begin{multline}
|\varOmega\rangle_{L}=\mathcal{G}(L)\Big(|0\rangle_{L}+\frac{g}{2}\sqrt{mL}|\{0\}\rangle_{L}+\sum_{I>0}K(\vartheta)N_{2}(\vartheta,L)|\{-I,I\}\rangle_{L}\\
+\sum_{I}\frac{g}{2}K(\vartheta)N_{3}(\vartheta,L)|\{-I,I,0\}\rangle_{L}+\frac{1}{2}\sum_{I\neq J}K(\vartheta_{1})K(\vartheta_{2})N_{4}(\vartheta_{1},\vartheta_{2},L)|\{-I,I,-J,J,0\}\rangle_{L}\Big)\;,\label{FinVolInitStateAppendix}
\end{multline}
}where the rapidities $\vartheta$ are the solutions of the appropriate
Bethe\textendash Yang equations with a constraint of zero overall
momentum. For the two-particle states the constrained Bethe-Yang equation
is 
\begin{equation}
\bar{Q}_{2}(\vartheta)=mL\sinh\vartheta+\delta(2\vartheta)=2\pi I\;,\label{barQ1}
\end{equation}
and the sum in (\ref{FinVolInitStateAppendix}) only runs over $I>0$
because the states $|-I,I\rangle_{L}$ and $|-I,I\rangle_{L}$ are
identical. The three-particle sector consists of states with rapidities
$\{-\vartheta,0,\vartheta\}$ where $\vartheta$ is determined by
the corresponding quantisation condition 
\begin{equation}
\bar{Q}_{3}(\vartheta)=mL\sinh\vartheta+\delta(\vartheta)+\delta(2\vartheta)=2\pi J\;,\label{barQ3-1}
\end{equation}
whereas for the four-particle case the quantisation condition for
the rapidities $\{-\vartheta_{1},\vartheta_{1},-\vartheta_{2},\vartheta_{2}\}$
is given by the system of equations 
\[
\begin{split}\bar{Q}_{4,1} & =mL\sinh\vartheta_{1}+\delta(\vartheta_{1}-\vartheta_{2})+\delta(\vartheta_{1}+\vartheta_{2})+\delta(2\vartheta_{1})=2\pi I_{1}\;,\\
\bar{Q}_{4,2} & =mL\sinh\vartheta_{2}+\delta(\vartheta_{2}-\vartheta_{1})+\delta(\vartheta_{1}+\vartheta_{2})+\delta(2\vartheta_{2})=2\pi I_{2}\;.
\end{split}
\]
The normalisation factors $N_{1}(L)$, $N_{2}(\theta,L)$, $N_{3}(\vartheta,L)$
and $N_{4}(\vartheta_{1},\vartheta_{2},L)$ in (\ref{FinVolInitStateAppendix})
are not determined by first principles and were calculated in \cite{OnePointFunctions}
up to finite size effects exponential decaying with the volume. For
the one and two-particle states one finds 
\begin{equation}
N_{1}(L)=\sqrt{mL}+\mathcal{O}(e^{-\mu L})\;,\qquad\qquad N_{2}(\vartheta,L)=\frac{\sqrt{\rho_{2}(\vartheta,-\vartheta)}}{\bar{\rho}_{2}(\vartheta)}+\mathcal{O}(e^{-\mu L})\;,\label{N1N2}
\end{equation}
where \textit{ 
\[
\bar{\rho}_{2}(\vartheta)=\frac{d\bar{Q}_{2}}{d\vartheta}=mL\cosh\vartheta+2\varphi(2\vartheta)\;.
\]
}Note that the total two-particle density $\rho_{2}$ satisfies 
\[
\rho_{2}(\vartheta,-\vartheta)=\rho_{1}(\vartheta)\bar{\rho}_{2}(\vartheta)\;,
\]
and so 
\begin{equation}
N_{2}(\vartheta,L)=\sqrt{\frac{\rho_{1}(\vartheta)}{\bar{\rho}_{2}(\vartheta)}}=1-\frac{\varphi(2\vartheta)}{mL\cosh\vartheta}+\mathcal{O}(1/L^{2})\;.\label{N2maskepp}
\end{equation}
For the three-particle state the normalisation of the three-particle
states is given by 
\[
N_{3}(\vartheta,L)=\frac{\sqrt{\rho_{3}(\vartheta,0,-\vartheta)}}{\bar{\rho}_{3}(\vartheta)}\;,
\]
where

\begin{equation}
\bar{\rho}_{3}(\vartheta)=\frac{d\bar{Q}_{3}}{d\vartheta}\;,
\end{equation}
and in the four-particle case the normalisation reads 
\[
N_{4}(\vartheta_{1},\vartheta_{2},L)=\frac{\sqrt{\rho_{4}(\vartheta_{1},-\vartheta_{1},\vartheta_{2},-\vartheta_{2})}}{\bar{\rho}_{4}(\vartheta_{1},\vartheta_{2})}\;,
\]
with 
\[
\bar{\rho}_{4}(\vartheta_{1},\vartheta_{2})=\det J\qquad\text{with}\qquad J_{ik}=\frac{\partial\bar{Q}_{4,i}}{\partial\vartheta_{k}}\;,\quad\quad i,k=1,2\;.
\]

\subsection{Form factors and their properties }

Form factors are matrix elements of (semi-)local operators with the
asymptotic states of the theory. We start with the infinite volume
case where they can be determined in terms of the so-called elementary
form factors 
\[
F_{m}^{\mathcal{O}}(\vartheta_{1},\dots,\vartheta_{m})=\langle0|\mathcal{O}(0)|\vartheta_{1},\dots,\vartheta_{m}\rangle\:.
\]
All other form factors 
\[
F_{mn}^{\mathcal{O}}(\vartheta_{1}^{'},\dots,\vartheta_{m}^{'}|\vartheta_{1},\dots,\vartheta_{n})=\langle\vartheta_{1}^{'},\dots,\vartheta_{m}^{'}|\mathcal{O}(0)|\vartheta_{1},\dots,\vartheta_{n}\rangle
\]
can be expressed in terms of the elementary form factors with the
help of the crossing relation 
\begin{multline*}
F_{mn}^{\mathcal{O}}(\vartheta_{1}^{'},\dots,\vartheta_{m}^{'}|\vartheta_{1},\dots,\vartheta_{n})=F_{m-1,n+1}^{\mathcal{O}}(\vartheta_{1}^{'},\dots,\vartheta_{m-1}^{'}|\vartheta_{m}^{'}+i\pi,\vartheta_{1},\dots,\vartheta_{n})\\
+\sum_{k=1}^{n}\Big(2\pi\delta(\vartheta_{m}^{'}-\vartheta_{k})\prod_{l=1}^{k-1}S(\vartheta_{l}-\vartheta_{k})\times F_{m-1,n-1}^{\mathcal{O}}(\vartheta_{1}^{'},\dots,\vartheta_{m-1}^{'}|\vartheta_{1},\dots,\vartheta_{k-1},\vartheta_{k+1}\dots,\vartheta_{n})\Big)\;.
\end{multline*}
In integrable models, the elementary form factors are meromorphic
functions of the rapidities and satisfy a series of form factor bootstrap
equations (for a review we refer to \cite{GiuseppeReview,SmirnovBook}).
In a theory with only one particle species and in the absence of bound
states the form factor equations are:

I. Lorentz invariance: 
\begin{equation}
F_{n}^{\mathcal{O}}(\vartheta_{1}+\Lambda,\dots,\vartheta_{n}+\Lambda)=e^{s\Lambda}F_{n}^{\mathcal{O}}(\vartheta_{1},\dots,\vartheta_{n})\;,\label{eq:FFLorentzSpin}
\end{equation}
where $s$ is the Lorentz spin of the operator $\mathcal{O}$. In
this work we only consider scalar operators corresponding to $s=0$.

II. Exchange: 
\begin{equation}
F_{n}^{\mathcal{O}}(\vartheta_{1},\dots,\vartheta_{k},\vartheta_{k+1},\dots,\vartheta_{n})=S(\vartheta_{k}-\vartheta_{k+1})F_{n}^{\mathcal{O}}(\vartheta_{1},\dots,\vartheta_{k+1},\vartheta_{k},\dots,\vartheta_{n})\;.\label{eq:exchangeaxiom}
\end{equation}

III. Cyclic permutation: 
\begin{equation}
F_{n}^{\mathcal{O}}(\vartheta_{1}+2i\pi,\vartheta_{2},\dots,\vartheta_{n})=e^{i\sigma}F_{n}^{\mathcal{O}}(\vartheta_{2},\dots,\vartheta_{n},\vartheta_{1})\;.\label{eq:cyclicaxiom}
\end{equation}

IV. Kinematical singularity 
\begin{equation}
-i\mathop{\textrm{Res}}_{\theta=\theta^{'}}F_{n+2}^{\mathcal{O}}(\vartheta+i\pi,\vartheta^{'},\vartheta_{1},\dots,\vartheta_{n})=\left(1-e^{i\sigma}\prod_{k=1}^{n}S(\vartheta-\vartheta_{k})\right)F_{n}^{\mathcal{O}}(\vartheta_{1},\dots,\vartheta_{n})\;,\label{eq:kinematicalaxiom}
\end{equation}
where $\sigma$ is the mutual locality index of the operator $\mathcal{O}$
with respect to the interpolating field $\phi$ that creates the particle
excitation and is defined via the condition 
\[
\mathcal{O}(x,t)\phi(y,t)=e^{2\pi i\sigma}\phi(y,t)\mathcal{O}(x,t)\;.
\]
Based on (\ref{eq:FFLorentzSpin}), (\ref{eq:exchangeaxiom}), (\ref{eq:cyclicaxiom})
and (\ref{eq:kinematicalaxiom}) form factors of operators can be
determined by the form factor bootstrap program. Form factors involving
$n$ particles of a single species with a diagonal scattering (such
as e.g. multi-$B_{1}$ form factors in sine\textendash Gordon theory)
can be written as 
\begin{eqnarray}
F_{n}^{\mathcal{O}}\left(\vartheta_{1},\dots,\vartheta_{n}\right) & =\left\langle 0\right|\mathcal{O}(0)|\vartheta_{1},\dots,\vartheta_{n}\rangle= & \mathcal{N}_{n}\frac{Q_{n}^{\mathcal{O}}\left(x_{1},\dots,x_{n}\right)}{\prod_{i<j}\left(x_{i}+x_{j}\right)}\prod_{i<j}f_{min}\left(\vartheta_{i}-\vartheta_{j}\right)\:,\label{eq:FFParametrization}
\end{eqnarray}
where $x_{i}=e^{\vartheta_{i}}$. The minimal form factors $f_{min}$
satisfying 
\[
f_{min}(-\vartheta)=S(\vartheta)f_{min}(\vartheta)\quad\text{and}\quad f_{min}(i\pi-\vartheta)=f_{min}(i\pi+\vartheta)
\]
enforces (\ref{eq:exchangeaxiom}), the $x_{i}$-dependence of the
rest satisfies (\ref{eq:cyclicaxiom}), the product in the denominators
ensures the presence of poles prescribed by (\ref{eq:kinematicalaxiom}),
and the operator content is encoded in $Q_{n}^{O}$, which is a symmetric
polynomial of the variables and for a nontrivial locality index $\sigma\neq0$
it contains an overall factor $\left(\prod x_{i}\right)^{min(\sigma,1-\sigma)}$.

In finite volume and when the sets of rapidities in the bra and ket
states are disjoint, the finite and infinite volume form factors can
be related \cite{FinVolFF1and2} as

\begin{equation}
_{L}\langle\{I'_{1}\dots I'_{k}\}|O|\{I{}_{1}\dots I{}_{l}\}\rangle_{L}=\frac{F_{k+l}^{O}\left(\vartheta_{1}^{'}+i\pi,...\vartheta_{k}^{'},+i\pi,\vartheta_{1}...\vartheta_{l}\right)}{\sqrt{\rho_{k}(\vartheta'_{1},\dots,\vartheta'_{k})\rho_{l}(\vartheta_{1},\dots,\vartheta_{l})}}+\mathcal{O}(e^{-\mu L})\:.\label{eq:FinVolFF}
\end{equation}
For the case of coinciding rapidities this relation must be modified
to account for disconnected contributions \cite{FinVolFF1and2}, but
we do not need the corresponding expressions here.

\section{The transverse field quantum Ising chain and its scaling limit \label{sec:IsingAppendix}}

\subsection{Definition of the model}

The transverse field quantum Ising model (TQIM) is defined by the
Hamiltonian

\begin{equation}
H=-J\sum_{i=1}^{N}\left(\sigma_{i}^{x}\sigma_{i+1}^{x}+h\sigma_{i}^{z}\right)\;,\label{IsingChainHamiltonian}
\end{equation}
where $\sigma_{i}^{\alpha}$ denotes the Pauli matrices at site $i$,
$J>0$, $h$ is the transverse field and the boundary conditions are
assumed to be periodic. By applying the Jordan-Wigner transformation,
the Hamiltonian (\ref{IsingChainHamiltonian}) can be mapped to spinless
Majorana fermions with dispersion relation \cite{IsingDiag1,IsingDiag2}

\begin{equation}
\varepsilon_{h}(k)=2J\sqrt{1+h^{2}-2h\cos k}\label{IsingDispRel}
\end{equation}
having an energy gap $\Delta=2J|1-h|$. The model possesses a quantum
critical point at $h=1$ separating the disordered or paramagnetic
phase (PM) for $h>1$ and the ordered, ferromagnetic phase (FM) for
$h<1$. In the disordered phase, the expectation value of the magnetisation
operator vanishes, while in the ferromagnetic phase its value is given
by 
\begin{equation}
\langle\sigma_{i}^{x}\rangle=\left(1-h^{2}\right)^{1/8}\:.\label{eq:ising_spont_magn}
\end{equation}
The Hilbert space of the model consists of two sectors with respect
to fermion number parity. In the Neveu\textendash Schwarz and Ramond
sectors states with even and odd number of fermions are present, respectively,
resulting in the quantisation condition for the wave numbers

\begin{equation}
\begin{split}k_{n}= & \frac{2\pi}{N}\left(n+\frac{1}{2}\right)\quad\text{Neveu-Schwarz}\\
p_{n}= & \frac{2\pi}{N}n\qquad\qquad\quad\text{Ramond}\;,
\end{split}
\end{equation}
where $n$ is a positive integer. In particular, the Fock space of
the model in the paramagnetic phase can be written as

\begin{equation}
\begin{split}|p_{1},...,p_{2m+1}\rangle= & \prod_{i=1}^{2m+1}a_{p_{i}}^{\dagger}|0\rangle_{R}^{PM}\quad p_{i}\subset\text{R\;,}\\
|k_{1},...,k_{2n}\rangle= & \prod_{i=1}^{2n}a_{k_{i}}^{\dagger}|0\rangle_{NS}^{PM}\qquad k_{i}\subset\text{NS}\,,
\end{split}
\end{equation}
and the ground state is the Neveu\textendash Schwarz vacuum, $|0\rangle_{NS}^{PM}$.
In the ferromagnetic phase, the zero momentum excitation has negative
energy, therefore the Ramond vacuum is redefined as $|0\rangle_{R}\rightarrow a_{0}^{\dagger}|0\rangle_{R}$
after which a particle-hole transformation is implemented on the zero-mode
$a_{0}^{\dagger}\rightarrow a_{0}$. The Fock space of the model in
the FM regime is hence

\begin{equation}
\begin{split}|p_{1},...,p_{2m}\rangle= & \prod_{i=1}^{2m}a_{p_{i}}^{\dagger}|0\rangle_{R}^{FM}\quad\quad p_{i}\subset\text{R\;,}\\
|k_{1},...,k_{2n}\rangle= & \prod_{i=1}^{2n}a_{k_{i}}^{\dagger}|0\rangle_{NS}^{FM}\qquad k_{i}\subset\text{NS}\,,
\end{split}
\end{equation}
where the ground state becomes degenerate in the thermodynamic limit
and the finite volume states corresponding to macroscopic magnetisation
are $\frac{1}{\sqrt{2}}(|0\rangle_{NS}^{FM}\pm|0\rangle_{R}^{FM})$.

\subsection{$FM\rightarrow PM$ quench and $\langle\sigma_{i}^{x}(t)\rangle$}

Performing a quench in the transverse field $h$, the pre- and post-quench
excitations can be related via a Bogoliubov transformation if the
initial state is the pre-quench vacuum. As a consequence, the squeezed-coherent
form of the initial state in the post quench basis (\ref{IntegrableQuench})
is guaranteed. Focusing on quenching from the ground state of the
FM phase to the PM phase, ( $h_{0}\rightarrow h$ with $h_{0}<1$
and $h>1$), one can write \cite{CalabreseEsslerFagotti1,CalabreseEsslerFagotti2,CalabreseEsslerFagotti3}

\begin{equation}
\begin{split}\frac{|0,h_{0}\rangle_{NS}^{FM}\pm|0,h_{0}\rangle_{R}^{FM}}{\sqrt{2}}= & \frac{1}{\sqrt{2}N_{NS}}\exp\left(i\sum_{k\in NS}K(k)a_{-k}^{\dagger}a_{k}^{\dagger}\right)|0,h\rangle_{NS}^{PM}\\
\pm & \frac{1}{\sqrt{2}N_{R}}\exp\left(i\sum_{p\in R\setminus\{0\}}K(p)a_{-p}^{\dagger}a_{p}^{\dagger}\right)a_{0}^{\dagger}|0,h\rangle_{R}^{PM},
\end{split}
\label{IsingFMInitState}
\end{equation}
where $N_{NS}$ and $N_{R}$ are normalisation constants and

\begin{equation}
|K(k)|^{2}=\frac{1-\cos\Delta_{k}}{1+\cos\Delta_{k}}\,,\label{IsingKSq}
\end{equation}
with

\begin{equation}
\cos\Delta_{k}=(2J)^{2}\frac{h_{0}h-(h+h_{0})\cos k+1}{\varepsilon_{h}(k)\varepsilon_{h_{0}}(k)}\;.\label{IsingCosDelta}
\end{equation}
The late time evolution of the magnetisation operator is given by
the expression \cite{CalabreseEsslerFagotti2}:

\begin{equation}
|\langle\sigma_{i}^{x}(t)\rangle|=\left(\mathcal{C}_{FP}^{x}\right)^{\frac{1}{2}}\left[1+\cos\left(2\varepsilon_{h}(k_{0})t+\alpha\right)+...\right]^{\frac{1}{2}}\exp\left[t\intop_{0}^{\pi}\frac{dk}{\pi}\varepsilon_{h}'(k)\ln|\cos\Delta_{k}|\right]\;,\label{SigmaTimeEvolLattice}
\end{equation}
where $k_{0}$ is a solution of the equation $\cos\Delta_{k}=0$,
$\alpha(h,h_{0})$ is an unknown constant and

\begin{equation}
\mathcal{C}_{FP}^{x}=\left[\frac{h\sqrt{1-h_{0}^{2}}}{h+h_{0}}\right]^{\frac{1}{2}}\:.
\end{equation}

\subsection{Continuum limit of the model and the $FM\rightarrow PM$ quench}

In the scaling limit of the TQIM, $J$ is sent to infinity together
with $h\rightarrow1$ such that the gap associated with the fermion
mass remains finite

\begin{equation}
M=2J|1-h|\;.
\end{equation}
In addition, the lattice spacing is sent to zero as $a=\frac{c}{2J}$,
where $c$ is the speed of light that we set to 1. It is easy to see
that the dispersion relation (\ref{IsingDispRel}) under scaling limit
transforms as

\begin{equation}
\varepsilon_{h}(ka)\rightarrow E(p)=\sqrt{M^{2}+p^{2}}\;,
\end{equation}
and the Hamiltonian (\ref{IsingChainHamiltonian}) scales to the Hamiltonian
of a massive Majorana fermion field theory

\begin{equation}
H=\frac{1}{2\pi}\int dx\frac{i}{2}\left(\psi(x)\partial_{x}\psi(x)-\bar{\psi}(x)\partial_{x}\bar{\psi}(x)\right)-iM\bar{\psi}(x)\psi(x)\;,
\end{equation}
with $\{\psi(x,t),\bar{\psi}(y,t)\}=2\pi\delta(x-y)$. The lattice
magnetisation operator $\sigma_{i}^{x}$ is related to the continuum
field $\sigma$ with the conformal normalisation via

\begin{equation}
\sigma(na)=\bar{s}J^{\frac{1}{8}}\sigma_{n}^{x}\;,
\end{equation}
with

\begin{equation}
\bar{s}=2^{\frac{1}{12}}e^{-\frac{1}{8}}\mathcal{A}^{\frac{3}{2}}\;,
\end{equation}
where $\mathcal{A}$=1.282427129... is Glaisher's constant.

In the following we perform the continuum limit for quantities relevant
for the $FM\rightarrow PM$ quench. It is important to note, however,
that unlike for quenches within the ferromagnetic phase, where explicit
calculations in the field theory framework \cite{SchurichtEssler}
and numerical studies \cite{Tibi} were carried out, for the $FM\rightarrow PM$
quench no such investigations have been done. On the other hand, performing
the naive scaling limit, the resulted QFT quantities make perfect
sense, hence expected to be the results for the field theory quench
problem. In the scaling limit sending $\delta h\rightarrow0$, we
write 
\begin{equation}
\begin{split}h=1+\delta h\;,\quad & h_{0}=1-\frac{M_{0}}{M}\delta h\;,\\
J=\frac{M}{2\delta h}\;,\quad & a=\frac{\delta h}{M}\;,
\end{split}
\end{equation}
which ensures that the dispersion relation in the post- and pre-quench
model is $\sqrt{M^{2}+p^{2}}$ and $\sqrt{M_{0}^{2}+p^{2}}$ respectively,
i.e. the mass in the PM and FM phase is $M$ and $M_{0}$ . Upon the
substitution $k=pa$, the continuum limit of (\ref{IsingKSq}) and
(\ref{IsingCosDelta}) are

\begin{equation}
|K(p)|^{2}=\frac{\sqrt{p^{2}+M^{2}}\sqrt{p^{2}+M_{0}^{2}}-p^{2}+MM_{0}}{\sqrt{p^{2}+M^{2}}\sqrt{p^{2}+M_{0}^{2}}+p^{2}-MM_{0}}\,,\label{IsingKSqQFT}
\end{equation}
and

\begin{equation}
\cos\Delta(p)=\frac{p^{2}-MM_{0}}{\sqrt{p^{2}+M^{2}}\sqrt{p^{2}+M_{0}^{2}}}\;.\label{IsingCosDeltaQFT}
\end{equation}
Introducing $\bar{\sigma}=\bar{s}M_{0}^{\frac{1}{8}}$, which is just
the pre-quench spontaneous magnetisation in the continuum theory obtained
from (\ref{eq:ising_spont_magn}), the scaling limit of (\ref{SigmaTimeEvolLattice})
reads

\begin{equation}
|\langle\sigma(t)\rangle|=\bar{\sigma}\frac{1}{2^{1/4}}\left[1+\cos\left(2\sqrt{M^{2}+MM_{0}}t+\alpha\right)+...\right]^{\frac{1}{2}}\exp\left[-M\:t\zeta\right]\;,
\end{equation}
where

\begin{equation}
-M\,\zeta=\frac{1}{\pi}\left\{ \frac{1}{2}\sqrt{M^{2}-M_{0}^{2}}\ln\left(\frac{M+\sqrt{M^{2}-M_{0}^{2}}}{M-\sqrt{M^{2}-M_{0}^{2}}}\right)-\sqrt{M^{2}+M\,M_{0}}\ln\left(\frac{\sqrt{M^{2}+M\,M_{0}}+M}{\sqrt{M^{2}+M\,M_{0}}-M}\right)\right\} \;
\end{equation}
for $M>M_{0}$ and

\begin{equation}
-M\,\zeta=\frac{1}{\pi}\left\{ \sqrt{M_{0}^{2}-M^{2}}\left(\tan^{-1}\left(\frac{M}{\sqrt{M_{0}^{2}-M^{2}}}\right)-\frac{\pi}{2}\right)-\sqrt{M^{2}+M\,M_{0}}\ln\left(\frac{\sqrt{M^{2}+M\,M_{0}}+M}{\sqrt{M^{2}+M\,M_{0}}-M}\right)\right\} 
\end{equation}
for $M<M_{0}$.

Note that $\zeta$ has a finite limit $-\ln2$ when $M_{0}=0$, but
its derivative is infinite at the origin as $M_{0}\rightarrow0$,
hence $\zeta$ is not an analytic expression for small values of $M_{0}$
.

\subsection{The singularity of the Ising $K$ function}

From (\ref{IsingKSqQFT}) it is easily seen that this function has
a $1/p^{2}$ singularity at the origin with the coefficient 

\[
\frac{4M_{0}^{2}M^{2}}{(M+M_{0})^{2}}\,.
\]
The appearance of this singularity is consistent with our consideration
presented in paper, as from the expansion of the initial state (\ref{IsingFMInitState})
it is seen that there is a zero-momentum particle in the Ramond sector.
In the following, we show that the pole strength in the $K$ or more
precisely in $|K|^{2}$ is $g^{4}/4$, if the one particle coupling
is $g$. For this goal, we first determine $g$ from the pole as 
\[
M^{2}\frac{g^{4}}{4}=\frac{4M_{0}^{2}M^{2}}{(M+M_{0})^{2}}
\]
 yielding 

\[
g=2\sqrt{\frac{M_{0}}{M+M_{0}}}
\]
as $p=M\sinh\vartheta$ and then show that the actual one-particle
coupling in (\ref{IsingFMInitState}) equals $g$. To calculate the
latter, we put the theory into finite volume, where from (\ref{FinVolInitState})
and (\ref{IsingFMInitState})

\begin{equation}
\sqrt{ML}\frac{g}{2}=\frac{N_{NS}}{N_{R}}
\end{equation}
must hold, where

\begin{equation}
\begin{split}N_{R}^{2}= & \prod_{n\subset\mathbb{N}^{+}}\left(1+\left|K\left(\frac{2\pi}{L}n\right)\right|^{2}\right)\\
N_{NS}^{2}= & \prod_{m\subset\mathbb{N}+\frac{1}{2}}\left(1+\left|K\left(\frac{2\pi}{L}m\right)\right|^{2}\right)\;.
\end{split}
\end{equation}
It is convenient to calculate the logarithm of the ratio, which we
write as

\begin{equation}
\begin{split}\ln\frac{N_{NS}^{2}}{N_{R}^{2}}= & \ln\prod_{n\subset\mathbb{N}^{+}}\frac{1+\left|K\left(\frac{2\pi}{L}(n-1/2)\right)\right|^{2}}{1+\left|K\left(\frac{2\pi}{L}(n)\right)\right|^{2}}\\
= & A+B\;,\\
A= & \ln\prod_{n\subset\mathbb{N}^{+}}\frac{1+\frac{x}{(n-\frac{1}{2})^{2}}}{1+\frac{x}{i^{2}}}\;,\\
B= & \ln\prod_{n\subset\mathbb{N}^{+}}\frac{1+\frac{x}{n^{2}}}{1+|K\left(\frac{2\pi}{L}n\right)|^{2}}\\
- & \ln\prod_{n\subset\mathbb{N}^{+}}\frac{1+\frac{x}{(n-\frac{1}{2})^{2}}}{1+|K\left(\frac{2\pi}{L}(n-\frac{1}{2})\right)|^{2}}\;,
\end{split}
\label{CsunyaLog}
\end{equation}
where

\[
x=\frac{4M^{2}M_{0}^{2}L^{2}}{(M+M_{0})^{2}(2\pi)^{2}}\;.
\]
When $ML\rightarrow0$, the expression denoted by $B$ in (\ref{CsunyaLog})
is zero, which can be easily seen by considering the Euler-Maclaurin
formula: for the two terms in $B$ the difference between the lower
endpoints for the integration and for the boundary terms consisting
of the functions and their derivatives goes as $1/L$ and the expressions
in the logarithms are smooth, well-behaved functions. The term $A$
in (\ref{CsunyaLog}) can explicitly be computed, therefore when $L\rightarrow\infty$,

\begin{equation}
\begin{split}\ln\frac{N_{NS}^{2}}{N_{R}^{2}}= & \ln\prod_{i\subset\mathbb{N}^{+}}\frac{1+\frac{x}{(i-\frac{1}{2})^{2}}}{1+\frac{x}{i^{2}}}\\
= & -\frac{1}{2}\left[-2\ensuremath{\ln\Gamma}\left(1-\sqrt{-x}\right)+2\ensuremath{\ln\Gamma}\left(\frac{1}{2}-\sqrt{-x}\right)\right.\\
- & \left.2\ensuremath{\ln\Gamma}\left(\sqrt{-x}+1\right)+2\ensuremath{\ln\Gamma}\left(\sqrt{-x}+\frac{1}{2}\right)\right]\\
+ & \ln\pi\:,
\end{split}
\label{LogGammas}
\end{equation}
and consequently, leading order behaviour of $\ln\frac{N_{NS}^{2}}{N_{R}^{2}}-\ln L$
from (\ref{LogGammas}) is $\ln(\pi\sqrt{x})$. Then, one finds that

\begin{equation}
\frac{N_{NS}}{N_{R}}=\sqrt{ML}\sqrt{\frac{M_{0}}{M_{0}+M}}\;,
\end{equation}
and that $g=2\sqrt{\frac{M_{0}}{M_{0}+M}}$ indeed holds.

\section{Phase quenches in the sine\textendash Gordon model and exponential
quenches \label{subsec:ExponentialQuenches}}

In this appendix we give more details of the calculation presented
in Section \ref{subsec:PhaseQuench}. Recall that the phase quenches
in the sine\textendash Gordon model consist of abruptly shifting the
sine\textendash Gordon field $\phi\rightarrow\phi+\delta/\beta$,
i.e. changing the phase in the model and regarding the pre-quench
vacuum as the initial state for the post-quench evolution. To obtain
the relation between the pre-quench and post-quench ground states,
consider the sine\textendash Gordon model as the perturbation of the
compactified free massless bosonic conformal field theory in finite
volume with the Hamiltonian

\begin{align}
H(\Phi) & =\int dx\frac{1}{2}:\left(\partial_{t}\Phi\right)^{2}+\left(\partial_{x}\Phi\right)^{2}:-\frac{\lambda}{2}\int dx\left(V_{1}+V_{-1}\right)\label{eq:pcft_Hamiltonian}\\
 & V_{a}=:e^{ia\beta\Phi}:\:,\nonumber 
\end{align}
where the $V_{a}$ are the so-called vertex operators (normal ordered
exponentials of the boson field). Note that the free bosonic part
of (\ref{eq:pcft_Hamiltonian}) commutes with the zero mode of the
conjugate momentum field $\Pi_{0}=\partial_{t}\Phi$, whereas due
to the canonical commutation relations 
\[
[\Phi(x,t),\Pi(y,t)]=i\delta(x-y)\;,
\]
one finds that the zero-mode 
\[
\Pi_{0}=\int dy\Pi(y,t)
\]
of the canonical momentum field satisfies

\begin{equation}
\ \exp\left(i\frac{\delta}{\beta}\Pi_{0}\right)\;:e^{ia\beta\Phi}:\;=\ :e^{ia(\beta\Phi+\delta)}:\;\exp\left(i\frac{\delta}{\beta}\Pi_{0}\right)\;.
\end{equation}
Hence for the ground state $H(\Phi)|0\rangle_{\Phi}=E_{0}|0\rangle_{\Phi}$,
we have

\begin{equation}
\begin{split}\exp\left(i\frac{\delta}{\beta}\Pi_{0}\right)H(\Phi)|0\rangle_{\Phi}=E_{0}\exp\left(i\frac{\delta}{\beta}\Pi_{0}\right)|0\rangle_{\Phi}\;,\\
H(\Phi+\delta/\beta)\exp\left(i\frac{\delta}{\beta}\pi_{0}\right)|0\rangle_{\Phi}=E_{0}\exp\left(i\frac{\delta}{\beta}\pi_{0}\right)|0\rangle_{\Phi}\:,
\end{split}
\label{eq:sGGS}
\end{equation}
from which 
\begin{equation}
\exp\left(i\frac{\delta}{\beta}\Pi_{0}\right)|0\rangle_{\Phi}=|0\rangle_{\Phi+\frac{\delta}{\beta}}\:.
\end{equation}
Therefore the initial state in the finite volume theory with PBC reads

\begin{equation}
|\Omega\rangle_{L}=\exp\left(i\frac{\delta}{\beta}\Pi_{0}\right)|0\rangle_{L}\:.
\end{equation}
For a form factor expansion for the overlaps in finite volume,

\begin{equation}
\vphantom{O}_{L}\langle\chi|\exp\left(i\frac{\delta}{\beta}\Pi_{0}\right)|0\rangle_{L}\;,\label{eq:OverlapDefApp}
\end{equation}
one can use the expression 
\[
\Pi_{0}=\int_{0}^{L}\partial_{t}\Phi(x)dx=i\int_{0}^{L}[H,\Phi(x)]dx\;,
\]
Expanding the exponential into a Taylor series, the general term is

\begin{equation}
\begin{split}\frac{i^{2l}}{l!}\mbox{\ensuremath{\left({\displaystyle \frac{\delta}{\beta}}\right)}}^{l}\sum_{\alpha_{1}}\ldots\sum_{\alpha_{l-1}}\int\left(\prod_{i}dx_{i}\right)\vphantom{O}_{L}\langle\chi|e^{i\hat{P}x_{1}}\Phi(0)e^{-i\hat{P}x_{1}}|\alpha_{1}\rangle_{L}(E_{\chi}-E_{\alpha_{1}})\times\\
\vphantom{O}_{L}\langle\alpha_{1}|e^{i\hat{P}x_{2}}\Phi(0)e^{-i\hat{P}x_{2}}|\alpha_{2}\rangle_{L}(E_{\alpha_{1}}-E_{\alpha_{2}})...\vphantom{O}_{L}\langle\alpha_{l-1}|e^{i\hat{P}x_{l}}\Phi(0)e^{-i\hat{P}x_{l}}|0\rangle_{L}(E_{\alpha_{l-1}}-E_{0})\;,
\end{split}
\end{equation}
in which $\hat{P}$ is the momentum operator and $\alpha_{i}$ indexes
a complete set of asymptotic eigenstates. Due to translation invariance,
the overlaps are only non-zero when the total momentum $p$ of the
state $|\chi\rangle$ is zero, and so only intermediate states with
zero total momentum contribute. We can then write

\begin{equation}
\begin{split}\frac{(-1)^{l}L^{l}}{l!}\mbox{\ensuremath{\left({\displaystyle \frac{\delta}{\beta}}\right)}}^{l}\tilde{\sum_{\alpha_{1}}}\ldots\tilde{\sum_{\alpha_{l-1}}}\vphantom{O}_{L}\langle\chi|\Phi(0)|\alpha_{1}\rangle_{L}(E_{\chi}-E_{\alpha_{1}})\times\\
\vphantom{O}_{L}\langle\alpha_{1}|\Phi(0)|\alpha_{2}\rangle_{L}(E_{\alpha_{1}}-E_{\alpha_{2}})...\vphantom{O}_{L}\langle\alpha_{l-1}|\Phi(0)|0\rangle_{L}(E_{\alpha_{l-1}}-E_{0})\;,
\end{split}
\label{eq:OverlapFFExpansionApp}
\end{equation}
where the tildes over the sums signal that only zero momentum states
are included.

Now for a state containing a zero-momentum first breather and to first
order in $\delta$, we can use (\ref{eq:OverlapFFExpansionApp}),
(\ref{N1N2}) and (\ref{eq:FinVolFF}) to obtain

\begin{equation}
\begin{split}\sqrt{m_{1}L}\frac{g}{2}= & _{L}\langle\{0\}|\Omega\rangle_{L}\\
= & (-1)L\mbox{\ensuremath{\left(\frac{\delta}{\beta}\right)}}{}_{L}\langle\{0\}|\Phi(0)|0\rangle_{L}m_{1\;,}
\end{split}
\end{equation}
with 

\[
_{L}\langle\{0\}|\Phi(0)|0\rangle_{L}=\frac{F_{B_{1}}^{*}}{\sqrt{m_{1}L}}\:,
\]
which gives

\begin{equation}
\frac{g}{2}=-{\displaystyle \frac{\delta}{\beta}}F_{B_{1}}^{*}\:,\label{eq:gper2-1}
\end{equation}
where $F$ denotes the form factor of the field $\Phi$ in the sine\textendash Gordon
theory which has only has non-vanishing matrix elements with states
composed first breathers when their total number is odd. In particular
\cite{Lukyanov},

\begin{equation}
F_{B_{1}}=F_{B_{1}}^{\text{*}}=\frac{\bar{\lambda}(\xi)\pi\xi}{\beta\sin\pi\xi}\:,\label{eq:FB1}
\end{equation}
with

\begin{equation}
\begin{split}\xi= & \frac{\beta^{2}}{\beta^{2}-8\pi}\;,\\
\bar{\lambda}(\xi)= & 2\cos\frac{\pi\xi}{2}\sqrt{2\sin\frac{\pi\xi}{2}}\exp\left(-\int_{0}^{\pi\xi}\frac{dt}{2\pi}\frac{t}{\sin t}\right)\;,
\end{split}
\end{equation}
and we remark that the form factors of the sine\textendash Gordon
field $\phi$ can be obtained from that of the vertex operators $:e^{i\alpha\phi}:$
by differentiating with respect to $\alpha$. For the pair overlap
the lowest non-trivial order is $\delta^{2}$ and one obtains

\begin{equation}
\begin{split}N_{2}(\vartheta,L)K(\vartheta)= & _{L}\langle I,-I|\Omega\rangle_{L}\\
= & \mbox{\ensuremath{\left({\displaystyle \frac{\delta}{\beta}}\right)}}^{2}\frac{L^{2}}{2}\tilde{\sum_{\alpha_{1}}}{}_{L}\langle I,-I|\Phi(0)|\alpha_{1}\rangle_{L}(2m_{1}\cosh\vartheta-E_{\alpha_{1}}){}_{L}\langle\alpha_{1}|\Phi(0)|0\rangle_{L}E_{\alpha_{1}}\\
= & \mbox{\ensuremath{\left({\displaystyle \frac{\delta}{\beta}}\right)}}^{2}\frac{L^{2}}{2}\sum_{n=1}^{\infty}\sum_{\{\theta_{k}\}}\frac{F_{B_{1},B_{1},A_{1}...A_{n}}(i\pi+\vartheta,i\pi-\vartheta,\theta_{1},...\theta_{n})}{\sqrt{\left(m_{1}L\cosh\vartheta\right)^{2}+\left(m_{1}L\cosh\vartheta\right)\,\varphi_{B_{1}B_{1}}(\vartheta)}\rho_{n}(\theta_{1},...\theta_{n})}\times\\
 & \left(2m_{1}\cosh\vartheta-\sum_{i=1}^{n}m_{i}\cosh\theta_{i}\right)F_{A_{1}...A_{n}}^{*}(\theta_{1},...\theta_{n})\left(\sum_{i=1}^{n}m_{i}\cosh\theta_{i}\right)\;,
\end{split}
\label{eq:KexpansionApp}
\end{equation}
where the particles $A_{k}$ are either breathers or solitons and
their is an implicit summation over all possible choice of their species.

Now we are interested in the singular behaviour of $K$ which can
only originate from those of the form factors. The infinite volume
limit itself is finite, since the numerator contains an explicit $L^{2}$,
and in the $n$-particle term the denomination contributes $1/L$
and also $1/L^{n}$ from the density factor, while rewriting the discrete
summation in terms of integrals results in a state density factor
of behaviour $L^{n-1}$, so the leading term is independent of $L$.

First let us focus on terms where all particles $A_{1}\dots A_{n}$
are first breathers. We now proceed to analyse the singularity of
these terms by setting $\vartheta$ to a value $\epsilon$ and to
consider the limit of small $\epsilon$. The form factor has a kinematical
singularity when some subset of $m$ particles among the $A_{1}\dots A_{n}$
has similarly small rapidities, which we take to be a multiple of
$\epsilon$. The dependence of the most singular term can be obtained
from the kinematical residue equation (\ref{eq:kinematicalaxiom})
but also taking into account that the form factor having a first order
zero when two rapidities coincide due to (\ref{eq:exchangeaxiom})
and $S(0)=-1$. There are three cases: 
\begin{itemize}
\item $m<n-1$: The behaviour of $F(i\pi+\vartheta,i\pi-\vartheta,\theta_{1},...\theta_{n})$
is $\epsilon^{1+(m-1)m/2-2m}$ , where the $1$ comes from the coincidence
of $i\pi-\vartheta$ and $i\pi+\vartheta$ , the $m(m-1)/2$ comes
from the coincidence of the $m$ rapidities in the set $\theta_{1},...\theta_{n}$,
and the $-2m$ are from the kinematical singularities ``activated''
in the limit. For $F^{*}(\theta_{1},...\theta_{n})$ one obtains a
factor $\epsilon^{(m-1)m/2}$, therefore the overall behaviour is
$\epsilon^{1+(m-1)m-2m}$ which is only singular for $m=1$ or $2.$ 
\item $m=n-1$: one must be aware that the condition of zero total momentum,
that sending $m=n-1$ rapidities to zero results in all the $n$ of
them going to zero, leading to the behaviour $\epsilon^{1+(n-1)n-2n}$
which is singular for $n=2$. 
\item $m=n$: same behaviour $\epsilon^{1+(n-1)n-2n}$ as for the case $m=n-1$,
resulting in $n=1$ or $n=2$. 
\end{itemize}
Note that whenever there is a singularity it is always of order one.
For the case when all the $A_{1}\dots A_{n}$ are first breathers,
$n$ must also be odd for the form factor not to vanish due to parity
invariance. The $n=1$ contribution is 
\begin{equation}
\mbox{\ensuremath{\left({\displaystyle \frac{\delta}{\beta}}\right)}}^{2}\frac{L^{2}}{2}\frac{F_{B_{1}B_{1}B_{1}}(i\pi+\vartheta,i\pi-\vartheta,0)(2m_{1}\cosh\vartheta-m_{1})F_{1}^{\text{*}}m_{1}}{\sqrt{\left(m_{1}L\cosh\vartheta\right)^{2}+\left(m_{1}L\cosh\vartheta\right)\,\varphi_{B_{1}B_{1}}(\vartheta)}m_{1}L}\;.\label{eq:KSingContribution-1}
\end{equation}
Using the form factor equations

\begin{equation}
F_{B_{1}B_{1}B_{1}}(i\pi+\vartheta,i\pi-\vartheta,0)\approx-\frac{4i}{\mathbf{\vartheta}}F_{B_{1}}\;,
\end{equation}
this contribution alone gives the expected pole contribution

\begin{equation}
\begin{split}K(\vartheta) & =-\mbox{\ensuremath{\left({\displaystyle \frac{\delta}{\beta}}\right)}}^{2}\frac{2i\left|F_{B_{1}}\right|^{2}}{2}\frac{1}{\vartheta}+\mathcal{O}(\vartheta^{0})\\
 & =-i\frac{g^{2}}{2}\frac{1}{\vartheta}\;.
\end{split}
\end{equation}
Now we demonstrate that contributions with $n\geq3$ are regular at
the origin $\vartheta=0$. Observe that in finite volume the rapidity
$\vartheta$ in finite volume is eventually quantised according to
\[
mL\sinh\vartheta+\delta(2\vartheta)=2\pi I
\]
with some half-integer quantum number $I,$ so it is always displaced
by an amount of order $1/L$ from the origin. Fixing $I$ results
in the parameter $\epsilon$ being essentially $1/L$, therefore the
singularity $1/\epsilon$ manifests itself as a divergence of (\ref{eq:KexpansionApp})
proportional to $L$ when $L$ goes to infinity with $I$ fixed. A
simple power counting then gives that the contribution is 
\[
\frac{L^{2}}{L^{n+1}}\times L\times L^{r}=L^{r+2-n}\;,
\]
where $L^{r}$ is the state density factor resulting for the $r$
particle rapidities among the $\theta_{1},...\theta_{n}$ whose summation
is left free once fixing the positions of those needed for the singularity
and also taking into account the zero-momentum constraint. Clearly
one obtains $r\leq n-2$ resulting in a cancellation of any divergence
for $L\rightarrow\infty$.

To finish this discussion, let us consider the case when the set of
particles $A_{1}\dots A_{n}$ contains other species (higher breathers
or solitons) as well. Let us suppose that the total number of first
breathers among $A_{1}\dots A_{n}$ is $k$ with $k<n$. The counting
of the degree of singularity only involves first breathers, so we
obtain that the only possible cases are again $k=1$ or $2$. However,
in the denominator of (\ref{eq:KexpansionApp}) now one has a density
$\rho_{n}$ with a behaviour $L^{n}$ with $n>k$, which leads to
a regular limit for $L\rightarrow\infty$.

\section{Some useful relations \label{sec:Some-useful-relations}}

Here we collect some useful formulae regarding the $K$ function and
form factors, identities from the theory of distributions and relations
for the stationary phase approximation (SPA) that are useful in the
text.

\subsection{The $K$ function}

The $K$ function possesses a singularity at the origin, if the one
particle coupling to the boundary denoted by $g$ is non-zero. In
this case, the singular term in $K$ is

\begin{equation}
K(\vartheta)\propto-i\frac{g^{2}}{2}\frac{1}{\vartheta}\:.
\end{equation}
Due to the relation $K(-\vartheta)=S(-2\vartheta)K(\vartheta)$ the
constant term in the expansion of $K$ is also expressible with $g$.
Writing $K$ as

\begin{equation}
K(\vartheta)=-i\frac{g^{2}}{2}\frac{1}{\vartheta}+K_{0}+K_{1}\vartheta+\ldots\:,
\end{equation}
and expanding

\begin{equation}
S(\vartheta)=-1-i\varphi(0)\vartheta+\frac{1}{2}\varphi^{2}(0)\vartheta^{2}+\ldots\:,
\end{equation}
one obtains

\begin{equation}
K_{0}=\frac{\varphi(0)g^{2}}{2}\:,
\end{equation}
hence 
\begin{equation}
K(\vartheta)=-i\frac{g^{2}}{2}\frac{1}{\vartheta}+\frac{\varphi(0)g^{2}}{2}+K_{1}\vartheta+\ldots\;.\label{eq:KExpansion}
\end{equation}
Also note that due to real analyticity $K(-\vartheta)=K^{*}(\vartheta),$
all even/odd coefficients in the expansion of $K$ around $\vartheta=0$
are purely real/imaginary, respectively.

\subsection{Form factor singularities and expansions}

Consider form factors for a single species, such as the the multi-$B_{1}$
form factors of the sine\textendash Gordon model. Using the form factor
equations (\ref{eq:exchangeaxiom}) and (\ref{eq:kinematicalaxiom})
one can derive the universal expression

\begin{equation}
F_{3}(i\pi,-\varepsilon,\varepsilon)=\frac{4i}{\varepsilon}F_{1}-4iF_{1}\frac{f''_{min}(0)}{f'_{min}(0)}+\mathcal{O}(\varepsilon)\;,\label{eq:F3Expansion}
\end{equation}
where

\begin{equation}
\frac{f''_{min}(0)}{f'_{min}(0)}=i\varphi(0)\;,\label{S1}
\end{equation}
i.e. the first derivative of the S-matrix at the origin. Another form
factor we need is $F_{5}(i\pi,-\varepsilon,\varepsilon,-\vartheta,\vartheta)$,
where $\vartheta$ is non-zero. Using (\ref{eq:FFParametrization}),
one can explicitly compute

\begin{equation}
F_{5}(i\pi,-\varepsilon,\varepsilon,-\vartheta,\vartheta)=\frac{4i}{\varepsilon}F_{3}(-\vartheta,\vartheta,0)+4F_{3}(-\vartheta,\vartheta,0)\varphi(0)+\mathcal{O}(\varepsilon)\label{F14Sing}
\end{equation}
which is universal as well.

For $F_{5}(i\pi+\vartheta+\varepsilon,i\pi-\vartheta-\varepsilon,-\vartheta,\vartheta,0)$
the most singular term is $O(\epsilon^{-2})$, however, we also need
the $\varepsilon^{-1}$ terms. Denoting the coefficient of the sub-leading
singularity by $F_{5}^{\varepsilon}(\vartheta)$, it can be shown
based on (\ref{eq:FFParametrization}) that

\begin{equation}
\begin{split}F_{5}(i\pi+\vartheta,i\pi-\vartheta,-\vartheta-\varepsilon,\vartheta+\varepsilon,0)= & \frac{1}{\varepsilon^{2}}\left(1-S(\vartheta)\right)\left(1-S(-\vartheta)\right)F_{1}\\
 & +\frac{1}{\varepsilon}F_{5}^{\varepsilon}(\vartheta)+\text{regular\;,}\\
F_{5}(i\pi+\vartheta,i\pi-\vartheta,\vartheta-\varepsilon,-\vartheta+\varepsilon,0)= & \frac{1}{\varepsilon^{2}}S(2\vartheta)\left(1-S(\vartheta)\right)\left(1-S(-\vartheta)\right)F_{1}\\
 & +\frac{1}{\varepsilon}S(2\vartheta)F_{5}^{\varepsilon}(-\vartheta)+\text{regular}\;,
\end{split}
\label{eq:F5e_def}
\end{equation}
where

\begin{equation}
F_{5}^{\varepsilon}(\vartheta)\approx\frac{8F_{1}}{\vartheta}\;,\qquad\vartheta\rightarrow0\:.
\end{equation}

\subsection{Some distribution identities}

Suppose that $f(x)$ is a well behaved function with vanishing at
infinity. Then it is well-known that

\begin{equation}
\int_{-\infty}^{\infty}dx\frac{f(x)}{x-x_{0}+i\epsilon}=P\int_{-\infty}^{\infty}dx\frac{f(x)}{x-x_{0}}-i\pi f(x_{0})\:,
\end{equation}
where $P\int$ denotes the principal value. This identity has the
following counterpart for second order singularity

\begin{equation}
\int_{-\infty}^{\infty}dx\frac{f(x)}{\left(x-x_{0}+i\epsilon\right)^{2}}=P\int_{-\infty}^{\infty}dx\frac{f'(x)}{x-x_{0}}-i\pi f'(x_{0})\:.
\end{equation}
One can also write

\begin{equation}
\int_{-\infty}^{\infty}dx\frac{f(x)}{\sinh\left(x-x_{0}+i\epsilon\right)}=P\int_{-\infty}^{\infty}dx\frac{f(x)}{\sinh\left(x-x_{0}\right)}-i\pi f(x_{0})\:,\label{DistSinh}
\end{equation}
and

\begin{equation}
\int_{-\infty}^{\infty}dxf(x)\frac{\cosh\left(x-x_{0}+i\epsilon\right)}{\sinh^{2}\left(x-x_{0}+i\epsilon\right)}=P\int_{-\infty}^{\infty}dx\frac{f'(x)}{\sinh\left(x-x_{0}\right)}-i\pi f'(x_{0})\:.\label{DistCoshSinh2}
\end{equation}
hold as well. A useful way of evaluating the principal value integral
is

\begin{equation}
P\int_{-\infty}^{\infty}dx\frac{f(x)}{x-x_{0}}=\int_{-\infty}^{\infty}dx\frac{f(x)-f(x_{0})/g(x-x_{0})}{x-x_{0}}\:,
\end{equation}
where $g(x)$ is an appropriate mask function that is even, grows
at the infinity and satisfies $g(0)=1$, a convenient choice being
$g(x)=\cosh x$.

\subsection{Stationary phase approximation\label{subsec:SPA}}

Consider the following integral with oscillatory argument: 
\[
\text{\ensuremath{\frac{1}{2\pi}}}\int dxf(x)e^{itg(x)}\,,
\]
in which $g(x)$ has one global extremum at $x_{0}$, and $f(x)$
is regular and decays fast enough for large $|x|$. The asymptotic
behaviour of this integral for large $t$ can then be evaluated as

\begin{equation}
\text{\ensuremath{\frac{1}{2\pi}}}\int_{-\infty}^{\infty}dxf(x)e^{itg(x)}=\frac{f(x_{0})e^{itg(x_{0})}e^{i\pi/4\text{sign}(g"(x_{0}))}}{\sqrt{2\pi t|g"(x_{0})|}}+\mathcal{O}(t^{-3/2})\;.\label{eq:StacPhase}
\end{equation}

\section{The finite volume 1-point function in the presence of boundaries
\label{sub:Eucl1point}}

The expectation value of a local operator in a finite volume with
boundaries 
\begin{equation}
\langle\mathcal{O}(x)\rangle^{B}=\frac{\langle B|\ e^{-Hx}\mathcal{O}(0)\ e^{-H(\mathcal{R}-x)}\ |B\rangle}{\langle B|e^{-H\mathcal{R}}|B\rangle}=\sum_{k,l}D_{kl}\label{ezkell-1-1}
\end{equation}
was calculated in \cite{OnePointFunctions} up to contributions $D_{kl}$
with $k+l\le4.$ We quote here the result:

\begin{equation}
\begin{split}D_{10} & =\frac{g_{B}}{2}F_{1}^{\mathcal{O}}e^{-mx}\;,\\
D_{20} & =\frac{1}{2}\int\frac{d\vartheta}{2\pi}K_{B}(\vartheta)F_{2}^{\mathcal{O}}(-\vartheta,\vartheta)e^{-2m\cosh\vartheta\ x}\;,\\
D_{30} & =\frac{1}{2}\int\frac{d\vartheta}{2\pi}K_{B}(\vartheta)\frac{g_{B}}{2}F_{3}^{\mathcal{O}}(-\vartheta,\vartheta,0)e^{-m(2\cosh\vartheta+1)\ x}\;,\\
D_{40} & =\frac{1}{8}\int\frac{d\vartheta_{1}}{2\pi}\frac{d\vartheta_{2}}{2\pi}K_{B}(\vartheta_{1})K_{B}(\vartheta_{2})F_{4}^{\mathcal{O}}(-\vartheta_{1},\vartheta_{1},-\vartheta_{2},\vartheta_{2})e^{-2m(\cosh\vartheta_{1}+\cosh\vartheta_{2})\ x}\;,\\
D_{01} & =\frac{g_{B}}{2}F_{1}^{\mathcal{O}}e^{-m(\mathcal{R}-x)}\;,\\
D_{02} & =\frac{1}{2}\int\frac{d\vartheta}{2\pi}K_{B}(\vartheta)F_{2}^{\mathcal{O}}(-\vartheta,\vartheta)e^{-2m\cosh\vartheta\ (\mathcal{R}-x)}\;,\\
D_{03} & =\frac{1}{2}\int\frac{d\vartheta}{2\pi}K_{B}(\vartheta)\frac{g_{B}}{2}F_{3}^{\mathcal{O}}(-\vartheta,\vartheta,0)e^{-m(2\cosh\vartheta+1)\ (\mathcal{R}-x)}\;,\\
D_{04} & =\frac{1}{8}\int\frac{d\vartheta_{1}}{2\pi}\frac{d\vartheta_{2}}{2\pi}K_{B}(\vartheta_{1})K_{B}(\vartheta_{2})F_{4}^{\mathcal{O}}(-\vartheta_{1},\vartheta_{1},-\vartheta_{2},\vartheta_{2})e^{-2m(\cosh\vartheta_{1}+\cosh\vartheta_{2})\ (\mathcal{R}-x)}\;,
\end{split}
\end{equation}
and 
\begin{align}
D_{11} & =\frac{g_{B}^{2}}{4}F_{2,s}^{\mathcal{O}}e^{-m\mathcal{R}}\;,\nonumber \\
D_{21} & =\frac{g_{B}}{4}\int\frac{d\vartheta}{2\pi}\left(F_{3}^{\mathcal{O}}(-\vartheta+i\pi,\vartheta+i\pi,0)K_{B}(\vartheta)e^{-2m\cosh\vartheta x-m(\mathcal{R}-x)}-\frac{2g_{B}^{2}F_{1}^{\mathcal{O}}\cosh\vartheta}{\sinh^{2}\vartheta}e^{-m(\mathcal{R}+x)}\right)\nonumber \\
 & +e^{-m(x+\mathcal{R})}g_{B}^{3}F_{1}^{\mathcal{O}}\frac{\varphi(0)}{4}\;,\nonumber \\
D_{12} & =\frac{g_{B}}{4}\int\frac{d\vartheta}{2\pi}\left(F_{3}^{\mathcal{O}}(i\pi,-\vartheta,\vartheta)K_{B}(\vartheta)e^{-2m\cosh\vartheta(\mathcal{R}-x)-mx}-\frac{2g_{B}^{2}F_{1}^{\mathcal{O}}\cosh\vartheta}{\sinh^{2}\vartheta}e^{-m(2\mathcal{R}-x)}\right)\nonumber \\
 & +e^{-m(2\mathcal{R}-x)}g_{B}^{3}F_{1}^{\mathcal{O}}\frac{\varphi(0)}{4}\;,\nonumber \\
D_{31} & =\frac{g_{B}^{2}}{8}e^{-m\mathcal{R}}\int\frac{d\vartheta}{2\pi}K_{B}(\vartheta)F_{4}^{\mathcal{O}}(-\vartheta+i\pi,\vartheta+i\pi,i\pi,0)e^{-2m\cosh\vartheta\ x}\;,\nonumber \\
D_{13} & =\frac{g_{B}^{2}}{8}e^{-m\mathcal{R}}\int\frac{d\theta}{2\pi}K_{B}(\vartheta)F_{4}^{\mathcal{O}}(-\vartheta+i\pi,\vartheta+i\pi,i\pi,0)e^{-2m\cosh\vartheta\ (\mathcal{R}-x)}\;,\nonumber \\
D_{22} & =\frac{1}{4}\int\frac{d\vartheta_{1}}{2\pi}\frac{d\vartheta_{2}}{2\pi}K_{B}(\vartheta_{1})K_{B}(\vartheta_{2})F_{4}^{\mathcal{O}}(-\vartheta_{1}+i\pi,\vartheta_{1}+i\pi,-\vartheta_{2},\vartheta_{2})e^{-2m\cosh\vartheta_{1}x-2m\cosh\vartheta_{2}(\mathcal{R}-x)}\\
+ & F_{2,s}^{\mathcal{O}}\int\frac{d\vartheta}{2\pi}\Big(K_{B}(-\vartheta)K_{B}(\vartheta)e^{-2m\cosh\vartheta\mathcal{R}}-\frac{g_{B}^{4}\cosh\vartheta}{4\sinh^{2}\vartheta}e^{-2m\mathcal{R}}\Big)+F_{2,s}^{\mathcal{O}}\frac{g_{B}^{4}}{8}e^{-2m\mathcal{R}}\varphi(0)\;,\nonumber 
\end{align}
where \textit{$F_{2,s}^{\mathcal{O}}=F_{2}^{\mathcal{O}}(i\pi,0)$
}\textit{\emph{and 
\[
\varphi(\vartheta)=-i\frac{\partial\log S(\vartheta)}{\partial\vartheta}\;.
\]
}}

\section{Evaluating $D_{12}$\label{sec:D12App}}

In order to analyse the time dependence of the term 
\begin{equation}
\begin{split}\frac{g}{2}\Re e\,\int_{-\infty}^{\infty}\frac{d\vartheta}{2\pi}\left\{ K(\vartheta)F_{3}^{\mathcal{O}}(i\pi,-\vartheta,\vartheta)e^{-imt\left(2\cosh\vartheta-1\right)}-2g^{2}\frac{\cosh\vartheta}{\sinh^{2}\vartheta}F_{1}^{\mathcal{O}}e^{-imt}\right\} \,,\end{split}
\end{equation}
we reintroduce the regulator $R$ enabling us to shift the contour
off the real axis where (as shown in \cite{OnePointFunctions}), the
contribution from the term 
\[
\frac{\cosh\vartheta}{\sinh^{2}\vartheta}
\]
vanishes, i.e., we end up with

\begin{equation}
\begin{split}\frac{g}{2}\Re e\,\int_{-\infty+i\varepsilon}^{\infty+i\varepsilon}\frac{d\vartheta}{2\pi}\left\{ K(\vartheta)F_{3}^{\mathcal{O}}(i\pi,-\vartheta,\vartheta)e^{-imt\left(2\cosh\vartheta-1\right)}e^{-mR/2(2\cosh\vartheta+1)}\right\} \,.\end{split}
\end{equation}
Using now \eqref{DistCoshSinh2}, we find

\begin{equation}
\begin{split}\begin{split}D_{12}(R)= & \frac{g}{2}\Re e\,\int_{-\infty+i\varepsilon}^{\infty+i\varepsilon}\frac{d\vartheta}{2\pi}\frac{\left(\sinh^{2}\vartheta K(\vartheta)F_{3}^{\mathcal{O}}(i\pi,-\vartheta,\vartheta)e^{-imt\left(2\cosh\vartheta-1\right)}e^{-mR/2(2\cosh\vartheta+1)}/\cosh\vartheta\right)'}{\sinh\vartheta}\\
D_{12}= & \frac{g}{2}\Re e\,\int_{-\infty}^{\infty}\frac{d\vartheta}{2\pi}\frac{\left(\sinh^{2}\vartheta K(\vartheta)F_{3}^{\mathcal{O}}(i\pi,-\vartheta,\vartheta)/\cosh\vartheta\right)e^{-imt\left(2\cosh\vartheta-1\right)}\left(-2imt\sinh\vartheta\right)}{\sinh\vartheta}\\
+ & \frac{g}{2}\Re e\,\int_{-\infty}^{\infty}\frac{d\vartheta}{2\pi}\frac{\left(\sinh^{2}\vartheta K(\vartheta)F_{3}^{\mathcal{O}}(i\pi,-\vartheta,\vartheta)e^{-imt\left(2\cosh\vartheta-1\right)}/\cosh\vartheta\right)'e^{-imt\left(2\cosh\vartheta-1\right)}}{\sinh\vartheta}\;,
\end{split}
\end{split}
\end{equation}
where we placed the integration contour back to the real axis as the
integrands are now free of poles for real rapidities and also got
rid off the regulator by setting its value to zero. Performing the
SPA \eqref{eq:StacPhase} results in 
\begin{equation}
\frac{g^{3}F_{1}^{\mathcal{O}}\left(\varphi^{2}(0)-2/3\right)}{2\sqrt{4\pi mt}}\Re e\,e^{-imt}e^{-i\pi/4}+g^{3}\sqrt{\frac{mt}{\pi}}\Re e\,F_{1}^{\mathcal{O}}e^{-imt}\frac{-\sqrt{2}-\sqrt{2}i}{2}\,,\qquad mt\gg1\,.\label{TimeDepg3Appendix}
\end{equation}
While the first term has the standard $\sim1/\sqrt{t}$ time dependence,
the second one behaves as $\sim\sqrt{t}$ for large time. This peculiar
finding is discussed in Sec. \ref{sub:paramres}.

\section{Evaluating $G_{5}$, part I. Notations, $D_{05}$, $D_{14}$ and
residue terms from $D_{23}$\label{sec:D23Res}}

In this and the following appendix we describe how to compute the
five-particle contribution $G_{5}$. According to the discussion in
Section \ref{sec:TimeDep} it can be obtained as (the real part of)
the infinite volume limit of 
\begin{align}
 & \tilde{D}_{05}+\tilde{D}_{14}+\tilde{D}_{23}\;,\\
 & \tilde{D}_{05}=C_{05}\;,\nonumber \\
 & \tilde{D}_{14}=C_{14}-Z_{1}C_{03}\;,\nonumber \\
 & \tilde{D}_{23}=C_{23}-Z_{1}C_{12}-(Z_{2}-Z_{1}^{2})C_{01}\;.\nonumber 
\end{align}
The calculation of $\tilde{D}_{05}$ and $\tilde{D}_{14}$ is easy;
we show explicitly the cancellation of terms with positive powers
of $mL$ and extract the time dependence based on stationary phase
approximation (SPA). It turns out that this time-dependence is only
of sub-leading order.

The calculation of $\tilde{D}_{23}$ is, however, so long and tedious
that we only extract terms accounting for the leading time-dependence
and then carefully argue why other terms produce only sub-leading
effects. In particular, in this Appendix we only focus on residue
terms and leave contributions resulting from a contour integration
for Appendix \ref{sec:CInt}. Finally, we provide numerical evidence
for the existence of the infinite volume limit demonstrating the cancellation
of terms behaving as $mL$ and $\left(mL\right)^{2}$ in Appendix
\ref{sec:Numerics}.

We use the following notation to abbreviate formulas involving form
form factors: whenever a vertical line is seen in the argument of
the form factors, rapidities in the argument refer to rapidity pairs,
and a single rapidity is indicated by putting it into brackets $\{\}$.
Rapidities to the right of the line are understood to be shifted by
$i\pi$. For example,

\[
F_{5}(\vartheta_{1}|\vartheta_{2},\{0\})=F_{5}(i\pi+\vartheta_{1},i\pi-\vartheta_{1},-\vartheta_{2},\vartheta_{2},0)\;,
\]
and

\[
F_{5}(|\vartheta_{1},\vartheta_{2},\{0\})=F_{5}(-\vartheta_{1},\vartheta_{1},-\vartheta_{2},\vartheta_{2},0)\;.
\]
In a similar spirit, we introduce

\begin{align*}
h(\vartheta_{1}|\vartheta_{2},\{0\})_{R} & =e^{imt(2\cosh\vartheta_{1}-2\cosh\vartheta_{2}-1)}e^{-mR/2(2\cosh\vartheta_{1}+2\cosh\vartheta_{2}+1)}\;,
\end{align*}
and

\begin{equation}
h(\vartheta_{1},\vartheta_{2},\{0\}|)_{R}=e^{imt(2\cosh\vartheta_{1}+2\cosh\vartheta_{2}+1)}e^{-mR/2(2\cosh\vartheta_{1}+2\cosh\vartheta_{2}+1)}\;,
\end{equation}
and

\begin{equation}
h(\vartheta_{1}|\vartheta_{2},\{0\})=h(\vartheta_{1}|\vartheta_{2},\{0\})_{R=0}\;.
\end{equation}

\subsection{Evaluation of $\tilde{D}_{05}=C_{05}$ and $\tilde{D}_{14}=C_{14}-Z_{1}C_{03}$}

Let us start with

\begin{equation}
\begin{split}\tilde{D}_{05}= & \frac{1}{2}\sum_{I\neq J}\frac{g}{2}K(\vartheta_{1})K(\vartheta_{2})N_{5}(\vartheta_{1},\vartheta_{2},L)\phantom{}_{L}\langle0|\mathcal{O}(0)|\{I,-I,J,-J,0\}\rangle_{L}\times\\
 & e^{-imt(2\cosh\vartheta_{1}+2\cosh\vartheta_{2}+1)}e^{-Rm/2\left(2\cosh\vartheta_{1}+2\cosh\vartheta_{2}+1\right)}
\end{split}
\end{equation}
which is free of divergences, therefore its infinite volume and $R\rightarrow0$
limit is simply

\begin{equation}
D_{05}=\frac{g}{8}\int_{-\infty}^{\infty}\frac{d\vartheta_{1}}{2\pi}\int_{-\infty}^{\infty}\frac{d\vartheta_{2}}{2\pi}K(\vartheta_{1})K(\vartheta_{2})F_{5}(0,-\vartheta_{1},\vartheta_{1},-\vartheta_{2},\vartheta_{2})e^{-imt(2\cosh\vartheta_{1}+2\cosh\vartheta_{2}+1)}\:.
\end{equation}
Concerning the long time asymptotics of this term, the stationary
points are $\vartheta_{1}=\vartheta_{2}=0$ where the product of the
form factor and the $K$ factors can be expanded in non-negative powers
$\vartheta_{1}^{n}\vartheta_{2}^{m}$. Applying a Gaussian approximation
$\int\frac{d\vartheta_{1}}{2\pi}\int\frac{d\vartheta_{2}}{2\pi}\vartheta_{1}^{n}\vartheta_{2}^{m}e^{-itm\left(1+2\vartheta_{1}^{2}+2\vartheta_{2}^{2}\right)}$
yields time dependence of the form $t^{-1-m-n}$ which we neglect.

Turning now to

\begin{equation}
\begin{split}C_{14}= & \frac{1}{2}\sum_{I\neq J}\frac{g}{2}K(\vartheta_{1})K(\vartheta_{2})N_{1}N_{4}(\vartheta_{1},\vartheta_{2},L)\phantom{}_{L}\langle\{0\}|O(0)|\{-I,I,J,-J\}\rangle_{L}\times\\
 & e^{-imt(2\cosh\vartheta_{1}+2\cosh\vartheta_{2}-1)}e^{-Rm/2\left(2\cosh\vartheta_{1}+2\cosh\vartheta_{2}+1\right)}\;,
\end{split}
\end{equation}
the corresponding form factor $F_{5}(i\pi,-\vartheta_{1},\vartheta_{1},-\vartheta_{2},\vartheta_{2})$
has a $\vartheta^{2}$ behaviour around the origin when $\vartheta_{1}=\vartheta_{2}=\vartheta$,
therefore it remains regular even when multiplied with the two $K$
functions. However when only one of the rapidities is close to zero
then it has a first order pole $F_{5}\propto\frac{1}{\vartheta}$,
hence taking into account the singularity of the appropriate $K$
function leads to a second order pole. Following the formalism introduced
in \cite{FiniteTCorr}, one can write the sum using contour integrals

\begin{equation}
\begin{split}C_{14}= & \frac{g}{4}\left(-\frac{1}{2}\right)\sum_{I\neq0}\oint_{C_{I}}\frac{d\vartheta_{1}}{2\pi}\left(-\frac{1}{2}\right)\sum_{J\neq0}\oint_{C_{J}}\frac{d\vartheta_{2}}{2\pi}K(\vartheta_{1})K(\vartheta_{2})\frac{F_{5}(i\pi,-\vartheta_{1},\vartheta_{1},-\vartheta_{2},\vartheta_{2})}{\left(e^{i\bar{Q}_{4,1}}+1\right)\left(e^{i\bar{Q}_{4,2}}+1\right)}\times\\
 & e^{-imt(2\cosh\vartheta_{1}+2\cosh\vartheta_{2}-1)}e^{-Rm/2\left(2\cosh\vartheta_{1}+2\cosh\vartheta_{2}+1\right)}\;,
\end{split}
\end{equation}
where the two-dimensional product contour $C_{I}\times C_{J}$ encircles
the solution of the Bethe-Yang equation determining $\vartheta_{1}$
and $\vartheta_{2}$ with quantum numbers $I$ and $J$ . To open
the contours it is necessary to subtract the residue terms when $\vartheta_{1}=0$
or $\vartheta_{2}=0$. Hence

\begin{equation}
\begin{split}C_{14} & =\frac{g}{16}\int_{-\infty+i\varepsilon}^{\infty+i\varepsilon}\frac{d\vartheta_{1}}{2\pi}\int_{-\infty+i\varepsilon}^{\infty+i\varepsilon}\frac{d\vartheta_{2}}{2\pi}K(\vartheta_{1})K(\vartheta_{2})\frac{F_{5}(\{0\}|\vartheta_{1},\vartheta_{2})}{\left(e^{i\bar{Q}_{4,1}}+1\right)\left(e^{i\bar{Q}_{4,2}}+1\right)}h(\{0\}|\vartheta_{1},\vartheta_{2})_{R}\\
 & +\frac{g}{4}\left(-\frac{1}{2}\right)\sum_{I\neq0}\oint_{C_{I}}\frac{d\vartheta_{1}}{2\pi}\left(\frac{1}{2}\right)\oint_{C_{0}}\frac{d\vartheta_{2}}{2\pi}K(\vartheta_{1})K(\vartheta_{2})\frac{F_{5}(\{0\}|\vartheta_{1},\vartheta_{2})}{\left(e^{i\bar{Q}_{4,1}}+1\right)\left(e^{i\bar{Q}_{4,2}}+1\right)}h(\{0\}|\vartheta_{1},\vartheta_{2})_{R}\\
 & +\frac{g}{4}\left(-\frac{1}{2}\right)\sum_{J\neq0}\oint_{C_{J}}\frac{d\vartheta_{2}}{2\pi}\left(\frac{1}{2}\right)\oint_{C_{0}}\frac{d\vartheta_{1}}{2\pi}K(\vartheta_{1})K(\vartheta_{2})\frac{F_{5}(\{0\}|\vartheta_{1},\vartheta_{2})}{\left(e^{i\bar{Q}_{4,1}}+1\right)\left(e^{i\bar{Q}_{4,2}}+1\right)}h(\{0\}|\vartheta_{1},\vartheta_{2})_{R}\;,
\end{split}
\label{eq:Dtilde14split}
\end{equation}
where we used that on the contours with imaginary parts $-i\varepsilon$
$e^{i\bar{Q}_{4,k}}\rightarrow\infty$ in the infinite volume limit.
Now we split $C_{14}$ according to the lines in (\ref{eq:Dtilde14split})
as $C_{14}=C_{14}^{int}+C{}_{14}^{res1}+C_{14}^{res2}$. Clearly,
$C_{14}^{res1}=C_{14}^{res2}=:C_{14}^{res}$ and the infinite volume
limit of $C_{14}^{int}$ is regular: 
\[
\lim_{L\rightarrow\infty}C_{14}^{int}=\frac{g}{16}\int_{-\infty+i\varepsilon}^{\infty+i\varepsilon}\frac{d\vartheta_{1}}{2\pi}\int_{-\infty+i\varepsilon}^{\infty+i\varepsilon}\frac{d\vartheta_{2}}{2\pi}K(\vartheta_{1})K(\vartheta_{2})F_{5}(\{0\}|\vartheta_{1},\vartheta_{2})h(\{0\}|\vartheta_{1},\vartheta_{2})_{R}\,,
\]
since $e^{i\bar{Q}_{4,k}}\rightarrow0$ in the infinite volume limit
on the contours with imaginary parts $+i\varepsilon$. The residue
contribution is

\begin{equation}
\begin{split}2C_{14}^{res} & =\frac{ig}{4}\left(-\frac{1}{2}\right)\sum_{I\neq0}\oint_{C_{I}}\frac{d\vartheta_{1}}{2\pi}\oint_{C_{0}}\frac{d\vartheta_{2}}{2\pi i}K(\vartheta_{1})K(\vartheta_{2})\frac{F_{5}(\{0\}|\vartheta_{1},\vartheta_{2})}{\left(e^{i\bar{Q}_{4,1}}+1\right)\left(e^{i\bar{Q}_{4,2}}+1\right)}h(\{0\}|\vartheta_{1},\vartheta_{2})_{R}\end{split}
\;.
\end{equation}
For $\vartheta_{2}\thickapprox0$ based on (\ref{F14Sing})

\begin{equation}
\begin{split}F_{5}(\{0\}|\vartheta_{1},\vartheta_{2})K(\vartheta_{2}) & =4iF_{3}(-\vartheta_{1},\vartheta,0)\left(\frac{1}{\vartheta_{2}}-i\varphi(0)\right)\left(\frac{-ig^{2}}{2\vartheta_{2}}+\frac{g^{2}}{2}\varphi(0)\right)+\text{regular}\\
 & =F_{3}(-\vartheta_{1},\vartheta,0)\left(\frac{2g^{2}}{\vartheta_{2}^{2}}\right)+\text{regular}\:,
\end{split}
\end{equation}
in the contour integral around the $\vartheta_{2}=0$ point the $1/\vartheta_{2}^{2}$
term acts as a differentiation on the other regular $\vartheta_{2}$
dependent factor, leading to

\begin{equation}
C_{14}^{res}=\frac{g^{3}}{16}\int_{-\infty}^{\infty}\frac{d\vartheta}{2\pi}F_{3}(|\{0\},\vartheta_{1})K(\vartheta_{1})h(\{0\}|\vartheta_{1},0)_{R}\left(mL+2\varphi(0)+2\varphi(\vartheta_{1})\right)+O\left(L^{-1}\right)\;.\label{eq:2D14res}
\end{equation}
Turning to $-Z_{1}C_{03}$ we have 
\[
Z_{1}=\frac{g^{2}}{4}mLe^{-mR}
\]
and 
\[
C_{03}=\frac{1}{2}\frac{g}{2}\int\frac{d\vartheta}{2\pi}F_{3}(0,-\vartheta_{1},\vartheta_{1})K(\vartheta_{1})h(\{0\}|\vartheta_{1},0)_{R}\,.
\]
Therefore the $O(L)$ term in (\ref{eq:2D14res}) is cancelled, resulting
in

\begin{equation}
\begin{split}D_{14} & =\lim_{L\rightarrow\infty}C_{14}-Z_{1}C_{03}\\
 & =\frac{g}{16}\int_{-\infty+i\varepsilon}^{\infty+i\varepsilon}\frac{d\vartheta_{1}}{2\pi}\int_{-\infty+i\varepsilon}^{\infty+i\varepsilon}\frac{d\vartheta_{2}}{2\pi}K(\vartheta_{1})K(\vartheta_{2})F_{5}(\{0\}|\vartheta_{1},\vartheta_{2})h(\{0\}|\vartheta_{1},\vartheta_{2})_{R}\\
 & +\frac{g^{3}}{16}\int_{-\infty}^{\infty}\frac{d\vartheta}{2\pi}F_{3}(|\{0\},\vartheta_{1})K(\vartheta_{1})h(\{0\}|\vartheta_{1},0)_{R}\left(2\varphi(0)+2\varphi(\vartheta_{1})\right)\;.
\end{split}
\end{equation}
Pulling the contours back to the real axis by sending $\varepsilon$
to zero, one can then send $R$ to zero as well with the final result

\begin{equation}
\begin{split}D_{14} & =\frac{g}{16}\int_{-\infty}^{\infty}\frac{d\vartheta_{1}}{2\pi}\int_{-\infty}^{\infty}\frac{d\vartheta_{2}}{2\pi}\left\{ \vphantom{\frac{F}{\left(e^{iQ}\right)}}K(\vartheta_{1})K(\vartheta_{2})F_{5}(i\pi,-\vartheta_{1},\vartheta_{1},-\vartheta_{2},\vartheta_{2})e^{-itm\left(2\cosh\vartheta_{1}+2\cosh\vartheta_{2}-1\right)}\right.\\
 & -2g^{2}F_{3}(0,-\vartheta_{2},\vartheta_{2})\frac{\cosh\vartheta_{1}}{\sinh^{2}\vartheta_{1}}K(\vartheta_{2})e^{-itm\left(2\cosh\vartheta_{2}+1\right)}\\
 & \left.-2g^{2}F_{3}(0,-\vartheta_{1},\vartheta_{1})\frac{\cosh\vartheta_{2}}{\sinh^{2}\vartheta_{2}}K(\vartheta_{1})e^{-itm\left(2\cosh\vartheta_{1}+1\right)}\right\} \\
 & +\frac{g^{3}}{16}\int_{-\infty}^{\infty}\frac{d\vartheta}{2\pi}F_{3}(0,-\vartheta_{1},\vartheta_{1})K(\vartheta_{1})e^{-itm\left(2\cosh\vartheta_{1}+1\right)}\left(2\varphi(0)+2\varphi(\vartheta_{1})\right)\;.
\end{split}
\end{equation}
Addressing the time dependence of this term, note that the structure
of $D_{14}$ is reminiscent of $D_{12}$ discussed in Section \ref{subsec:TimeDepG4}\textbf{
}resulting in a $\sqrt{t}$ type behaviour again due to a mechanism
analogous to parametric resonance. The difference from the case of
$D_{12}$ is that the oscillations are of the form $\cos3mt$ instead
of $\cos mt$. Since our primary focus is on one-particle oscillations,
we do not discuss this term further here.

\subsection{Evaluation of $D_{23}=C_{23}-Z_{1}C_{12}-(Z_{2}-Z_{1}^{2})C_{01}$}

Consider

\begin{equation}
\begin{split}C_{23} & =\frac{g}{2}\sum_{I\geq0}\sum_{J\geq0}N_{2}N_{3}K(-\vartheta_{1})K(\vartheta_{2})\phantom{}_{L}\langle\{I,-I\}|\mathcal{O}(0)|\{J,-J,0\}\rangle_{L}\text{\ensuremath{\times}}\\
 & e^{imt(2\cosh\vartheta_{1}-2\cosh\vartheta_{2}-1)}e^{-Rm/2\left(2\cosh\vartheta_{1}+2\cosh\vartheta_{2}+1\right)}\\
 & =\frac{g}{2}\sum_{I\geq0}\sum_{J\geq0}K(-\vartheta_{1})K(\vartheta_{2})\frac{F_{5}(i\pi+\vartheta_{1},i\pi-\vartheta_{1},-\vartheta_{2},\vartheta_{2},0)}{\bar{\rho}_{2}(\vartheta_{1})\bar{\rho}_{3}(\vartheta_{2})}\times\\
 & e^{imt(2\cosh\vartheta_{1}-2\cosh\vartheta_{2}-1)}e^{-Rm/2\left(2\cosh\vartheta_{1}+2\cosh\vartheta_{2}+1\right)}+\mathcal{O}(e^{-\mu L})\;,
\end{split}
\end{equation}
where $I$ and $J$ are the quantum numbers specifying $\vartheta_{1}$
and $\vartheta_{2}$, i.e. 
\[
\bar{Q}_{2}(\vartheta_{1})=2\pi I\;,\qquad\bar{Q}_{3}(\vartheta_{2})=2\pi J\;,
\]
where $\bar{Q}_{2}$ and $\bar{Q}_{3}$ are defined in (\ref{barQ1})
and (\ref{barQ3-1}), respectively. The density factors are given
by 
\[
\bar{\rho}_{2}(\vartheta_{1})=\frac{\partial\bar{Q}_{2}(\vartheta_{1})}{\partial\vartheta_{1}}\;,\qquad\bar{\rho}_{3}(\vartheta_{1})=\frac{\partial\bar{Q}_{3}(\vartheta_{2})}{\partial\vartheta_{2}}\;.
\]
Note that $I$ takes half-integer, while $J$ takes integer values,
according to the discussion in Appendix \ref{subsec:Multi-particle-states-in}.
The expression 
\[
F_{5}(i\pi+\vartheta_{1},i\pi-\vartheta_{1},-\vartheta_{2},\vartheta_{2},0)K^{*}(\vartheta_{1})K(\vartheta_{2})
\]
is singular when $\vartheta_{1}=\vartheta_{2}$ or when $\vartheta_{1}=0$
and $\vartheta_{2}$ is finite, and these singularities are of second
order. One can first write the sum over $J$ as a contour integral:

\begin{equation}
\begin{split}C_{23} & =\frac{g}{2}\left(\frac{1}{2}\right)^{2}\sum_{J\neq0}\sum_{I\neq0}\oint_{C_{J}}\frac{d\vartheta_{2}}{2\pi}h(\vartheta_{1}|\vartheta_{2},\{0\})_{R}\,K^{*}(\vartheta_{1})K(\vartheta_{2})\frac{F_{5}(\vartheta_{1}|\vartheta_{2},\{0\})}{\bar{\rho}_{2}(\vartheta_{1})\left(e^{i\bar{Q}_{3}(\vartheta_{2})}-1\right)}\:,\end{split}
\label{C23Discrete}
\end{equation}
where $C_{J}$ surrounds the positions $\vartheta_{2}$ corresponding
to $J$. Opening the contour leads to

\begin{equation}
\begin{split}C_{23} & =\frac{g}{2}\left(\frac{1}{2}\right)^{2}\sum_{I\neq0}\int_{-\infty+i\varepsilon}^{\infty+i\varepsilon}\frac{d\vartheta_{2}}{2\pi}h(\vartheta_{1}|\vartheta_{2},\{0\})_{R}\,K(-\vartheta_{1})K(\vartheta_{2})\frac{F_{5}(\vartheta_{1}|\vartheta_{2},\{0\})}{\bar{\rho}_{2}(\vartheta_{1})\left(e^{i\bar{Q}_{3}(\vartheta_{2})}-1\right)}\\
 & +\frac{g}{2}\left(\frac{1}{2}\right)^{2}\sum_{I\neq0}\int_{-\infty-i\varepsilon}^{\infty-i\varepsilon}\frac{d\vartheta_{2}}{2\pi}h(\vartheta_{1}|\vartheta_{2},\{0\})_{R}\,K(-\vartheta_{1})K(\vartheta_{2})\frac{F_{5}(\vartheta_{1}|\vartheta_{2},\{0\})}{\bar{\rho}_{2}(\vartheta_{1})\left(e^{i\bar{Q}_{3}(\vartheta_{2})}-1\right)}\\
 & -i\frac{g}{2}\left(\frac{1}{2}\right)^{2}\sum_{I\neq0}\oint_{C_{\vartheta_{1}}}\frac{d\vartheta_{2}}{2\pi i}h(\vartheta_{1}|\vartheta_{2},\{0\})_{R}\,K(-\vartheta_{1})K(\vartheta_{2})\frac{F_{5}(\vartheta_{1}|\vartheta_{2},\{0\})}{\bar{\rho}_{2}(\vartheta_{1})\left(e^{i\bar{Q}_{3}(\vartheta_{2})}-1\right)}\\
 & -i\frac{g}{2}\left(\frac{1}{2}\right)^{2}\sum_{I\neq0}\oint_{C_{-\vartheta_{1}}}\frac{d\vartheta_{2}}{2\pi i}h(\vartheta_{1}|\vartheta_{2},\{0\})_{R}\,K(-\vartheta_{1})K(\vartheta_{2})\frac{F_{5}(\vartheta_{1}|\vartheta_{2},\{0\})}{\bar{\rho}_{2}(\vartheta_{1})\left(e^{i\bar{Q}_{3}(\vartheta_{2})}-1\right)}\:,
\end{split}
\label{C23IntAndResPre}
\end{equation}
where in the first two terms we perform the $mL\rightarrow\infty$
only for the rapidities $\vartheta_{2}$, from which only the integration
along the contour above the real axis survives. We thus have

\begin{equation}
\begin{split}C_{23} & =\frac{g}{2}\left(\frac{1}{2}\right)^{2}\sum_{I\neq0}\int_{-\infty+i\varepsilon}^{\infty+i\varepsilon}\frac{d\vartheta_{2}}{2\pi}h(\vartheta_{1}|\vartheta_{2},\{0\})_{R}\,K(-\vartheta_{1})K(\vartheta_{2})\frac{F_{5}(\vartheta_{1}|\vartheta_{2},\{0\})}{\bar{\rho}_{2}(\vartheta_{1})}\\
 & -i\frac{g}{2}\left(\frac{1}{2}\right)^{2}\sum_{I\neq0}\oint_{C_{\vartheta_{1}}}\frac{d\vartheta_{2}}{2\pi i}h(\vartheta_{1}|\vartheta_{2},\{0\})_{R}\,K(-\vartheta_{1})K(\vartheta_{2})\frac{F_{5}(\vartheta_{1}|\vartheta_{2},\{0\})}{\bar{\rho}_{2}(\vartheta_{1})\left(e^{i\bar{Q}_{3}(\vartheta_{2})}-1\right)}\\
 & -i\frac{g}{2}\left(\frac{1}{2}\right)^{2}\sum_{I\neq0}\oint_{C_{-\vartheta_{1}}}\frac{d\vartheta_{2}}{2\pi i}h(\vartheta_{1}|\vartheta_{2},\{0\})_{R}\,K(-\vartheta_{1})K(\vartheta_{2})\frac{F_{5}(\vartheta_{1}|\vartheta_{2},\{0\})}{\bar{\rho}_{2}(\vartheta_{1})\left(e^{i\bar{Q}_{3}(\vartheta_{2})}-1\right)}\:.
\end{split}
\label{C23IntAndRes}
\end{equation}
This contour manipulation was checked numerically using known form
factor solutions and comparing (\ref{C23Discrete}) and (\ref{C23IntAndRes})
for finite $R$. The terms in the three lines in eqn. (\ref{C23IntAndRes})
are written in short as 
\[
C_{23}=C_{23}^{int}+C_{23}^{res1A}+C_{23}^{res1B}\,.
\]

\subsubsection{Time dependence from residue term $C_{23}^{res1A}$ and $C_{23}^{res1B}$}

Consider the residue terms

\begin{equation}
\begin{split}C_{23}^{res1A}= & -i\frac{g}{2}\left(\frac{1}{2}\right)^{2}\sum_{I\neq0}\oint_{C_{\vartheta_{1}}}\frac{d\vartheta_{2}}{2\pi i}h(\vartheta_{1}|\vartheta_{2},\{0\})_{R}\,K(-\vartheta_{1})K(\vartheta_{2})\frac{F_{5}(\vartheta_{1}|\vartheta_{2},\{0\})}{\bar{\rho}_{2}(\vartheta_{1})\left(e^{i\bar{Q}_{3}(\vartheta_{2})}-1\right)}\\
= & -i\frac{g}{2}\left(\frac{1}{2}\right)^{2}F_{1}\sum_{I\neq0}\frac{K(-\vartheta_{1})\left(1-S(\vartheta_{1})\right)\left(1-S(-\vartheta_{1})\right)h(\vartheta_{1}|\vartheta_{1},\{0\})_{R}}{\bar{\rho}_{2}(\vartheta_{1})}\times\\
 & \left\{ K(\vartheta_{1})\frac{\left(-2imt-Rm\right)\sinh\vartheta_{1}}{\left(S(\vartheta_{1})-1\right)}+\frac{K'(\vartheta_{1})}{\left(S(\vartheta_{1})-1\right)}\right.\\
 & \left.-\frac{K(\vartheta_{1})S(\vartheta_{1})i\left(mL\cosh\vartheta_{1}+2\varphi(2\vartheta_{1})+\varphi(\vartheta_{1})\right)}{\left(S(\vartheta_{1})-1\right)^{2}}\right\} \\
- & i\frac{g}{2}\left(\frac{1}{2}\right)^{2}F_{1}\sum_{I\neq0}\frac{K(-\vartheta_{1})K(\vartheta_{1})F_{5}^{\varepsilon}(\vartheta_{1})h(\vartheta_{1}|\vartheta_{1},\{0\})_{R}}{\left(S(\vartheta_{1})-1\right)\bar{\rho}_{2}(\vartheta_{1})}\;,\\
\\
\end{split}
\end{equation}
and

\begin{equation}
\begin{split}C_{23}^{res1B}= & -i\frac{g}{2}\left(\frac{1}{2}\right)^{2}\sum_{I\neq0}\oint_{C_{-\vartheta_{1}}}\frac{d\vartheta_{2}}{2\pi i}h(\vartheta_{1}|\vartheta_{2},\{0\})_{R}\,K(-\vartheta_{1})K(\vartheta_{2})\frac{F_{5}(\vartheta_{1}|\vartheta_{2},\{0\})}{\bar{\rho}_{2}(\vartheta_{1})\left(e^{i\bar{Q}_{3}(\vartheta_{2})}-1\right)}\\
= & -i\frac{g}{2}\left(\frac{1}{2}\right)^{2}F_{1}\sum_{I\neq0}\frac{K(-\vartheta_{1})S(2\vartheta_{1})\left(1-S(\vartheta_{1})\right)\left(1-S(-\vartheta_{1})\right)h(\vartheta_{1}|\vartheta_{1},\{0\})_{R}}{\bar{\rho}_{2}(\vartheta_{1})}\times\\
 & \left\{ K(-\vartheta_{1})\frac{(-1)\left(-2imt-Rm\right)\sinh\vartheta_{1}}{\left(S(-\vartheta_{1})-1\right)}+\frac{K'(-\vartheta_{1})}{\left(S(-\vartheta_{1})-1\right)}\right.\\
 & \left.-\frac{K(-\vartheta_{1})S(-\vartheta_{1})i\left(mL\cosh\vartheta_{1}+2\varphi(2\vartheta_{1})+\varphi(\vartheta_{1})\right)}{\left(S(-\vartheta_{1})-1\right)^{2}}\right\} \\
 & -i\frac{g}{2}\left(\frac{1}{2}\right)^{2}F_{1}\sum_{I\neq0}\frac{K(-\vartheta_{1})K(-\vartheta_{1})F_{5}^{\varepsilon}(-\vartheta_{1})S(2\vartheta_{1})h(\vartheta_{1}|\vartheta_{1},\{0\})_{R}}{\left(S(-\vartheta_{1})-1\right)\bar{\rho}_{2}(\vartheta_{1})}\;,\\
\\
\end{split}
\end{equation}
where (\ref{eq:F5e_def}) was made use of. Note, that both the second
and first order singularities are to be taken into account. Adding
$C_{23}^{res1A}$ and $C_{23}^{res1B}$, one has

\begin{equation}
\begin{split}C_{23}^{res1}= & -i\frac{g}{2}\left(\frac{1}{2}\right)^{2}F_{1}\sum_{I\neq0}|K(\vartheta_{1})|^{2}\frac{\left(-2i\Im m\,S(\vartheta_{1})\right)\left(-2imt-Rm\right)\sinh\vartheta_{1}}{\bar{\rho}_{2}(\vartheta_{1})}h(\vartheta_{1}|\vartheta_{1},\{0\})_{R}\\
 & -i\frac{g}{2}\left(\frac{1}{2}\right)^{2}F_{1}\sum_{I\neq0}2\frac{|K(\vartheta_{1})|^{2}i\left(mL\cosh\vartheta_{1}+2\varphi(2\vartheta_{1})+\varphi(\vartheta_{1})\right)h(\vartheta_{1}|\vartheta_{1},\{0\})_{R}}{\bar{\rho}_{2}(\vartheta_{1})}\\
 & +i\frac{g}{2}\left(\frac{1}{2}\right)^{2}F_{1}\sum_{I\neq0}\frac{K(\vartheta_{1})K'(-\vartheta_{1})\left(1-S(\vartheta_{1})\right)+K(-\vartheta_{1})K'(\vartheta_{1})\left(1-S(-\vartheta_{1})\right)}{\bar{\rho}_{2}(\vartheta_{1})}h(\vartheta_{1}|\vartheta_{1},\{0\})_{R}\\
 & -i\frac{g}{2}\left(\frac{1}{2}\right)^{2}\sum_{I\neq0}\frac{|K(\vartheta_{1})|^{\text{2}}h(\vartheta_{1}|\vartheta_{1},\{0\})_{R}}{\bar{\rho}_{2}(\vartheta_{1})}\left(\frac{F_{5}^{\varepsilon}(\vartheta_{1})}{\left(S(\vartheta_{1})-1\right)}+\frac{F_{5}^{\varepsilon}(-\vartheta_{1})}{\left(S(-\vartheta_{1})-1\right)}\right)\;,
\end{split}
\end{equation}
or after some manipulations

\begin{equation}
\begin{split}C_{23}^{res1}= & -\frac{g}{2}\left(\frac{1}{2}\right)^{2}F_{1}\sum_{I\neq0}|K(\vartheta_{1})|^{2}\frac{\left(2\Im m\,S(\vartheta_{1})\right)\left(-2imt-Rm\right)\sinh\vartheta_{1}}{\bar{\rho}_{2}(\vartheta_{1})}h(\vartheta_{1}|\vartheta_{1},\{0\})_{R}\\
 & +\frac{g}{2}\left(\frac{1}{2}\right)^{2}F_{1}\sum_{I\neq0}2\frac{|K(\vartheta_{1})|^{2}\left(mL\cosh\vartheta_{1}+2\varphi(2\vartheta_{1})+\varphi(\vartheta_{1})\right)h(\vartheta_{1}|\vartheta_{1},\{0\})_{R}}{\bar{\rho}_{2}(\vartheta_{1})}\\
 & -\frac{g}{2}\left(\frac{1}{2}\right)^{2}F_{1}\sum_{I\neq0}\frac{\Im m\,S(\vartheta_{1})\left(|K(\vartheta_{1})|^{2}\right)'+2\varphi(2\vartheta_{1})|K(\vartheta_{1})|^{2}\text{Re}\left(1-S(\vartheta_{1})\right)}{\bar{\rho}_{2}(\vartheta_{1})}h(\vartheta_{1}|\vartheta_{1},\{0\})_{R}\\
 & -i\frac{g}{2}\left(\frac{1}{2}\right)^{2}\sum_{I\neq0}\frac{|K(\vartheta_{1})|^{\text{2}}h(\vartheta_{1}|\vartheta_{1},\{0\})_{R}}{\bar{\rho}_{2}(\vartheta_{1})}\left(\frac{F_{5}^{\varepsilon}(\vartheta_{1})}{\left(S(\vartheta_{1})-1\right)}+\frac{F_{5}^{\varepsilon}(-\vartheta_{1})}{\left(S(-\vartheta_{1})-1\right)}\right)\;,
\end{split}
\end{equation}
The four terms of $C_{23}^{res1}$ have singularities at $\vartheta_{1}=0$.
However, these singularities can only produce terms with positive
powers in $mL$ but no non-trivial time dependence since $h(\vartheta_{1}|\vartheta_{1},\{0\})_{R}$
contains no $\vartheta_{1}$ dependent function multiplied by $t$.
We therefore have a single secular term, namely

\begin{equation}
\begin{split}C_{23}^{res/sec}= & \frac{g}{2}F_{1}e^{-imt}\left(imt\right)\int_{-\infty}^{\infty}\frac{d\vartheta}{2\pi}|K(\vartheta)|^{2}\Im m\,S(\vartheta)\sinh\vartheta e^{-Rm/2\left(4\cosh\vartheta+1\right)}\\
 & +\frac{g}{2}F_{1}e^{-imt}\left(Rm/2\right)\int_{-\infty}^{\infty}\frac{d\vartheta}{2\pi}|K(\vartheta)|^{2}\Im m\,S(\vartheta)\sinh\vartheta e^{-Rm/2\left(4\cosh\vartheta+1\right)}\;,
\end{split}
\end{equation}
and taking the limit $R\rightarrow0$ results in 
\begin{equation}
\begin{split}C_{23}^{res/sec}= & \frac{g}{2}F_{1}e^{-imt}\left(imt\right)\int_{-\infty}^{\infty}\frac{d\vartheta}{2\pi}|K(\vartheta)|^{2}\Im m\,S(\vartheta)\sinh\vartheta\end{split}
\;.
\end{equation}

\section{Evaluating $G_{5}$, part II. Contour integral terms from $D_{23}$\label{sec:CInt}}

In this appendix we evaluate the contour integral from (\ref{C23IntAndResPre})
which reads

\begin{equation}
C_{23}^{int}=\frac{g}{2}\left(\frac{1}{2}\right)^{2}\sum_{I\neq0}\int_{-\infty+i\varepsilon}^{\infty+i\varepsilon}\frac{d\vartheta_{2}}{2\pi}h(\vartheta_{1}|\vartheta_{2},\{0\})_{R}\,K(-\vartheta_{1})K(\vartheta_{2})\frac{F_{5}(\vartheta_{1}|\vartheta_{2},\{0\})}{\bar{\rho}_{2}(\vartheta_{1})}\:.\label{eq:CInt23}
\end{equation}
Let us start with a summary of our method first. Manipulating (\ref{eq:CInt23})
leads to various terms of which only (\ref{eq:CInt23BI2}) and (\ref{CIntBI-II})
give rise to interesting time dependence. The origin of these terms
was made clear at the beginning of this Appendix, but their actual
evaluation needs further non trivial integral manipulations discussed
in Appendix \ref{sec:EvaluatingKernel}. The largest part of this
Appendix is dedicated to showing that apart from (\ref{eq:CInt23BI2})
and (\ref{CIntBI-II}) no other term yields any interesting time dependence.

Starting from (\ref{eq:CInt23}) we can subtract and add back the
singularities of the five particle form factors. Using (\ref{eq:F5e_def}),
this leads to

\begin{equation}
\begin{split}C_{23}^{int}= & \frac{g}{2}\left(\frac{1}{2}\right)^{2}\sum_{I\neq0}\int_{-\infty+i\varepsilon}^{\infty+i\varepsilon}\frac{d\vartheta_{2}}{2\pi}\left\{ \vphantom{\frac{F}{\left(e^{iQ}\right)}}\frac{h(\vartheta_{1}|\vartheta_{2},\{0\})_{R}\,K(-\vartheta_{1})K(\vartheta_{2})}{\bar{\rho}_{2}(\vartheta_{1})}\left[\vphantom{\frac{\frac{OOOO}{OO}}{\frac{OOOO}{OO}}}F_{5}(\vartheta_{1}|\vartheta_{2},\{0\})\right.\right.\\
 & \left.\left.-\Omega(\vartheta_{1})F_{1}\left(\frac{\cosh(\vartheta_{2}-\vartheta_{1})}{\sinh^{2}(\vartheta_{2}-\vartheta_{1})}+\frac{S(2\vartheta_{1})\cosh(\vartheta_{2}+\vartheta_{1})}{\sinh^{2}(\vartheta_{2}+\vartheta_{1})}\right)-\frac{F_{5}^{\varepsilon}(\vartheta_{1})}{\sinh(\vartheta_{2}-\vartheta_{1})}-\frac{S(2\vartheta_{1})F_{5}^{\varepsilon}(-\vartheta_{1})}{\sinh(\vartheta_{2}+\vartheta_{1})}\right]\right\} \\
+ & \frac{g}{2}\left(\frac{1}{2}\right)^{2}\sum_{I\neq0}\int_{-\infty+i\varepsilon}^{\infty+i\varepsilon}\frac{d\vartheta_{2}}{2\pi}\left\{ \vphantom{\frac{F}{\left(e^{iQ}\right)}}\frac{h(\vartheta_{1}|\vartheta_{2},\{0\})_{R}K(-\vartheta_{1})K(\vartheta_{2})}{\bar{\rho}_{2}(\vartheta_{1})}\times\right.\\
 & \left.\Omega(\vartheta_{1})F_{1}\left(\frac{\cosh(\vartheta_{2}-\vartheta_{1})}{\sinh^{2}(\vartheta_{2}-\vartheta_{1})}+\frac{S(2\vartheta_{1})\cosh(\vartheta_{2}+\vartheta_{1})}{\sinh^{2}(\vartheta_{2}+\vartheta_{1})}\right)\right\} \\
+ & \frac{g}{2}\left(\frac{1}{2}\right)^{2}\sum_{I\neq0}\int_{-\infty+i\varepsilon}^{\infty+i\varepsilon}\frac{d\vartheta_{2}}{2\pi}\left\{ \frac{h(\vartheta_{1}|\vartheta_{2},\{0\})_{R}\,K(-\vartheta_{1})K(\vartheta_{2})}{\bar{\rho}_{2}(\vartheta_{1})}\left(\frac{F_{5}^{\varepsilon}(\vartheta_{1})}{\sinh(\vartheta_{2}-\vartheta_{1})}+\frac{S(2\vartheta_{1})F_{5}^{\varepsilon}(-\vartheta_{1})}{\sinh(\vartheta_{2}+\vartheta_{1})}\right)\right\} \\
= & C_{23}^{intA}+C_{23}^{intBI}+C_{23}^{intBII}\;,
\end{split}
\end{equation}
where $\Omega(\vartheta)=\left(1-S(\vartheta)\right)\left(1-S(-\vartheta)\right)$.

To make the rather technical evaluation more transparent, we introduce
an additional simplification which does not affect the end result.
Apart from (\ref{eq:CInt23BI2}) and (\ref{CIntBI-II}), many other
terms also possess singularities that in principle could lead to non-trivial
time-dependence, but cancel each other in the end. These singularities
emerge from the region where $\vartheta_{1}$ is around zero. But
at zero rapidity $S(0)=-1$, and all other quantities behave similar
to the Ising model which has a constant $S=-1$ everywhere, resulting
in $K(\vartheta)$ being an odd function of $\vartheta$. A good example
is the sub-leading singularity of the form factor $F_{5}^{\varepsilon}(\vartheta_{1})$
defined in (\ref{eq:F5e_def}), which for small $\vartheta_{1}$ behaves
as 
\[
\frac{8F_{1}}{\vartheta_{1}}\:,
\]
whereas its Ising counterpart is exactly 
\begin{equation}
F_{5}^{\varepsilon}(\vartheta_{1})=\frac{8F_{1}}{\sinh\vartheta_{1}}\,.\label{eq:F5eIsing}
\end{equation}
Therefore to keep the reasoning as simple as possible, from now on
we perform our calculations for the Ising scattering matrix $S=-1$
and show the cancellation of certain singularities. It turns out that
in the only nontrivial time dependent terms (\ref{eq:CInt23BI2Second})
and (\ref{CIntBI-II}) the original $S$ matrix can be easily restored.

\subsection{Term $C_{23}^{intBI}$ and its descendants}

$C_{23}^{intBI}$ reads

\begin{equation}
\begin{split}C_{23}^{intBI}= & \frac{g}{2}\left(\frac{1}{2}\right)^{2}\sum_{I\neq0}\int_{-\infty+i\varepsilon}^{\infty+i\varepsilon}\frac{d\vartheta_{2}}{2\pi}\left\{ \vphantom{\frac{\frac{e^{iQ}}{e^{iQ}}}{\left(\frac{e^{iQ}}{e^{iQ}}\right)}}h(\vartheta_{1}|\vartheta_{2},\{0\})_{R}\,K(-\vartheta_{1})K(\vartheta_{2})\times\right.\\
 & \left.\frac{4F_{1}\left(\frac{\cosh(\vartheta_{2}-\vartheta_{1})}{\sinh^{2}(\vartheta_{2}-\vartheta_{1})}-\frac{\cosh(\vartheta_{2}+\vartheta_{1})}{\sinh^{2}(\vartheta_{2}+\vartheta_{1})}\right)}{\bar{\rho}_{2}(\vartheta_{1})}\right\} \;.
\end{split}
\end{equation}
To proceed we focus on the integral with respect to $\vartheta_{2}$
and separate the singularities in 
\[
K(\vartheta_{2})\frac{\cosh(\vartheta_{2}-\vartheta_{1})}{\sinh^{2}(\vartheta_{2}-\vartheta_{1})}
\]
to prepare for application of the identities of distribution theory.
Using the shorthand $s(\vartheta_{2})=K(\vartheta_{2})$ and 
\[
c(\vartheta_{2})=\frac{\cosh(\vartheta_{2}-\vartheta_{1})}{\sinh^{2}(\vartheta_{2}-\vartheta_{1})}\;,
\]
where $s(\vartheta_{2})$ is singular in $0$ and $c(\vartheta_{2})$
at $\vartheta_{2}=\vartheta_{1}$ one can write the singular terms
as

\begin{equation}
\begin{split}s(\vartheta_{2})c(\vartheta_{2})= & \left(\left(s(\vartheta_{2})-s(\vartheta_{1})\right)+s(\vartheta_{1})\right)\left(\left(c(\vartheta_{2})-c(0)\right)+c(0)\right)\\
 & \left(s(\vartheta_{2})-s(\vartheta_{1})\right)\left(c(\vartheta_{2})-c(0)\right)+s(\vartheta_{1})c(\vartheta_{2})+c(0)s(\vartheta_{2})-c(0)s(\vartheta_{1})\:,
\end{split}
\end{equation}
from which for fixed, non zero $\vartheta_{1}$, the first term $\left(s(\vartheta_{2})-s(\vartheta_{1})\right)\left(c(\vartheta_{2})-c(0)\right)$
is singular only in $\vartheta_{2}=\vartheta_{1}$ and this singularity
is milder than $1/x^{2}$, the second term $s(\vartheta_{1})c(\vartheta_{2})$
is singular only in $\vartheta_{2}=\vartheta_{1}$ which is of type
$1/x^{2}$, whereas the third term $c(0)s(\vartheta_{2})$ is singular
at the origin with $1/x$ behaviour and the last term $c(0)s(\vartheta_{1})$
is regular. The terms corresponding to this separation are denoted
by $C_{23}^{intBI1},C_{23}^{intBI2},C_{23}^{intBI3}$ and $C_{23}^{intBI4}$.

Considering the first term and using (\ref{DistSinh}), we have

\begin{equation}
\begin{split}C_{23}^{intBI1}= & \frac{g}{2}\left(\frac{1}{2}\right)^{2}\sum_{I\neq0}\int_{-\infty+i\varepsilon}^{\infty+i\varepsilon}\frac{d\vartheta_{2}}{2\pi}\left\{ \frac{h(\vartheta_{1}|\vartheta_{2},\{0\})_{R}\,K(-\vartheta_{1})\left(K(\vartheta_{2})-K(\vartheta_{1})\right)}{\bar{\rho}_{2}(\vartheta_{1})}4F_{1}\times\right.\\
 & \left.\qquad\qquad\qquad\qquad\qquad\times\left(\frac{\cosh(\vartheta_{2}-\vartheta_{1})}{\sinh^{2}(\vartheta_{2}-\vartheta_{1})}-\frac{\cosh\vartheta_{1}}{\sinh^{2}\vartheta_{1}}\right)\right\} \\
- & \frac{g}{2}\left(\frac{1}{2}\right)^{2}\sum_{I\neq0}\int_{-\infty+i\varepsilon}^{\infty+i\varepsilon}\frac{d\vartheta_{2}}{2\pi}\left\{ \frac{h(\vartheta_{1}|\vartheta_{2},\{0\})_{R}\,K(-\vartheta_{1})\left(K(\vartheta_{2})-K(-\vartheta_{1})\right)}{\bar{\rho}_{2}(\vartheta_{1})}4F_{1}\times\right.\\
 & \left.\qquad\qquad\qquad\qquad\qquad\times\left(\frac{\cosh(\vartheta_{2}+\vartheta_{1})}{\sinh^{2}(\vartheta_{2}+\vartheta_{1})}-\frac{\cosh\vartheta_{1}}{\sinh^{2}\vartheta_{1}}\right)\right\} \\
= & -i\frac{\pi}{2\pi}\frac{g}{2}\left(\frac{1}{2}\right)^{2}\sum_{I\neq0}h(\vartheta_{1}|\vartheta_{1},\{0\})_{R}\,K(-\vartheta_{1})K^{'}|_{\vartheta_{1}}\frac{4F_{1}}{\bar{\rho}_{2}(\vartheta_{1})}\\
+ & \frac{g}{2}\left(\frac{1}{2}\right)^{2}\sum_{I\neq0}\int_{-\infty}^{\infty}\frac{d\vartheta_{2}}{2\pi}\frac{4F_{1}}{\bar{\rho}_{2}}\frac{K(-\vartheta_{1})}{\sinh(\vartheta_{2}-\vartheta_{1})}\left\{ \vphantom{\frac{\frac{OO}{OO}}{\frac{OO}{OO}}}h(\vartheta_{1}|\vartheta_{2},\{0\})_{R}\left(K(\vartheta_{2})-K(\vartheta_{1})\right)\times\right.\\
 & \left.\times\left(\frac{\cosh(\vartheta_{2}-\vartheta_{1})}{\sinh^{2}(\vartheta_{2}-\vartheta_{1})}-\frac{\cosh\vartheta_{1}}{\sinh^{2}\vartheta_{1}}\right)\sinh(\vartheta_{2}-\vartheta_{1})-\frac{h(\vartheta_{1}|\vartheta_{1},\{0\})_{R}\,K^{'}|_{\vartheta_{1}}}{\cosh(\vartheta_{2}-\vartheta_{1})}\right\} \\
+ & \vartheta_{1}\longleftrightarrow-\vartheta_{1}\;,\\
\\
\end{split}
\end{equation}
which equals

\begin{equation}
\begin{split} & \frac{g}{2}\left(\frac{1}{2}\right)^{2}\sum_{I\neq0}\int_{-\infty}^{\infty}\frac{d\vartheta_{2}}{2\pi}\frac{4F_{1}}{\bar{\rho}_{2}(\vartheta_{1})}\frac{K(-\vartheta_{1})}{\sinh\vartheta_{2}-\vartheta_{1}}\left\{ \vphantom{\frac{\frac{OOOOOO}{OO}}{\frac{OOOOOO}{OO}}}h(\vartheta_{1}|\vartheta_{2},\{0\})_{R}\left(K(\vartheta_{2})-K(\vartheta_{1})\right)\times\right.\\
 & \left.\times\left(\frac{\cosh(\vartheta_{2}-\vartheta_{1})}{\sinh^{2}(\vartheta_{2}-\vartheta_{1})}-\frac{\cosh\vartheta_{1}}{\sinh^{2}\vartheta_{1}}\right)\sinh(\vartheta_{2}-\vartheta_{1})-\frac{h(\vartheta_{1}|\vartheta_{1},\{0\})_{R}\,K^{'}|_{\vartheta_{1}}}{\cosh(\vartheta_{2}-\vartheta_{1})}\right\} \;,\\
 & +\vartheta_{1}\longleftrightarrow-\vartheta_{1}\;.
\end{split}
\end{equation}
One can split it further as

\begin{equation}
\begin{split}C_{23}^{intBI1}= & C_{23}^{intBI1a}+C_{23}^{intBI1b}\\
C_{23}^{intBI1a}= & \frac{g}{2}\left(\frac{1}{2}\right)^{2}\sum_{I\neq0}\int_{-\infty}^{\infty}\frac{d\vartheta_{2}}{2\pi}4F_{1}\frac{K(-\vartheta_{1})}{\bar{\rho}_{2}(\vartheta_{1})}\frac{h(\vartheta_{1}|\vartheta_{2},\{0\})_{R}}{\sinh\vartheta_{2}-\vartheta_{1}}\times\\
 & \times\left[\left(K(\vartheta_{2})-K(\vartheta_{1})\right)\left(\frac{\cosh(\vartheta_{2}-\vartheta_{1})}{\sinh^{2}(\vartheta_{2}-\vartheta_{1})}-\frac{\cosh\vartheta_{1}}{\sinh^{2}\vartheta_{1}}\right)\sinh(\vartheta_{2}-\vartheta_{1})-\frac{K'(\vartheta_{1})}{\cosh(\vartheta_{2}-\vartheta_{1})}\right]\\
+ & \frac{g}{2}\left(\frac{1}{2}\right)^{2}\sum_{I\neq0}\int_{-\infty}^{\infty}\frac{d\vartheta_{2}}{2\pi}4F_{1}\frac{K(\vartheta_{1})}{\bar{\rho}_{2}(\vartheta_{1})}\frac{h(\vartheta_{1}|\vartheta_{2},\{0\})_{R}}{\sinh\vartheta_{2}+\vartheta_{1}}\times\\
 & \text{\ensuremath{\times}}\left[\left(K(\vartheta_{2})-K(-\vartheta_{1})\right)\left(\frac{\cosh(\vartheta_{2}+\vartheta_{1})}{\sinh^{2}(\vartheta_{2}+\vartheta_{1})}-\frac{\cosh\vartheta_{1}}{\sinh^{2}\vartheta_{1}}\right)\sinh(\vartheta_{2}-\vartheta_{1})-\frac{K'(-\vartheta_{1})}{\cosh(\vartheta_{2}+\vartheta_{1})}\right]\\
C_{23}^{intBI1b}= & \frac{g}{2}\left(\frac{1}{2}\right)^{2}\sum_{I\neq0}\int_{-\infty}^{\infty}\frac{d\vartheta_{2}}{2\pi}4F_{1}\frac{K(-\vartheta_{1})}{\bar{\rho}_{2}(\vartheta_{1})}\frac{\left[h(\vartheta_{1}|\vartheta_{2},\{0\})_{R}-h(\vartheta_{1}|\vartheta_{1},\{0\})_{R}\right]K'(\vartheta_{1})}{\cosh(\vartheta_{2}-\vartheta_{1})\sinh(\vartheta_{2}-\vartheta_{1})}\\
+ & \frac{g}{2}\left(\frac{1}{2}\right)^{2}\sum_{I\neq0}\int_{-\infty}^{\infty}\frac{d\vartheta_{2}}{2\pi}4F_{1}\frac{K(\vartheta_{1})}{\bar{\rho}_{2}(\vartheta_{1})}\frac{\left[h(\vartheta_{1}|\vartheta_{2},\{0\})_{R}-h(\vartheta_{1}|\vartheta_{1},\{0\})_{R}\right]K'(-\vartheta_{1})}{\cosh(\vartheta_{2}+\vartheta_{1})\sinh(\vartheta_{2}+\vartheta_{1})}\;.
\end{split}
\end{equation}
The second term $C_{23}^{intBI2}$ reads as follows: 
\begin{equation}
\begin{split}C_{23}^{intBI2}= & \frac{g}{2}\left(\frac{1}{2}\right)^{2}\sum_{I\neq0}\int_{-\infty+i\varepsilon}^{\infty+i\varepsilon}\frac{d\vartheta_{2}}{2\pi}h(\vartheta_{1}|\vartheta_{2},\{0\})_{R}\frac{|K(\vartheta_{1})|^{2}}{\bar{\rho}_{2}(\vartheta_{1})}4F_{1}\left(\frac{\cosh(\vartheta_{2}-\vartheta_{1})}{\sinh^{2}(\vartheta_{2}-\vartheta_{1})}\right)\\
+ & \frac{g}{2}\left(\frac{1}{2}\right)^{2}\sum_{I\neq0}\int_{-\infty+i\varepsilon}^{\infty+i\varepsilon}\frac{d\vartheta_{2}}{2\pi}h(\vartheta_{1}|\vartheta_{2},\{0\})_{R}\frac{|K(\vartheta_{1})|^{2}}{\bar{\rho}_{2}(\vartheta_{1})}4F_{1}\left(\frac{\cosh(\vartheta_{2}+\vartheta_{1})}{\sinh^{2}(\vartheta_{2}+\vartheta_{1})}\right)\\
= & \frac{g}{2}\left(\frac{1}{2}\right)^{2}\sum_{I\neq0}\int_{-\infty+i\varepsilon}^{\infty+i\varepsilon}\frac{d\vartheta_{2}}{2\pi}h(\vartheta_{1}|\vartheta_{2},\{0\})_{R}\frac{|K(\vartheta_{1})|^{2}}{\bar{\rho}_{2}(\vartheta_{1})}4F_{1}\left(\frac{\cosh(\vartheta_{2}-\vartheta_{1})}{\sinh^{2}(\vartheta_{2}-\vartheta_{1})}+\frac{\cosh(\vartheta_{2}+\vartheta_{1})}{\sinh^{2}(\vartheta_{2}+\vartheta_{1})}\right)\\
= & -i\frac{\pi}{2\pi}\frac{g}{2}\left(\frac{1}{2}\right)^{2}\sum_{I\neq0}h(\vartheta_{1}|\vartheta_{1},\{0\})_{R}\frac{|K(\vartheta_{1})|^{2}}{\bar{\rho}_{2}(\vartheta_{1})}4F_{1}\left(-2imt-Rm\right)\left(\sinh\vartheta_{1}+\sinh(-\vartheta_{1})\right)\\
+ & \left(-2imt-Rm\right)\frac{g}{2}\left(\frac{1}{2}\right)^{2}\sum_{I\neq0}\int_{-\infty}^{\infty}\frac{d\vartheta_{2}}{2\pi}\frac{|K(\vartheta_{1})|^{2}}{\bar{\rho}_{2}(\vartheta_{1})}\frac{4F_{1}}{\sinh(\vartheta_{2}-\vartheta_{1})}\times\\
 & \qquad\qquad\qquad\times\left[h(\vartheta_{1}|\vartheta_{2},\{0\})_{R}\,\sinh\vartheta_{2}-\frac{h(\vartheta_{1}|\vartheta_{1},\{0\})_{R}\,\sinh\vartheta_{1}}{\cosh(\vartheta_{2}-\vartheta_{1})}\right]\\
+ & \left(-2imt-Rm\right)\frac{g}{2}\left(\frac{1}{2}\right)^{2}\sum_{I\neq0}\int_{-\infty}^{\infty}\frac{d\vartheta_{2}}{2\pi}\frac{|K(\vartheta_{1})|^{2}}{\bar{\rho}_{2}(\vartheta_{1})}\frac{4F_{1}}{\sinh(\vartheta_{2}+\vartheta_{1})}\times\\
 & \qquad\qquad\qquad\times\left[h(\vartheta_{1}|\vartheta_{2},\{0\})_{R}\,\sinh\vartheta_{2}-\frac{h(\vartheta_{1}|\vartheta_{1},\{0\})_{R}\,\sinh(-\vartheta_{1})}{\cosh(\vartheta_{2}+\vartheta_{1})}\right]\;,
\end{split}
\end{equation}
where (\ref{DistCoshSinh2}) was used. Thus,

\begin{equation}
\begin{split}C_{23}^{intBI2}= & \left(-2imt-Rm\right)\frac{g}{2}\left(\frac{1}{2}\right)^{2}\sum_{I\neq0}\int_{-\infty}^{\infty}\frac{d\vartheta_{2}}{2\pi}\frac{|K(\vartheta_{1})|^{2}}{\bar{\rho}_{2}(\vartheta_{1})}\frac{4F_{1}}{\sinh(\vartheta_{2}-\vartheta_{1})}\times\\
 & \qquad\qquad\qquad\times\left[h(\vartheta_{1}|\vartheta_{2},\{0\})_{R}\,\sinh\vartheta_{2}-\frac{h(\vartheta_{1}|\vartheta_{1},\{0\})_{R}\,\sinh\vartheta_{1}}{\cosh(\vartheta_{2}-\vartheta_{1})}\right]\\
+ & \left(-2imt-Rm\right)\frac{g}{2}\left(\frac{1}{2}\right)^{2}\sum_{I\neq0}\int_{-\infty}^{\infty}\frac{d\vartheta_{2}}{2\pi}\frac{|K(\vartheta_{1})|^{2}}{\bar{\rho}_{2}(\vartheta_{1})}\frac{4F_{1}}{\sinh(\vartheta_{2}+\vartheta_{1})}\times\\
 & \qquad\qquad\qquad\times\left[h(\vartheta_{1}|\vartheta_{2},\{0\})_{R}\,\sinh\vartheta_{2}-\frac{h(\vartheta_{1}|\vartheta_{1},\{0\})_{R}\,\sinh(-\vartheta_{1})}{\cosh(\vartheta_{2}+\vartheta_{1})}\right]\;.
\end{split}
\label{eq:CInt23BI2}
\end{equation}
Turning to the third contribution and using (\ref{DistCoshSinh2}),
we have

\begin{equation}
\begin{split}C_{23}^{intBI3}= & \frac{g}{2}\left(\frac{1}{2}\right)^{2}\sum_{I\neq0}\int_{-\infty+i\varepsilon}^{\infty+i\varepsilon}\frac{d\vartheta_{2}}{2\pi}h(\vartheta_{1}|\vartheta_{2},\{0\})_{R}\,K(-\vartheta_{1})K(\vartheta_{2})\frac{4F_{1}\frac{\cosh\vartheta_{1}}{\sinh^{2}\vartheta_{1}}}{\bar{\rho}_{2}(\vartheta_{1})}\\
- & \frac{g}{2}\left(\frac{1}{2}\right)^{2}\sum_{I\neq0}\int_{-\infty+i\varepsilon}^{\infty+i\varepsilon}\frac{d\vartheta_{2}}{2\pi}h(\vartheta_{1}|\vartheta_{2},\{0\})_{R}\,K(-\vartheta_{1})K(\vartheta_{2})\frac{4F_{1}\frac{\cosh\vartheta_{1}}{\sinh^{2}\vartheta_{1}}}{\bar{\rho}_{2}(\vartheta_{1})}\\
= & 0\;.
\end{split}
\end{equation}
The $4$th term reads

\begin{equation}
\begin{split}C_{23}^{intBI4}= & -2\frac{g}{2}\left(\frac{1}{2}\right)^{2}\sum_{I\neq0}\int_{-\infty}^{\infty}\frac{d\vartheta_{2}}{2\pi}\frac{h(\vartheta_{1}|\vartheta_{2},\{0\})_{R}\,|K(\vartheta_{1})|^{2}}{\bar{\rho}_{2}(\vartheta_{1})}4F_{1}\frac{\cosh\vartheta_{1}}{\sinh^{2}\vartheta_{1}}\end{split}
\:.
\end{equation}

\subsection{Term $C_{23}^{intBII}$ and its descendants}

\label{subsec:CintTerm1andDescendants}

Consider now 
\begin{equation}
\begin{split}C_{23}^{intBII}= & \frac{g}{2}\left(\frac{1}{2}\right)^{2}\sum_{I\neq0}\int_{-\infty+i\varepsilon}^{\infty+i\varepsilon}\frac{d\vartheta_{2}}{2\pi}\left\{ h(\vartheta_{1}|\vartheta_{2},\{0\})_{R}K(-\vartheta_{1})K(\vartheta_{2})\frac{\left[\frac{F_{5}^{\varepsilon}(\vartheta_{1})}{\sinh(\vartheta_{2}-\vartheta_{1})}-\frac{F_{5}^{\varepsilon}(-\vartheta_{1})}{\sinh(\vartheta_{2}+\vartheta_{1})}\right]}{\bar{\rho}_{2}(\vartheta_{1})}\right\} \end{split}
\;.
\end{equation}
Similarly to $C_{23}^{intBI}$ , we first separate the singularities
using the shorthand $s(\vartheta_{2})=K(\vartheta_{2})$ and 
\[
c(\vartheta_{2})=\frac{1}{\sinh(\vartheta_{2}-\vartheta_{1})}\;,
\]
where $s(\vartheta_{2})$ is singular at $\vartheta_{2}=0$ and $c(\vartheta_{2})$
at $\vartheta_{2}=\vartheta_{1}$, to write

\begin{equation}
\begin{split}s(\vartheta_{2})c(\vartheta_{2})= & \left(\left(s(\vartheta_{2})-s(\vartheta_{1})\right)+s(\vartheta_{1})\right)\left(\left(c(\vartheta_{2})-c(0)\right)+c(0)\right)\\
 & \left(s(\vartheta_{2})-s(\vartheta_{1})\right)\left(c(\vartheta_{2})-c(0)\right)+s(\vartheta_{1})c(\vartheta_{2})+c(0)s(\vartheta_{2})-c(0)s(\vartheta_{1})\:,
\end{split}
\end{equation}
from which for fixed, non zero $\vartheta_{1}$, the first term $\left(s(\vartheta_{2})-s(\vartheta_{1})\right)\left(c(\vartheta_{2})-c(0)\right)$
is regular in $\vartheta_{2}$, the second term $s(\vartheta_{1})c(\vartheta_{2})$
is singular only in $\vartheta_{2}=\vartheta_{1}$ which is of type
$1/x$, whereas the third term $c(0)s(\vartheta_{2})$ is singular
at the origin with $1/x$ behaviour and the last term $c(0)s(\vartheta_{1})$
is regular. The terms corresponding to this separation are denoted
by $C_{23}^{intBII1},C_{23}^{intBII2},C_{23}^{intBII3}$ and $C_{23}^{intBII4}$
from which we have the first term

\begin{equation}
\begin{split}C_{23}^{intBII1}= & \frac{g}{2}\left(\frac{1}{2}\right)^{2}\sum_{I\neq0}\int_{-\infty}^{\infty}\frac{d\vartheta_{2}}{2\pi}\frac{h(\vartheta_{1}|\vartheta_{2},\{0\})_{R}\,K(-\vartheta_{1})\left(K(\vartheta_{2})-K(\vartheta_{1})\right)}{\bar{\rho}_{2}(\vartheta_{1})}\times\\
 & \left(\frac{1}{\sinh\vartheta_{2}-\vartheta_{1}}+\frac{1}{\sinh\vartheta_{1}}\right)F_{5}^{\varepsilon}(\vartheta_{1})\\
- & \frac{g}{2}\left(\frac{1}{2}\right)^{2}\sum_{I\neq0}\int_{-\infty}^{\infty}\frac{d\vartheta_{2}}{2\pi}\frac{h(\vartheta_{1}|\vartheta_{2},\{0\})_{R}\,K(-\vartheta_{1})\left(K(\vartheta_{2})-K(-\vartheta_{1})\right)}{\bar{\rho}_{2}(\vartheta_{1})}\times\\
 & \left(\frac{1}{\sinh\vartheta_{2}+\vartheta_{1}}-\frac{1}{\sinh\vartheta_{1}}\right)F_{5}^{\varepsilon}(-\vartheta_{1})
\end{split}
\end{equation}
which is regular, and the second term

\begin{equation}
\begin{split}C_{23}^{intBII2}= & \frac{g}{2}\left(\frac{1}{2}\right)^{2}\sum_{I\neq0}\int_{-\infty+i\varepsilon}^{\infty+i\varepsilon}\frac{d\vartheta_{2}}{2\pi}\frac{h(\vartheta_{1}|\vartheta_{2},\{0\})_{R}\,K(-\vartheta_{1})K(\vartheta_{1})}{\bar{\rho}_{2}(\vartheta_{1})}\frac{F_{5}^{\varepsilon}(\vartheta_{1})}{\sinh\left(\vartheta_{2}-\vartheta_{1}\right)}\\
- & \frac{g}{2}\left(\frac{1}{2}\right)^{2}\sum_{I\neq0}\int_{-\infty+i\varepsilon}^{\infty+i\varepsilon}\frac{d\vartheta_{2}}{2\pi}\frac{h(\vartheta_{1}|\vartheta_{2},\{0\})_{R}\,K(-\vartheta_{1})K(-\vartheta_{1})}{\bar{\rho}_{2}(\vartheta_{1})}\frac{F_{5}^{\varepsilon}(-\vartheta_{1})}{\sinh\left(\vartheta_{2}+\vartheta_{1}\right)}\\
= & -i\frac{\pi}{2\pi}\frac{g}{2}\left(\frac{1}{2}\right)^{2}\sum_{I\neq0}\frac{h(\vartheta_{1}|\vartheta_{1},\{0\})_{R}\,|K(\vartheta_{1})|^{2}}{\bar{\rho}_{2}(\vartheta_{1})}\left(8/\sinh\vartheta_{1}+8/\sinh(-\vartheta_{1})\right)F_{1}\\
+ & \frac{g}{2}\left(\frac{1}{2}\right)^{2}\sum_{I\neq0}\frac{|K(\vartheta_{1})|^{2}}{\bar{\rho}_{2}(\vartheta_{1})}\frac{8F_{1}}{\sinh\vartheta_{1}}\times\\
 & \times\int_{-\infty}^{\infty}\frac{d\vartheta_{2}}{2\pi}\left(\frac{h(\vartheta_{1}|\vartheta_{2},\{0\})_{R}-h(\vartheta_{1}|\vartheta_{1},\{0\})_{R}/\cosh(\vartheta_{2}-\vartheta_{1})}{\sinh(\vartheta_{2}-\vartheta_{1})}\right)\\
+ & \frac{g}{2}\left(\frac{1}{2}\right)^{2}\sum_{I\neq0}\frac{|K(\vartheta_{1})|^{2}}{\bar{\rho}_{2}(\vartheta_{1})}\frac{8F_{1}}{\sinh\left(-\vartheta_{1}\right)}\times\\
 & \times\int_{-\infty}^{\infty}\frac{d\vartheta_{2}}{2\pi}\left(\frac{h(\vartheta_{1}|\vartheta_{2},\{0\})_{R}-h(\vartheta_{1}|\vartheta_{1},\{0\})_{R}/\cosh(\vartheta_{2}+\vartheta_{1})}{\sinh(\vartheta_{2}+\vartheta_{1})}\right)\;,
\end{split}
\end{equation}
where we used (\ref{eq:F5eIsing}) and (\ref{DistSinh}). This can
be further simplified to

\begin{equation}
\begin{split}C_{23}^{intBII2}= & C_{23}^{intBII2a}+C_{23}^{intBII2b}\;,\\
C_{23}^{intBII2a}= & \frac{g}{2}\left(\frac{1}{2}\right)^{2}\sum_{I\neq0}\int_{-\infty}^{\infty}\frac{d\vartheta_{2}}{2\pi}\frac{|K(\vartheta_{1})|^{2}}{\bar{\rho}_{2}(\vartheta_{1})}\left(\frac{h(\vartheta_{1}|\vartheta_{2},\{0\})_{R}\,(1-1/\cosh(\vartheta_{2}-\vartheta_{1}))}{\sinh(\vartheta_{2}-\vartheta_{1})}\right)8F_{1}/\sinh\vartheta_{1}\\
+ & \frac{g}{2}\left(\frac{1}{2}\right)^{2}\sum_{I\neq0}\int_{-\infty}^{\infty}\frac{d\vartheta_{2}}{2\pi}\frac{|K(\vartheta_{1})|^{2}}{\bar{\rho}_{2}(\vartheta_{1})}\left(\frac{h(\vartheta_{1}|\vartheta_{2},\{0\})_{R}\,(1-1/\cosh(\vartheta_{2}+\vartheta_{1}))}{\sinh(\vartheta_{2}+\vartheta_{1})}\right)8F_{1}/\sinh(-\vartheta_{1})\;,\\
\\
C_{23}^{intBII2b}= & \frac{g}{2}\left(\frac{1}{2}\right)^{2}\sum_{I\neq0}\int_{-\infty}^{\infty}\frac{d\vartheta_{2}}{2\pi}\frac{|K(\vartheta_{1})|^{2}}{\bar{\rho}_{2}(\vartheta_{1})}\left(\frac{h(\vartheta_{1}|\vartheta_{2},\{0\})_{R}-h(\vartheta_{1}|\vartheta_{1},\{0\})_{R}}{\sinh(\vartheta_{2}-\vartheta_{1})\cosh\vartheta_{2}-\vartheta_{1}}\right)8F_{1}/\sinh\vartheta_{1}\\
+ & \frac{g}{2}\left(\frac{1}{2}\right)^{2}\sum_{I\neq0}\int_{-\infty}^{\infty}\frac{d\vartheta_{2}}{2\pi}\frac{|K(\vartheta_{1})|^{2}}{\bar{\rho}_{2}(\vartheta_{1})}\left(\frac{h(\vartheta_{1}|\vartheta_{2},\{0\})_{R}-h(\vartheta_{1}|\vartheta_{1},\{0\})_{R}}{\sinh(\vartheta_{2}+\vartheta_{1})\cosh(\vartheta_{2}+\vartheta_{1})}\right)8F_{1}/\sinh(-\vartheta_{1})\;.
\end{split}
\end{equation}
Now consider

\begin{equation}
\begin{split}C_{23}^{intBII3}= & \frac{g}{2}\left(\frac{1}{2}\right)^{2}\sum_{I\neq0}\int_{-\infty+i\varepsilon}^{\infty+i\varepsilon}\frac{d\vartheta_{2}}{2\pi}\frac{h(\vartheta_{1}|\vartheta_{2},\{0\})_{R}\,K(-\vartheta_{1})K(\vartheta_{2})}{\bar{\rho}_{2}(\vartheta_{1})}\frac{F_{5}^{\varepsilon}(\vartheta_{1})}{\sinh(-\vartheta_{1})}\\
- & \frac{g}{2}\left(\frac{1}{2}\right)^{2}\sum_{I\neq0}\int_{-\infty+i\varepsilon}^{\infty+i\varepsilon}\frac{d\vartheta_{2}}{2\pi}\frac{h(\vartheta_{1}|\vartheta_{2},\{0\})_{R}\,K(-\vartheta_{1})K(\vartheta_{2})}{\bar{\rho}_{2}(\vartheta_{1})}\frac{F_{5}^{\varepsilon}(-\vartheta_{1})}{\sinh\vartheta_{1}}\\
= & 0\;.
\end{split}
\end{equation}
Finally,

\begin{equation}
\begin{split}C_{23}^{intBII4}= & -\frac{g}{2}\left(\frac{1}{2}\right)^{2}\sum_{I\neq0}\int_{-\infty}^{\infty}\frac{d\vartheta_{2}}{2\pi}\frac{h(\vartheta_{1}|\vartheta_{2},\{0\})_{R}\,|K(\vartheta_{1})|^{2}}{\bar{\rho}_{2}(\vartheta_{1})}\frac{F_{5}^{\varepsilon}(\vartheta_{1})}{\sinh(-\vartheta_{1})}\\
- & \frac{g}{2}\left(\frac{1}{2}\right)^{2}\sum_{I\neq0}\int_{-\infty}^{\infty}\frac{d\vartheta_{2}}{2\pi}\frac{h(\vartheta_{1}|\vartheta_{2},\{0\})_{R}\,|K(\vartheta_{1})|^{2}}{\bar{\rho}_{2}(\vartheta_{1})}\frac{F_{5}^{\varepsilon}(-\vartheta_{1})}{\sinh\vartheta_{1}}\\
= & \frac{g}{2}\left(\frac{1}{2}\right)^{2}\sum_{I\neq0}\int_{-\infty}^{\infty}\frac{d\vartheta_{2}}{2\pi}\frac{h(\vartheta_{1}|\vartheta_{2},\{0\})_{R}\,|K(\vartheta_{1})|^{2}}{\bar{\rho}_{2}(\vartheta_{1})}\frac{16F_{1}}{\sinh^{2}\vartheta_{1}}\quad.\\
\\
\end{split}
\end{equation}

\subsection{The term $C^{intA}$ \label{subsec:CintTerm2andDescendants}}

As a function of $\vartheta_{2}$ this term has no singularity at
the origin so we can pull the contour back to the real axis at once:

\begin{equation}
\begin{split}C_{23}^{intA}= & \frac{g}{2}\left(\frac{1}{2}\right)^{2}\sum_{I\neq0}\int_{-\infty+i\varepsilon}^{\infty+i\varepsilon}\frac{d\vartheta_{2}}{2\pi}\left\{ \vphantom{\frac{F}{\left(e^{iQ}\right)}}\frac{h(\vartheta_{1}|\vartheta_{2},\{0\})_{R}\,K(-\vartheta_{1})K(\vartheta_{2})}{\bar{\rho}_{2}(\vartheta_{1})}\left[\vphantom{\frac{OOOO}{OOOO}}F_{5}(\vartheta_{1}|\vartheta_{2},\{0\})\right.\right.\\
 & \left.\left.-4F_{1}\left(\frac{\cosh(\vartheta_{2}-\vartheta_{1})}{\sinh^{2}(\vartheta_{2}-\vartheta_{1})}-\frac{\cosh(\vartheta_{2}+\vartheta_{1})}{\sinh^{2}(\vartheta_{2}+\vartheta_{1})}\right)-\frac{F_{5}^{\varepsilon}(\vartheta_{1})}{\sinh(\vartheta_{2}-\vartheta_{1})}+\frac{F_{5}^{\varepsilon}(-\vartheta_{1})}{\sinh(\vartheta_{2}+\vartheta_{1})}\right]\right\} \\
\\
= & \frac{g}{2}\left(\frac{1}{2}\right)^{2}\sum_{I\neq0}\int_{-\infty}^{\infty}\frac{d\vartheta_{2}}{2\pi}\left\{ \vphantom{\frac{F}{\left(e^{iQ}\right)}}\frac{h(\vartheta_{1}|\vartheta_{2},\{0\})_{R}\,K(-\vartheta_{1})K(\vartheta_{2})}{\bar{\rho}_{2}(\vartheta_{1})}\left[\vphantom{\frac{OOOO}{OOOO}}F_{5}(\vartheta_{1}|\vartheta_{2},\{0\})\right.\right.\\
 & \left.\left.-4F_{1}\left(\frac{\cosh(\vartheta_{2}-\vartheta_{1})}{\sinh^{2}(\vartheta_{2}-\vartheta_{1})}-\frac{\cosh(\vartheta_{2}+\vartheta_{1})}{\sinh^{2}(\vartheta_{2}+\vartheta_{1})}\right)-\frac{F_{5}^{\varepsilon}(\vartheta_{1})}{\sinh(\vartheta_{2}-\vartheta_{1})}+\frac{F_{5}^{\varepsilon}(-\vartheta_{1})}{\sinh(\vartheta_{2}+\vartheta_{1})}\right]\right\} \;,\\
\\
\end{split}
\end{equation}
thus a stationary phase evaluation using (\ref{eq:StacPhase}) yields
a $1/\sqrt{t}$ behaviour multiplied by the function value in $\vartheta_{2}=0$.
This expression as a function of $\vartheta_{1}$ has singularities
of 4th and 2nd order, but the 4th order ones just cancel due to 
\begin{align*}
 & \lim_{\vartheta_{1}\rightarrow0}K(-\vartheta_{1})4\lim_{\vartheta_{2}\rightarrow0}\frac{F_{1}K(\vartheta_{2})\sinh\vartheta_{2}}{\sinh\vartheta_{2}}\left(\frac{\cosh(\vartheta_{2}-\vartheta_{1})}{\sinh^{2}(\vartheta_{2}-\vartheta_{1})}-\frac{\cosh(\vartheta_{2}+\vartheta_{1})}{\sinh^{2}(\vartheta_{2}+\vartheta_{1})}\right)=\lim_{\vartheta_{1}\rightarrow0}g^{4}2\frac{\cosh^{2}\vartheta_{1}+1}{\sinh^{4}\vartheta_{1}}F_{1}\;,\\
 & \lim_{\vartheta_{1}\rightarrow0}K(-\vartheta_{1})\lim_{\vartheta_{2}\rightarrow0}\frac{K(\vartheta_{2})\sinh\vartheta_{2}}{\sinh\vartheta_{2}}\left(-\frac{F_{5}^{\varepsilon}(\vartheta_{1})}{\sinh(\vartheta_{2}-\vartheta_{1})}+\frac{F_{5}^{\varepsilon}(\vartheta_{1})}{\sinh(\vartheta_{2}+\vartheta_{1})}\right)=\lim_{\vartheta_{1}\rightarrow0}-2g^{4}2\frac{\cosh\vartheta_{1}}{\sinh^{4}\vartheta_{1}}F_{1}
\end{align*}
as $F_{5}^{\varepsilon}(\vartheta_{1})\propto\frac{8F_{1}}{\vartheta_{1}}$
around the origin. Hence in $\vartheta_{1}$ no 4th order singularity
is present and when the sum over $I$ is converted to an integral
the remaining 2nd order singularity can only produce terms of the
type $mL$ but no higher power of $L$. These are expected to be cancelled
by terms from $Z_{1}D_{12}$.

\subsection{Singularities and their cancellation from $C_{23}^{intBI}$ and $C_{23}^{intBII}$\label{subsec:E4FourthOrderSing}}

There are some terms that have an integral regular at $\vartheta_{2}=0$
hence the SPA (\ref{eq:StacPhase}) can be directly applied yielding
a $1/\sqrt{t}$ factor. But the resulting $\vartheta_{1}$ dependent
prefactor has a dangerous $1/\vartheta_{1}^{4}$ singularity which
must cancel for the volume dependence to be regular when combined
with $Z_{1}D_{12}$.

\subsubsection{$4$th order singularities from $C_{23}^{intBII1}$, $C_{23}^{intBII4}$
and their cancellation}

Now we turn to $C_{23}^{intBII1}$ and $C_{23}^{intBII4}$

\begin{equation}
\begin{split}C_{23}^{intBII1}= & \frac{g}{2}\left(\frac{1}{2}\right)^{2}\sum_{I\neq0}\int_{-\infty}^{\infty}\frac{d\vartheta_{2}}{2\pi}\frac{h(\vartheta_{1}|\vartheta_{2},\{0\})_{R}\,K(-\vartheta_{1})\left(K(\vartheta_{2})-K(\vartheta_{1})\right)F_{5}^{\varepsilon}(\vartheta_{1})}{\bar{\rho}_{2}(\vartheta_{1})}\times\\
 & \qquad\qquad\qquad\qquad\times\left(\frac{1}{\sinh(\vartheta_{2}-\vartheta_{1})}+\frac{1}{\sinh\vartheta_{1}}\right)\\
- & \frac{g}{2}\left(\frac{1}{2}\right)^{2}\sum_{I\neq0}\int_{-\infty}^{\infty}\frac{d\vartheta_{2}}{2\pi}\frac{h(\vartheta_{1}|\vartheta_{2},\{0\})_{R}\,K(-\vartheta_{1})\left(K(\vartheta_{2})-K(-\vartheta_{1})\right)F_{5}^{\varepsilon}(-\vartheta_{1})}{\bar{\rho}_{2}(\vartheta_{1})}\times\\
 & \qquad\qquad\qquad\qquad\times\left(\frac{1}{\sinh(\vartheta_{2}+\vartheta_{1})}-\frac{1}{\sinh\vartheta_{1}}\right)\;,\\
C_{23}^{intBII4}= & \frac{g}{2}\left(\frac{1}{2}\right)^{2}\sum_{I\neq0}\int_{-\infty}^{\infty}\frac{d\vartheta_{2}}{2\pi}\frac{h(\vartheta_{1}|\vartheta_{2},\{0\})_{R}\,|K(\vartheta_{1})|^{2}}{\bar{\rho}_{2}(\vartheta_{1})}\frac{F_{5}^{\varepsilon}(\vartheta_{1})-F_{5}^{\varepsilon}(-\vartheta_{1})}{\sinh\vartheta_{1}}\;,\\
\\
\end{split}
\end{equation}
from which we have after the $\vartheta_{2}$ integration applying
the SPA, (\ref{eq:StacPhase})

\begin{equation}
\begin{split}C_{23}^{intBII1} & \approx\frac{g}{8}\sum_{I\neq0}h(\vartheta_{1}|0,\{0\})_{R}\,\left(\frac{ig^{2}/2}{\sinh^{2}\vartheta_{1}}\cosh\vartheta_{1}\right)8F_{1}\frac{K(-\vartheta_{1})-K(\vartheta_{1})}{\bar{\rho}_{2}\sinh\vartheta_{1}}/\sqrt{4\pi mt}\\
 & \approx-\frac{g}{8}\vartheta_{1}^{-4}h(\vartheta_{1}|0,\{0\})_{R}\,(g^{4}/4)16F_{1}/\sqrt{4\pi mt}\;,\\
\\
C_{23}^{intBII4} & \approx\frac{g}{8}\sum_{I\neq0}h(\vartheta_{1}|0,\{0\})_{R}\,K(-\vartheta_{1})K(\vartheta_{1})\frac{1}{\bar{\rho}_{2}}\frac{F_{5}^{\varepsilon}(\vartheta_{1})-F_{5}^{\varepsilon}(-\vartheta_{1})}{\sinh\vartheta_{1}}/\sqrt{4\pi mt}\\
 & \approx\frac{g}{8}\vartheta_{1}^{-4}(g^{4}/4)16F_{1}h(\vartheta_{1}|0,\{0\})_{R}\,/\sqrt{4\pi mt}\;,
\end{split}
\end{equation}
hence the 4th order singularity vanishes.

\subsubsection{$4$th order singularities from $C_{23}^{intBI4}$, $C_{23}^{intBI1a}$
and their cancellation}

We begin by $C_{23}^{intBI4}$ and its integration with respect to
$\vartheta_{2}$ using the SPA, (\ref{eq:StacPhase}):

\begin{equation}
\begin{split}C_{23}^{intBI4}= & -2\frac{g}{2}\left(\frac{1}{2}\right)^{2}\sum_{I\neq0}\int_{-\infty}^{\infty}\frac{d\vartheta_{2}}{2\pi}\frac{h(\vartheta_{1}|\vartheta_{2},\{0\})_{R}\,|K(\vartheta_{1})|^{2}}{\bar{\rho}_{2}(\vartheta_{1})}4F_{1}\frac{\cosh\vartheta_{1}}{\sinh^{2}\vartheta_{1}}\;,\end{split}
\end{equation}
from which, neglecting the sum, the following behaviour is obtained
for $\vartheta_{1}\approx0$

\begin{equation}
C_{23}^{intBI4}\approx-2\frac{g}{8}\vartheta_{1}^{-4}h(\vartheta_{1}|0,\{0\})_{R}\,g^{4}\frac{F_{1}}{\bar{\rho}_{2}(\vartheta_{1})}/\sqrt{4\pi mt}\;,
\end{equation}
whereas for $C_{23}^{intBI1a}$, which reads

\begin{equation}
\begin{split}C_{23}^{intBI1a}= & \frac{g}{2}\left(\frac{1}{2}\right)^{2}\sum_{I\neq0}\int_{-\infty}^{\infty}\frac{d\vartheta_{2}}{2\pi}4F_{1}\frac{K(-\vartheta_{1})}{\bar{\rho}_{2}(\vartheta_{1})}\frac{h(\vartheta_{1}|\vartheta_{2},\{0\})_{R}}{\sinh\vartheta_{2}-\vartheta_{1}}\times\\
 & \times\left[\left(K(\vartheta_{2})-K(\vartheta_{1})\right)\left(\frac{\cosh(\vartheta_{2}-\vartheta_{1})}{\sinh^{2}(\vartheta_{2}-\vartheta_{1})}-\frac{\cosh\vartheta_{1}}{\sinh^{2}\vartheta_{1}}\right)\sinh(\vartheta_{2}-\vartheta_{1})-\frac{K'(\vartheta_{1})}{\cosh(\vartheta_{2}-\vartheta_{1})}\right]\\
+ & \frac{g}{2}\left(\frac{1}{2}\right)^{2}\sum_{I\neq0}\int_{-\infty}^{\infty}\frac{d\vartheta_{2}}{2\pi}4F_{1}\frac{K(\vartheta_{1})}{\bar{\rho}_{2}(\vartheta_{1})}\frac{h(\vartheta_{1}|\vartheta_{2},\{0\})_{R}}{\sinh\vartheta_{2}+\vartheta_{1}}\times\\
 & \text{\ensuremath{\times}}\left[\left(K(\vartheta_{2})-K(-\vartheta_{1})\right)\left(\frac{\cosh(\vartheta_{2}+\vartheta_{1})}{\sinh^{2}(\vartheta_{2}+\vartheta_{1})}-\frac{\cosh\vartheta_{1}}{\sinh^{2}\vartheta_{1}}\right)\sinh(\vartheta_{2}+\vartheta_{1})-\frac{K'(-\vartheta_{1})}{\cosh(\vartheta_{2}+\vartheta_{1})}\right]\;,
\end{split}
\end{equation}
we have after integration on $\vartheta_{2}$ applying the SPA, (\ref{eq:StacPhase})

\begin{equation}
\begin{split}C_{23}^{intBI1a}\approx & \frac{g}{8}\left\{ 4F_{1}\frac{K(-\vartheta_{1})}{\bar{\rho}_{2}(\vartheta_{1})}h(\vartheta_{1}|0,\{0\})_{R}\,\left(\left(-ig^{2}/2\right)(-1)\frac{\cosh^{2}\vartheta_{1}+1}{\sinh^{3}\left(-\vartheta_{1}\right)}+\frac{K'(\vartheta_{1})}{\sinh\vartheta_{1}\cosh\vartheta_{1}}\right)+\right.\\
 & \left.+4F_{1}\frac{K(\vartheta_{1})}{\bar{\rho}_{2}(\vartheta_{1})}h(\vartheta_{1}|0,\{0\})_{R}\,\left(\left(-ig^{2}/2\right)(-1)\frac{\cosh^{2}\vartheta_{1}+1}{\sinh^{3}\vartheta_{1}}-\frac{K'(-\vartheta_{1})}{\sinh\vartheta_{1}\cosh\vartheta_{1}}\right)\right\} /\sqrt{4\pi mt}\\
\\
= & \frac{g}{8}\left\{ 4F_{1}\frac{K(-\vartheta_{1})}{\bar{\rho}_{2}(\vartheta_{1})}h(\vartheta_{1}|0,\{0\})_{R}\,\left(-ig^{2}/2\right)\left(\frac{\cosh^{2}\vartheta_{1}+1}{\sinh^{3}\vartheta_{1}}+\frac{-1}{\sinh^{3}\vartheta_{1}\cosh\vartheta_{1}}\right)+\right.\\
 & \left.+4F_{1}\frac{K(\vartheta_{1})}{\bar{\rho}_{2}(\vartheta_{1})}h(\vartheta_{1}|0,\{0\})_{R}\,\left(-ig^{2}/2\right)\left(-\frac{\cosh^{2}\vartheta_{1}+1}{\sinh^{3}\vartheta_{1}}+\frac{1}{\sinh^{3}\vartheta_{1}\cosh\vartheta_{1}}\right)\right\} /\sqrt{4\pi mt}\\
\\
\approx & 2\frac{g}{8}\vartheta_{1}^{-4}F_{1}\frac{1}{\bar{\rho}_{2}(\vartheta_{1})}h(\vartheta_{1}|0,\{0\})_{R}\,g^{4}/\sqrt{4\pi mt}\\
\\
\end{split}
\end{equation}
that cancels the 4th order singularity from $C_{23}^{intBI4}$.

\subsubsection{Singularities from $C_{23}^{intBII2a}$ }

We start with $C_{23}^{intBII2a}$ and show that after the $\vartheta_{2}$
integration with SPA (\ref{eq:StacPhase}), no 4th order singularity
remains. $C_{23}^{intBII2a}$ reads

\begin{equation}
\begin{split}C_{23}^{intBII2a}= & +\frac{g}{2}\left(\frac{1}{2}\right)^{2}\sum_{I\neq0}\int_{-\infty}^{\infty}\frac{d\vartheta_{2}}{2\pi}\frac{|K(\vartheta_{1})|^{2}}{\bar{\rho}_{2}(\vartheta_{1})}\left(\frac{h(\vartheta_{1}|\vartheta_{2},\{0\})_{R}\,(1-1/\cosh(\vartheta_{2}-\vartheta_{1}))}{\sinh(\vartheta_{2}-\vartheta_{1})}\right)8F_{1}/\sinh\vartheta_{1}\\
 & +\frac{g}{2}\left(\frac{1}{2}\right)^{2}\sum_{I\neq0}\int_{-\infty}^{\infty}\frac{d\vartheta_{2}}{2\pi}\frac{|K(\vartheta_{1})|^{2}}{\bar{\rho}_{2}(\vartheta_{1})}\left(\frac{h(\vartheta_{1}|\vartheta_{2},\{0\})_{R}\,(1-1/\cosh(\vartheta_{2}+\vartheta_{1}))}{\sinh(\vartheta_{2}+\vartheta_{1})}\right)8F_{1}/\sinh(-\vartheta_{1})
\end{split}
\end{equation}
and is regular in $\vartheta_{2}=0$. Hence performing the SPA (\ref{eq:StacPhase})
for $C_{23}^{intBII2a}$, it is seen that the $\vartheta_{1}$ dependence
is $1/\vartheta_{1}^{2}$.

\subsection{Terms with non trivial time dependence}

As seen in (\ref{subsec:E4FourthOrderSing}), a large number of terms
originating from (\ref{eq:CInt23}) involving integration with respect
to $\vartheta_{2}$ can be evaluated directly using the SPA (\ref{eq:StacPhase}).
Once the SPA is performed, these terms have singularities in $\vartheta_{1}$,
which are of either $4$th or second order, and those of $4$th order
behaviour cancel each other, whereas the second order singularities
cannot produce time dependence stronger than $O(t^{0})$.

For some terms, however, the $\vartheta_{2}$ integration cannot be
easily performed. The first example is provided by (\ref{eq:CInt23BI2})
which we write again as

\begin{equation}
\begin{split}C_{23}^{intBI2}= & \left(-2imt-Rm\right)\frac{g}{2}\left(\frac{1}{2}\right)^{2}\sum_{I\neq0}\int_{-\infty}^{\infty}\frac{d\vartheta_{2}}{2\pi}\frac{|K(\vartheta_{1})|^{2}}{\bar{\rho}_{2}(\vartheta_{1})}\frac{4F_{1}}{\sinh(\vartheta_{2}-\vartheta_{1})}\times\\
 & \qquad\qquad\qquad\times\left[h(\vartheta_{1}|\vartheta_{2},\{0\})_{R}\,\sinh\vartheta_{2}-\frac{h(\vartheta_{1}|\vartheta_{1},\{0\})_{R}\,\sinh\vartheta_{1}}{\cosh(\vartheta_{2}-\vartheta_{1})}\right]\\
+ & \left(-2imt-Rm\right)\frac{g}{2}\left(\frac{1}{2}\right)^{2}\sum_{I\neq0}\int_{-\infty}^{\infty}\frac{d\vartheta_{2}}{2\pi}\frac{|K(\vartheta_{1})|^{2}}{\bar{\rho}_{2}(\vartheta_{1})}\frac{4F_{1}}{\sinh(\vartheta_{2}+\vartheta_{1})}\times\\
 & \qquad\qquad\qquad\times\left[h(\vartheta_{1}|\vartheta_{2},\{0\})_{R}\,\sinh\vartheta_{2}-\frac{h(\vartheta_{1}|\vartheta_{1},\{0\})_{R}\,\sinh(-\vartheta_{1})}{\cosh(\vartheta_{2}+\vartheta_{1})}\right]\;,
\end{split}
\label{eq:CInt23BI2Second}
\end{equation}
and split into $C_{23}^{intBI2a}+C_{23}^{intBI2b}$ as

\begin{equation}
\begin{split}C_{23}^{intBI2a}= & \left(-2imt-Rm\right)\frac{g}{2}\left(\frac{1}{2}\right)^{2}\sum_{I\neq0}\int_{-\infty}^{\infty}\frac{d\vartheta_{2}}{2\pi}\frac{|K(\vartheta_{1})|^{2}}{\bar{\rho}_{2}(\vartheta_{1})}\frac{4F_{1}h(\vartheta_{1}|\vartheta_{2},\{0\})_{R}\,\left(\sinh\vartheta_{2}-\frac{\sinh\vartheta_{1}}{\cosh(\vartheta_{2}-\vartheta_{1})}\right)}{\sinh(\vartheta_{2}-\vartheta_{1})}\\
+ & \left(-2imt-Rm\right)\frac{g}{2}\left(\frac{1}{2}\right)^{2}\sum_{I\neq0}\int_{-\infty}^{\infty}\frac{d\vartheta_{2}}{2\pi}\frac{|K(\vartheta_{1})|^{2}}{\bar{\rho}_{2}(\vartheta_{1})}\frac{4F_{1}h(\vartheta_{1}|\vartheta_{2},\{0\})_{R}\,\left(\sinh\vartheta_{2}-\frac{\sinh-\vartheta_{1}}{\cosh(\vartheta_{2}+\vartheta_{1})}\right)}{\sinh(\vartheta_{2}+\vartheta_{1})}\;,\\
C_{23}^{intBI2b}= & \left(-2imt-Rm\right)\frac{g}{2}\left(\frac{1}{2}\right)^{2}\sum_{I\neq0}\int_{-\infty}^{\infty}\frac{d\vartheta_{2}}{2\pi}\frac{|K(\vartheta_{1})|^{2}}{\bar{\rho}_{2}(\vartheta_{1})}\frac{4F_{1}\sinh\vartheta_{1}}{\sinh(\vartheta_{2}-\vartheta_{1})\cosh(\vartheta_{2}-\vartheta_{1})}\times\\
 & \qquad\qquad\qquad\qquad\qquad\times\left[h(\vartheta_{1}|\vartheta_{2},\{0\})_{R}-h(\vartheta_{1}|\vartheta_{1},\{0\})_{R}\right]\\
+ & \left(-2imt-Rm\right)\frac{g}{2}\left(\frac{1}{2}\right)^{2}\sum_{I\neq0}\int_{-\infty}^{\infty}\frac{d\vartheta_{2}}{2\pi}\frac{|K(\vartheta_{1})|^{2}}{\bar{\rho}_{2}(\vartheta_{1})}\frac{4F_{1}\sinh(-\vartheta_{1})}{\sinh(\vartheta_{2}+\vartheta_{1})\cosh(\vartheta_{2}+\vartheta_{1})}\times\\
 & \qquad\qquad\qquad\qquad\qquad\times\left[h(\vartheta_{1}|\vartheta_{2},\{0\})_{R}-h(\vartheta_{1}|\vartheta_{1},\{0\})_{R}\right]\quad.
\end{split}
\label{eq:CInt23BI2ab}
\end{equation}
$C_{23}^{intBI1b}$ and $C_{23}^{intBII2b}$ provide the second example,
which is written again for better transparency as

\begin{equation}
\begin{split}C_{23}^{intBI1b}= & \frac{g}{2}\left(\frac{1}{2}\right)^{2}\sum_{I\neq0}\int_{-\infty}^{\infty}\frac{d\vartheta_{2}}{2\pi}4F_{1}\frac{K(-\vartheta_{1})}{\bar{\rho}_{2}(\vartheta_{1})}\frac{\left[h(\vartheta_{1}|\vartheta_{2},\{0\})_{R}-h(\vartheta_{1}|\vartheta_{1},\{0\})_{R}\right]K'(\vartheta_{1})}{\cosh(\vartheta_{2}-\vartheta_{1})\sinh(\vartheta_{2}-\vartheta_{1})}\\
+ & \frac{g}{2}\left(\frac{1}{2}\right)^{2}\sum_{I\neq0}\int_{-\infty}^{\infty}\frac{d\vartheta_{2}}{2\pi}4F_{1}\frac{K(\vartheta_{1})}{\bar{\rho}_{2}(\vartheta_{1})}\frac{\left[h(\vartheta_{1}|\vartheta_{2},\{0\})_{R}-h(\vartheta_{1}|\vartheta_{1},\{0\})_{R}\right]K'(-\vartheta_{1})}{\cosh(\vartheta_{2}+\vartheta_{1})\sinh(\vartheta_{2}+\vartheta_{1})}\;,\\
\\
C_{23}^{intBII2b}= & \frac{g}{2}\left(\frac{1}{2}\right)^{2}\sum_{I\neq0}\int_{-\infty}^{\infty}\frac{d\vartheta_{2}}{2\pi}\frac{|K(\vartheta_{1})|^{2}}{\bar{\rho}_{2}(\vartheta_{1})}\left(\frac{h(\vartheta_{1}|\vartheta_{2},\{0\})_{R}-h(\vartheta_{1}|\vartheta_{1},\{0\})_{R}}{\sinh(\vartheta_{2}-\vartheta_{1})\cosh(\vartheta_{2}-\vartheta_{1})}\right)8F_{1}/\sinh\vartheta_{1}\\
+ & \frac{g}{2}\left(\frac{1}{2}\right)^{2}\sum_{I\neq0}\int_{-\infty}^{\infty}\frac{d\vartheta_{2}}{2\pi}\frac{|K(\vartheta_{1})|^{2}}{\bar{\rho}_{2}(\vartheta_{1})}\left(\frac{h(\vartheta_{1}|\vartheta_{2},\{0\})_{R}-h(\vartheta_{1}|\vartheta_{1},\{0\})_{R}}{\sinh(\vartheta_{2}+\vartheta_{1})\cosh(\vartheta_{2}+\vartheta_{1})}\right)8F_{1}/\sinh(-\vartheta_{1})\;,\\
\\
\end{split}
\end{equation}
which we add and denote by $C_{23}^{intBI-II}$ . Therefore,

\begin{equation}
\begin{split}C_{23}^{intBI-II}= & C_{23}^{intBI1b}+C_{23}^{intBII2b}\\
= & \frac{g}{2}\left(\frac{1}{2}\right)^{2}\sum_{I\neq0}\int_{-\infty}^{\infty}\frac{d\vartheta_{2}}{2\pi}4F_{1}\frac{K(-\vartheta_{1})}{\bar{\rho}_{2}(\vartheta_{1})}\frac{\left[h(\vartheta_{1}|\vartheta_{2},\{0\})_{R}-h(\vartheta_{1}|\vartheta_{1},\{0\})_{R}\right]\left(\frac{2K(\vartheta_{1})}{\sinh\vartheta_{1}}+K'(\vartheta_{1})\right)}{\cosh(\vartheta_{2}-\vartheta_{1})\sinh(\vartheta_{2}-\vartheta_{1})}\\
+ & \frac{g}{2}\left(\frac{1}{2}\right)^{2}\sum_{I\neq0}\int_{-\infty}^{\infty}\frac{d\vartheta_{2}}{2\pi}4F_{1}\frac{K(\vartheta_{1})}{\bar{\rho}_{2}(\vartheta_{1})}\frac{\left[h(\vartheta_{1}|\vartheta_{2},\{0\})_{R}-h(\vartheta_{1}|\vartheta_{1},\{0\})_{R}\right]\left(\frac{2K(-\vartheta_{1})}{\sinh(-\vartheta_{1})}+K'(-\vartheta_{1})\right)}{\cosh(\vartheta_{2}+\vartheta_{1})\sinh(\vartheta_{2}+\vartheta_{1})}\;.\\
\\
\\
\end{split}
\label{CIntBI-II}
\end{equation}
Both $C_{23}^{intBI2b}$ and $C_{23}^{intBI-II}$ defined in (\ref{CIntBI-II})
involve the same integral kernel

\begin{equation}
\int_{-\infty}^{\infty}\frac{d\vartheta_{2}}{2\pi}\frac{h(\vartheta_{1}|\vartheta_{2},\{0\})_{R}-h(\vartheta_{1}|\vartheta_{1},\{0\})_{R}}{\cosh(\vartheta_{2}-\vartheta_{1})\sinh(\vartheta_{2}-\vartheta_{1})}
\end{equation}
which is a function of $t,\vartheta_{1}$ and is evaluated in the
next Appendix.

\subsection{Summary}

We have shown that from $C_{23}^{int}$, i.e. the contour integral,
apart from $C_{23}^{intBI2}$ defined in (\ref{eq:CInt23BI2}), (\ref{eq:CInt23BI2Second})
and $C_{23}^{intBI-II}$ defined in (\ref{CIntBI-II}), only terms
with dependence $mL$ contribute to one-particle oscillations. The
$\sqrt{t}mL$ term resulting from (\ref{eq:CInt23BI2}) (cf. eqn.
(\ref{eq:sqrttML_term})) is expected to be cancelled by the denominator
of (\ref{tervszerint}) through $Z_{1}D_{12}$, and the only surviving
time dependence comes from $C_{23}^{intBI2}$ and $C_{23}^{intBI-II}$
which are analysed in the next section.

\section{Evaluation of the integral kernel and time dependence\label{sec:EvaluatingKernel}}

In this section we evaluate

\begin{equation}
\begin{split}C_{23}^{intBI2a}= & \left(-2imt-Rm\right)\frac{g}{2}\left(\frac{1}{2}\right)^{2}\sum_{I\neq0}\int_{-\infty}^{\infty}\frac{d\vartheta_{2}}{2\pi}\frac{|K(\vartheta_{1})|^{2}}{\bar{\rho}_{2}(\vartheta_{1})}\frac{\Omega(\vartheta_{1})F_{1}h(\vartheta_{1}|\vartheta_{2},\{0\})_{R}}{\sinh(\vartheta_{2}-\vartheta_{1})}\times\\
 & \qquad\qquad\qquad\qquad\qquad\qquad\times\left(\sinh\vartheta_{2}-\frac{\sinh\vartheta_{1}}{\cosh(\vartheta_{2}-\vartheta_{1})}\right)\\
+ & \left(-2imt-Rm\right)\frac{g}{2}\left(\frac{1}{2}\right)^{2}\sum_{I\neq0}\int_{-\infty}^{\infty}\frac{d\vartheta_{2}}{2\pi}\frac{|K(\vartheta_{1})|^{2}}{\bar{\rho}_{2}(\vartheta_{1})}\frac{\Omega(\vartheta_{1})F_{1}h(\vartheta_{1}|\vartheta_{2},\{0\})_{R}}{\sinh(\vartheta_{2}+\vartheta_{1})}\\
 & \qquad\qquad\qquad\qquad\qquad\qquad\times\left(\sinh\vartheta_{2}-\frac{\sinh-\vartheta_{1}}{\cosh(\vartheta_{2}+\vartheta_{1})}\right)\;,\\
C_{23}^{intBI2b}= & \left(-2imt-Rm\right)\frac{g}{2}\left(\frac{1}{2}\right)^{2}\sum_{I\neq0}\int_{-\infty}^{\infty}\frac{d\vartheta_{2}}{2\pi}\frac{|K(\vartheta_{1})|^{2}}{\bar{\rho}_{2}(\vartheta_{1})}\frac{\Omega(\vartheta_{1})F_{1}\sinh\vartheta_{1}}{\sinh(\vartheta_{2}-\vartheta_{1})\cosh(\vartheta_{2}-\vartheta_{1})}\times\\
 & \qquad\qquad\qquad\qquad\qquad\qquad\times\left[h(\vartheta_{1}|\vartheta_{2},\{0\})_{R}-h(\vartheta_{1}|\vartheta_{1},\{0\})_{R}\right]\\
+ & \left(-2imt-Rm\right)\frac{g}{2}\left(\frac{1}{2}\right)^{2}\sum_{I\neq0}\int_{-\infty}^{\infty}\frac{d\vartheta_{2}}{2\pi}\frac{|K(\vartheta_{1})|^{2}}{\bar{\rho}_{2}(\vartheta_{1})}\frac{\Omega(\vartheta_{1})F_{1}\sinh(-\vartheta_{1})}{\sinh(\vartheta_{2}+\vartheta_{1})\cosh(\vartheta_{2}+\vartheta_{1})}\times\\
 & \qquad\qquad\qquad\qquad\qquad\qquad\times\left[h(\vartheta_{1}|\vartheta_{2},\{0\})_{R}-h(\vartheta_{1}|\vartheta_{1},\{0\})_{R}\right]\;,
\end{split}
\label{eq:CInt23BI2abSecond}
\end{equation}
and

\begin{equation}
\begin{split}C_{23}^{intBI-II}= & \frac{g}{2}\left(\frac{1}{2}\right)^{2}\sum_{I\neq0}\int_{-\infty}^{\infty}\frac{d\vartheta_{2}}{2\pi}\frac{K(-\vartheta_{1})}{\bar{\rho}_{2}(\vartheta_{1})}\frac{\left(K(\vartheta_{1})F_{5}^{\varepsilon}(\vartheta_{1})+\Omega(\vartheta_{1})F_{1}K'(\vartheta_{1})\right)}{\cosh(\vartheta_{2}-\vartheta_{1})\sinh(\vartheta_{2}-\vartheta_{1})}\times\\
 & \qquad\qquad\qquad\qquad\qquad\qquad\times\left[h(\vartheta_{1}|\vartheta_{2},\{0\})_{R}-h(\vartheta_{1}|\vartheta_{1},\{0\})_{R}\right]\\
+ & \frac{g}{2}\left(\frac{1}{2}\right)^{2}\sum_{I\neq0}\int_{-\infty}^{\infty}\frac{d\vartheta_{2}}{2\pi}\frac{K(\vartheta_{1})}{\bar{\rho}_{2}(\vartheta_{1})}\frac{\left(K(-\vartheta_{1})F_{5}^{\varepsilon}(-\vartheta_{1})+\Omega(\vartheta_{1})F_{1}K'(-\vartheta_{1})\right)}{\cosh(\vartheta_{2}+\vartheta_{1})\sinh(\vartheta_{2}+\vartheta_{1})}\times\\
 & \qquad\qquad\qquad\qquad\qquad\qquad\times\left[h(\vartheta_{1}|\vartheta_{2},\{0\})_{R}-h(\vartheta_{1}|\vartheta_{1},\{0\})_{R}\right]\;.
\end{split}
\label{CIntBI-IISecond}
\end{equation}

Here we restored the $S$ matrix dependence and $\Omega(\vartheta_{1})=\left(1-S(\vartheta_{1})\right)\left(1-S(-\vartheta_{1})\right)$
compared to (\ref{eq:CInt23BI2}). For the integration over $\vartheta_{2}$
the SPA (\ref{eq:StacPhase}) cannot be directly applied to the second
term in (\ref{eq:CInt23BI2abSecond}) and to (\ref{CIntBI-IISecond}).

$C_{23}^{intBI2a}$ , $C_{23}^{intBI2b}$ and $C_{23}^{intBI-II}$
are even functions of $\vartheta_{2}$, therefore we can define the
following integral kernels

\begin{equation}
\begin{split}Ker^{a}(\vartheta_{1},t,R)= & e^{imt}\Omega(\vartheta_{1})\int_{-\infty}^{\infty}\frac{d\vartheta_{2}}{2\pi}\frac{\left[h(\vartheta_{1}|\vartheta_{2},\{0\})_{R}\,\left(\sinh\vartheta_{2}-\frac{\sinh\vartheta_{1}}{\cosh(\vartheta_{2}-\vartheta_{1})}\right)\right]}{\sinh(\vartheta_{2}-\vartheta_{1})}\end{split}
\label{KerA-1}
\end{equation}
and

\begin{equation}
\begin{split}Ker(\vartheta_{1},t,R)= & e^{imt}\Omega(\vartheta_{1})\int_{-\infty}^{\infty}\frac{d\vartheta_{2}}{2\pi}\frac{\left[h(\vartheta_{1}|\vartheta_{2},\{0\})_{R}-h(\vartheta_{1}|\vartheta_{1},\{0\})_{R}\right]\sinh\vartheta_{1}}{\sinh(\vartheta_{2}-\vartheta_{1})\cosh(\vartheta_{2}-\vartheta_{1})}\end{split}
\:,\label{Ker}
\end{equation}
which are even functions of $\vartheta_{1}$. Then $Ker_{a}$ appears
in $C_{23}^{intBI2a}$ in (\ref{eq:CInt23BI2abSecond}), while $Ker$
in $C_{23}^{intBI2b}$ in (\ref{eq:CInt23BI2abSecond}) and in $C_{23}^{intBI-II}$,
(\ref{CIntBI-IISecond}). For $Ker_{a}$, the long time limit can
be calculated by applying the SPA (\ref{eq:StacPhase}) resulting
in

\begin{equation}
\begin{split}Ker_{stac}^{a}(\vartheta_{1},t,R)= & e^{imt}\Omega(\vartheta_{1})\frac{1}{\sqrt{2\pi2mt}}\frac{h(\vartheta_{1},0,\{0\})_{R}\,e^{-i\pi/4}}{\cosh\vartheta_{1}}\\
= & \frac{\Omega(\vartheta_{1})}{\sqrt{\pi mt}}\frac{e^{2imt(\cosh\vartheta_{1}-1)}e^{-i\pi/4}}{2\cosh\vartheta_{1}}e^{-mR\left(\cosh\vartheta_{1}+3/2\right)}\;.
\end{split}
\end{equation}
As for $Ker$ the SPA, (\ref{eq:StacPhase}) cannot be directly applied,
we proceed as follows: we first differentiate the integrand with respect
to $t$ and apply the SPA which becomes now possible:

\begin{equation}
\begin{split} & \Omega(\vartheta_{1})\frac{d}{dt}\int_{-\infty}^{\infty}\frac{d\vartheta_{2}}{2\pi}\frac{e^{imt}\left[h(\vartheta_{1}|\vartheta_{2},\{0\})_{R}-h(\vartheta_{1}|\vartheta_{1},\{0\})_{R}\right]\sinh\vartheta_{1}}{\sinh(\vartheta_{2}-\vartheta_{1})\cosh(\vartheta_{2}-\vartheta_{1})}\\
= & \Omega(\vartheta_{1})\int_{-\infty}^{\infty}\frac{d\vartheta_{2}}{2\pi}\frac{e^{imt}h(\vartheta_{1},\vartheta_{2},\{0\})_{R}\,2im\left(\cosh\vartheta_{2}-\cosh\vartheta_{1}\right)\sinh\vartheta_{1}}{\sinh(\vartheta_{2}-\vartheta_{1})\cosh(\vartheta_{2}-\vartheta_{1})}\\
= & \Omega(\vartheta_{1})\frac{e^{2imt(\cosh\vartheta_{1}-1)}e^{-i\pi/4}}{\sqrt{\pi mt}\cosh\vartheta_{1}}2im\left(\cosh\vartheta_{1}-1\right)e^{-mR\left(\cosh\vartheta_{1}+3/2\right)}\;.
\end{split}
\end{equation}
Now integrating with respected to $t$ one ends up with the Fresnel
sine and cosine functions, denoted here by $F_{S}$ and $F_{C}$ respectively:

\begin{equation}
\begin{split}Ker_{stac}(\vartheta_{1},t,R)= & \Omega(\vartheta_{1})\frac{\sqrt{\cosh\vartheta_{1}-1)}}{\cosh\vartheta_{1}}e^{-mR\left(\cosh\vartheta_{1}+3/2\right)}\\
 & \times\left\{ \frac{\sqrt{2}}{2}\left(F_{S}\left(\sqrt{\frac{4mt(\cosh(\vartheta_{1})-1)}{\pi}}\right)-F_{C}\left(\sqrt{\frac{4mt(\cosh(\vartheta_{1})-1)}{\pi}}\right)\right)\right.\\
 & \left.-\frac{\sqrt{2}}{2}i\left(F_{C}\left(\sqrt{\frac{4mt(\cosh(\vartheta_{1})-1)}{\pi}}\right)+F_{S}\left(\sqrt{\frac{4mt(\cosh(\vartheta_{1})-1)}{\pi}}\right)\right)\right.\\
 & \left.+if(\vartheta_{1},R)\vphantom{\left(\sqrt{\frac{4t(\cosh(x)-1)}{\pi}}\right)}\right\} \;,
\end{split}
\label{KerFresnel}
\end{equation}
where $f$ is an integration constant that is independent of time,
but is a function of $\vartheta_{1}$ and $R$. We do not determine
the precise form of $f(\vartheta_{1},R)$, only quote its $R=0$ limit:

\begin{equation}
2\frac{\sqrt{(\cosh\vartheta_{1}-1)}}{\cosh\vartheta_{1}}f(\vartheta_{1},0)=\left(\frac{\sqrt{2(\cosh\vartheta_{1}-1)}}{\cosh\vartheta_{1}}-\sqrt{\sinh^{2}\vartheta_{1}}\right)\label{eq:f}
\end{equation}
which is determined by noticing that the $t\rightarrow\infty$ limit
for $Ker(\vartheta_{1},t,R)$ is proportional to $|\sinh\vartheta_{1}|$.
Note that $\frac{\sqrt{\cosh\vartheta_{1}-1)}}{\cosh\vartheta_{1}}f(\vartheta_{1},R)\approx\vartheta_{1}^{2}$
around the origin, i.e. its second derivative is continuous at the
origin. Hence integrating it with $|K|^{2}$ gives a finite and well-defined
result. In (\ref{CIntBI-IISecond}), however, $Ker(\vartheta_{1},t,R)$
is integrated with a $1/\sinh^{4}(\vartheta_{1})$ type of function
due to $K(\vartheta_{1})F_{5}^{\varepsilon}(\vartheta_{1})+F_{1}\Omega(\vartheta_{1})K'(\vartheta_{1})$,
where the non-analytic behaviour of $f(\vartheta_{1})$ must be carefully
handled. The origin of the non-analytic term can be summarised as
follows: using the SPA, (\ref{eq:StacPhase}), a term proportional
to $1/\sqrt{t}$ is obtained, but in the asymptotic expansion of the
oscillatory integral, terms proportional to $t^{-1/2+n}$ are also
present. For any finite $t$, these lead to analytic behaviour as
expected from (\ref{Ker}), but in the $t\rightarrow\infty$ limit
keeping only the leading terms, non-analyticity can emerge.

\subsection{Time dependence from $Ker_{a}$ and $C_{23}^{intBI2a}$.}

Substituting $Ker_{stac}^{a}$ into (\ref{eq:CInt23BI2abSecond}),
the discrete sum to evaluate reads

\begin{equation}
\frac{g}{2}F_{1}e^{-imt}(-imt-Rm/2)\sum_{I\neq0}\frac{|K(\vartheta_{1})|^{2}Ker_{stac}^{a}(\vartheta_{1},t,R)}{\bar{\rho}_{2}(\vartheta_{1})}
\end{equation}
that has a $1/\vartheta_{1}^{2}$ singularity at the origin which
we treat with a contour integral representation:

\begin{equation}
\begin{split} & \frac{g}{2}F_{1}e^{-imt}(-imt-Rm/2)\sum_{I\neq0}\frac{|K(\vartheta_{1})|^{2}Ker_{stac}^{a}(\vartheta_{1},t,R)}{\bar{\rho}_{2}(\vartheta_{1})}\\
= & -\frac{g}{2}F_{1}e^{-imt}(-imt-Rm/2)\sum_{I\neq0}\oint_{C_{I}}\frac{d\vartheta_{1}}{2\pi}\frac{|K(\vartheta_{1})|^{2}Ker_{stac}^{a}(\vartheta_{1},t,R)}{e^{i\bar{Q}_{2}(\vartheta_{1})}+1}\\
= & \frac{g}{2}F_{1}e^{-imt}(-imt-Rm/2)\int_{-\infty+i\varepsilon}^{\infty+i\varepsilon}\frac{d\vartheta_{1}}{2\pi}|K(\vartheta_{1})|^{2}Ker_{stac}^{a}(\vartheta_{1},t,R)\\
+ & i\frac{g}{2}F_{1}e^{-imt}(-imt-Rm/2)\oint_{0}\frac{d\vartheta_{1}}{2\pi i}\frac{|K(\vartheta_{1})|^{2}Ker_{stac}^{a}(\vartheta_{1},t,R)}{e^{i\bar{Q}_{2}(\vartheta_{1})}+1}\;.
\end{split}
\end{equation}
As $Ker^{a}(\vartheta_{1},t,R)$ is even in $\vartheta_{1}$, its
derivative vanishes at the origin, and so only the derivative of $e^{i\bar{Q}_{2}(\vartheta_{1})}+1$
gives a pole contribution resulting in

\begin{equation}
\begin{split} & i\frac{g}{2}F_{1}e^{-imt}(-imt-Rm/2)\oint_{0}\frac{d\vartheta_{1}}{2\pi i}\frac{|K(\vartheta_{1})|^{2}Ker_{stac}^{a}(\vartheta_{1},t,R)}{e^{i\bar{Q}_{2}}+1}\\
= & \frac{g}{2}F_{1}e^{-imt}(-imt-Rm/2)\frac{g^{4}Ker_{stac}^{a}(0,t,0)}{16}mL\\
\propto & \sqrt{t}mL\;,
\end{split}
\label{eq:sqrttML_term}
\end{equation}
and hence expected to be cancelled by the appropriate counter-term
from $Z_{1}D_{12}$. The contour integral in the $L\rightarrow\infty$
limit reduces to the upper contour and can be rewritten using (\ref{DistCoshSinh2})
as

\begin{equation}
\begin{split} & \frac{g}{2}F_{1}e^{-imt}(-imt-Rm/2)\int_{-\infty+i\varepsilon}^{\infty+i\varepsilon}\frac{d\vartheta_{1}}{2\pi}|K(\vartheta_{1})|^{2}Ker_{stac}^{a}(\vartheta_{1},t,R)\\
= & \frac{g}{2}F_{1}e^{-imt}(-imt-Rm/2)\int_{-\infty+i\varepsilon}^{\infty+i\varepsilon}\frac{d\vartheta_{1}}{2\pi}\left(|K(\vartheta_{1})|^{2}\text{\ensuremath{\frac{sinh^{2}\vartheta_{1}}{\cosh\vartheta_{1}}}}Ker_{stac}^{a}(\vartheta_{1},t,R)\right)\frac{\cosh\vartheta_{1}}{\sinh^{2}\vartheta_{1}}\\
= & \frac{g}{2}F_{1}e^{-imt}(-imt-Rm/2)\int_{-\infty}^{\infty}\frac{d\vartheta_{1}}{2\pi}\left(|K(\vartheta_{1})|^{2}\text{\ensuremath{\frac{sinh^{2}\vartheta_{1}}{\cosh\vartheta_{1}}}}Ker_{stac}^{a}(\vartheta_{1},t,R)\right)^{'}\frac{1}{\sinh\vartheta_{1}}\;,
\end{split}
\end{equation}
which is integrable since $Ker_{stac}^{a}(\vartheta_{1},t,R)$ is
an even and regular function with respect to $\vartheta_{1}$ . Then
the derivative of $|K(\vartheta_{1})|^{2}\text{\ensuremath{\frac{sinh^{2}\vartheta_{1}}{\cosh\vartheta_{1}}}}$
gives a $\sqrt{t}$ contribution which we are not interested in at
the moment. The contribution linear in $t$ comes from

\begin{equation}
\begin{split} & \frac{g}{2}F_{1}e^{-imt}(-imt)\int_{-\infty}^{\infty}\frac{d\vartheta_{1}}{2\pi}|K(\vartheta_{1})|^{2}\tanh(\vartheta_{1})\left(Ker_{stac}^{a}(\vartheta_{1},t,0)\right)^{'}\\
= & \frac{g}{2}F_{1}e^{-imt}\int_{-\infty}^{\infty}\frac{d\vartheta_{1}}{2\pi}|K(\vartheta_{1})|^{2}\tanh^{2}(\vartheta_{1})\frac{\Omega(\vartheta_{1})\left(-imt\right)e^{-i\pi/4}e^{2imt(\cosh(\vartheta_{1})-1)}\left(2imt-\text{sech}(\vartheta_{1})\right)}{2\sqrt{\pi mt}}\\
+ & \frac{g}{2}F_{1}e^{-imt}\int_{-\infty}^{\infty}\frac{d\vartheta_{1}}{2\pi}|K(\vartheta_{1})|^{2}\tanh(\vartheta_{1})\frac{\left(-imt\right)}{\sqrt{\pi mt}}\frac{e^{2imt(\cosh\vartheta_{1}-1)}e^{-i\pi/4}}{\cosh\vartheta_{1}}\Omega'(\vartheta_{1})\\
= & \frac{g}{2}F_{1}e^{-imt}\frac{g^{4}}{4}\frac{\left(-2imt\right)}{\sqrt{\pi mt4mt\pi}}\left(2imt-1+\varphi^{2}(0)\right)\\
= & \frac{g}{2}F_{1}e^{-imt}\frac{g^{4}}{4}\frac{\left(2mt+i(1-\varphi^{2}(0))\right)}{\pi}\;.
\end{split}
\end{equation}

\subsection{Time dependence from $C_{23}^{intBI2b}$ via $Ker$: imaginary part }

As $Ker\approx\vartheta_{1}^{2}$ around the origin, one can naively
try to evaluate the integral

\begin{equation}
\frac{g}{2}F_{1}e^{-imt}\int_{-\infty}^{\infty}\frac{d\vartheta_{1}}{2\pi}|K(\vartheta_{1})|^{2}\Im m\,(-imt-Rm/2)Ker_{stac}(\vartheta_{1},t,R)\;,
\end{equation}
However, as time $t$ grows, due to the asymptotics of the Fresnel
function ($F_{S}\rightarrow\frac{1}{2}$ and $F_{C}\rightarrow\frac{1}{2}$),
the interval in which $Ker_{b}\approx\vartheta_{1}^{2}$ holds shrinks
as $[-\frac{1}{\sqrt{t}},\frac{1}{\sqrt{t}}]$, outside of which $Ker_{b}\approx0$
since $\Im m\,Ker_{b}$ includes the difference of $F_{C}$ and $F_{S}$.
On the other hand, at the endpoints of the intervals the integrand
behaves as $t/\vartheta_{1}\rightarrow t\sqrt{t}$, so one expects
that the integral is linear in $t$ and its coefficient is determined
by the small $\vartheta$ behaviour of the $K$ function. One can
check this assumption numerically and conclude that in the long time
limit

\begin{equation}
\begin{split} & \frac{g}{2}F_{1}e^{-imt}i\int_{-\infty}^{\infty}\frac{d\vartheta_{1}}{2\pi}|K(\vartheta_{1})|^{2}\Im m\,(-imt-Rm/2)Ker_{stac}(\vartheta_{1},t,R)\\
= & \frac{g}{2}F_{1}e^{-imt}\frac{g^{4}}{4}\frac{1}{2}imt\;,
\end{split}
\end{equation}

\subsection{Time dependence from $C_{23}^{intBI2b}$ via $Ker$: real part and
logarithmic anomaly}

Similarly to the imaginary part, $\Re e\,Ker\approx\vartheta_{1}^{2}$
around the origin, hence one can once again try to evaluate the integral
directly

\begin{equation}
\frac{g}{2}F_{1}e^{-imt}\int_{-\infty}^{\infty}\frac{d\vartheta_{1}}{2\pi}|K(\vartheta_{1})|^{2}\Re e\,(-imt-Rm/2)Ker_{stac}(\vartheta_{1},t,R)\;,\label{KerLog}
\end{equation}
Again, as the time $t$ grows, $Ker\approx\vartheta_{1}^{2}$ holds
in the interval $[-\frac{1}{\sqrt{t}},\frac{1}{\sqrt{t}}]$ , while
at the endpoints of the intervals the integrand behaves as $t/\vartheta_{1}\rightarrow t\sqrt{t}$,
so one expects a term linearly dependent on $t$ whose coefficient
is given by the small $\vartheta$ behaviour of the $K$ function.
However, outside this interval the overall behaviour of the integrand
is now of the type $1/\vartheta_{1}$ accounting for a logarithmic
dependence and a $t\ln t$ type behaviour with a coefficient related
to the the small $\vartheta$ behaviour of the $K$ again. Differentiating
$Ker_{stac}$ with respect to $t$ and neglecting $(-imt-Rm/2)$ in
(\ref{KerLog}), the SPA (\ref{eq:StacPhase}) can be directly applied
resulting in a $1/t$ term whose coefficient can be identified with
that of the logarithmic term. Explicit calculation shows that

\begin{equation}
\begin{split}\frac{g}{2}F_{1}e^{-imt}mt\left(\frac{g^{4}}{4}\left(-\frac{\log(mt)}{\pi}\right)+\mathcal{K}\right)\;,\end{split}
\end{equation}
where

\begin{equation}
\begin{split}\mathcal{K}=\lim_{t\rightarrow\infty} & \left\{ -\frac{1}{2}\int_{-\infty}^{\infty}\frac{d\vartheta}{2\pi}\Omega(\vartheta)|K(\vartheta)|^{2}\left[\frac{\sqrt{2(\cosh\vartheta-1)}}{\cosh\vartheta}\times\right.\right.\\
 & \left.\left(F_{C}\left(\sqrt{\frac{4mt(\cosh(\vartheta)-1)}{\pi}}\right)+F_{S}\left(\sqrt{\frac{4mt(\cosh(\vartheta)-1)}{\pi}}\right)-1\right)+\sqrt{\sinh^{2}\vartheta}\vphantom{F_{C}\left(\sqrt{\frac{4t(\cosh(\vartheta)-1)}{\pi}}\right)}\right]\\
 & \left.+\frac{g^{4}}{4}\left(\frac{\log(mt)}{\pi}\right)\right\} \;.
\end{split}
\end{equation}

\subsection{Time dependence from $C_{23}^{intBI-II}$}

To calculate the time-dependence from $C_{23}^{intBI-II}$ we cannot
use $Ker_{stac}$ directly, i.e. the long-time approximation (\ref{KerFresnel})
of (\ref{Ker}) in the sum

\begin{equation}
\frac{g}{2}F_{1}e^{-imt}\sum_{I\neq0}\frac{|K(\vartheta_{1})|^{2}Ker(\vartheta_{1},t,R)}{\bar{\rho}_{2}(\vartheta_{1})}\frac{\frac{F_{5}^{\varepsilon}(\vartheta_{1})}{F_{1}\Omega(\vartheta)}+\frac{K'(\vartheta_{1})}{K(\vartheta_{1})}-\frac{F_{5}^{\varepsilon}(-\vartheta_{1})}{F_{1}\Omega(\vartheta)}-\frac{K'(-\vartheta_{1})}{K(-\vartheta_{1})}}{4\sinh\vartheta_{1}}\;.\label{CIntBI-IISum}
\end{equation}
The reason is that the singularity of $|K(\vartheta_{1})|^{2}\frac{\frac{F_{5}^{\varepsilon}(\vartheta_{1})}{\Omega(\vartheta)}+F_{1}\frac{K'(\vartheta_{1})}{K(\vartheta_{1})}-\frac{F_{5}^{\varepsilon}(-\vartheta_{1})}{\Omega(\vartheta)}-F_{1}\frac{K'(-\vartheta_{1})}{K(-\vartheta_{1})}}{4\sinh\vartheta_{1}}$
is of order four, whereas for any finite $t$, $Ker_{stac}$ (and
also $Ker$) behave as $\vartheta_{1}^{2}$ around the origin and
the resulting 2nd order singularity is sensitive to the non-analyticity
of $f$ in the long-time approximation of $Ker_{stac}(\vartheta_{1},t,R)$
in (\ref{KerFresnel}).

With $Ker(\vartheta_{1},t,R)$, which is an analytic function for
any finite $t$ without such a singular behaviour, one can formally
express the time dependence using the contour manipulations and eventually
(\ref{DistCoshSinh2}), yielding

\begin{equation}
\begin{split} & \frac{g}{2}F_{1}e^{-imt}\int_{-\infty}^{\infty}\frac{d\vartheta}{2\pi}\frac{\left(|K(\vartheta)|^{2}Ker(\vartheta,t,0)\left(\frac{F_{5}^{\varepsilon}(\vartheta)}{F_{1}\Omega(\vartheta)}+\frac{K'(\vartheta)}{K(\vartheta_{1})}-\frac{F_{5}^{\varepsilon}(-\vartheta)}{F_{1}\Omega(\vartheta)}-\frac{K'(-\vartheta)}{K(-\vartheta)}\right)\tanh\vartheta\right)'}{4\sinh\vartheta}\\
 & +\frac{g}{2}F_{1}e^{-imt}\frac{g^{4}}{4}\frac{mLKer(\vartheta,t,0)''}{8}+....\;,
\end{split}
\label{CIntBI-IIintegralRep}
\end{equation}
where $Ker(\vartheta,t,0)''$ is the second derivative with respect
to $\vartheta_{1}$ and $...$ refers $(mL)^{0}$ terms with no time
dependence. $Ker(\vartheta,t,0)''$ can be approximated using the
SPA, (\ref{eq:StacPhase}) giving $\frac{\pi}{2}\left(-1-i\right)\sqrt{mt}$,
hence the pole term yields an allowed $mL\sqrt{mt}$ factor. The way
to actually evaluate $C_{23}^{intBI-II}$ is done by performing numerically
the integration in (\ref{Ker}) and its derivative needed for (\ref{CIntBI-IIintegralRep})
or to calculate only (\ref{Ker}) and perform the sum in (\ref{CIntBI-IISum}).

After performing the numerical integral, we evaluated the sum for
the Ising case and came to the conclusion that the leading order time
dependence has the form

\begin{equation}
\frac{g}{2}F_{1}e^{-imt}\left(\Upsilon_{1}mt+i\Upsilon_{2}\log mt\right)\:,
\end{equation}
for large $mt$. It turns out, however, that for the linear term one
can even keep the real part from $Ker_{stac}$ in (\ref{KerFresnel})
which has a milder behaviour than $f$ in (\ref{eq:f}) and the contour
integral manipulations can formally be performed. The end of the analysis
by generalising the Ising result is

\begin{equation}
\Upsilon_{1}mt=\int_{-\infty}^{\infty}\frac{d\vartheta}{2\pi}\frac{\left(|K(\vartheta)|^{2}\Re e\,Ker_{stac}(\vartheta,t,R)\,\Re e\,\left(\frac{F_{5}^{\varepsilon}(\vartheta)}{F_{1}\Omega(\vartheta)}+\frac{K'(\vartheta)}{K(\vartheta_{1})}-\frac{F_{5}^{\varepsilon}(-\vartheta)}{F_{1}\Omega(\vartheta)}-\frac{K'(-\vartheta)}{K(-\vartheta)}\right)\tanh\vartheta\right)'}{4\sinh\vartheta}\;.\label{CIntBI-IILinearTimeDep}
\end{equation}
Note, that for the Ising model $\frac{F_{5}^{\varepsilon}(\vartheta)}{\Omega(\vartheta)}+F_{1}\frac{K'(\vartheta)}{K(\vartheta_{1})}-\frac{F_{5}^{\varepsilon}(-\vartheta)}{\Omega(\vartheta)}-F_{1}\frac{K'(-\vartheta)}{K(-\vartheta)}$
is a real function, whereas for an arbitrary interacting IQFT it has
an imaginary part as well.

As argued at the beginning of Appendix (\ref{sec:CInt}), the singular
structure at the origin of $\frac{F_{5}^{\varepsilon}(\vartheta)}{\Omega(\vartheta)}+F_{1}\frac{K'(\vartheta)}{K(\vartheta_{1})}-\frac{F_{5}^{\varepsilon}(-\vartheta)}{\Omega(\vartheta)}-F_{1}\frac{K'(-\vartheta)}{K(-\vartheta)}$
is the same for the Ising an for any interacting theory, and since
the source of time dependence is attributed to the singularity, we
can keep the real, i.e. singular part in (\ref{CIntBI-IILinearTimeDep}).
As a consequence, it is possible to extract the linear time dependence
from (\ref{CIntBI-IILinearTimeDep}), and one is allowed to use again
the Ising model, where

\begin{equation}
\frac{F_{5}^{\varepsilon}(\vartheta)}{\Omega(\vartheta)}+F_{1}\frac{K'(\vartheta)}{K(\vartheta_{1})}-\frac{F_{5}^{\varepsilon}(-\vartheta)}{\Omega(\vartheta)}-F_{1}\frac{K'(-\vartheta)}{K(-\vartheta)}=\frac{4F_{1}}{\sinh\vartheta_{1}}-2F_{1}\frac{\cosh\vartheta_{1}}{\sinh\vartheta_{1}}\;.
\end{equation}
For $\Upsilon_{1}$ it is useful to differentiate (\ref{CIntBI-IILinearTimeDep})
with respect to $t$ because for the resulting function the SPA (\ref{eq:StacPhase})
can be applied yielding a constant and a $1/t$ type term. Elementary
calculation shows that the former term is

\begin{equation}
\Upsilon_{1}=\frac{g^{4}}{4}\frac{1}{\pi}\;.
\end{equation}
The details of the numerical study can be found in the next Appendix.

\section{Numerical checks for the calculations\label{sec:Numerics}}

Our quite lengthy and tedious analytic calculations were extensively
cross-checked using numerics out of which we discuss and present here
three important parts: the cancellation of $mL$ terms in $\tilde{D}_{23}$,
the time dependence from the term $C^{intBI-II}$ (\ref{CIntBI-II}),
and match between the time dependence of $D_{23}$ and our analytic
predictions (\ref{D23(t)}). Beyond these, we also numerically verified
other parts of the calculations, such as: numerically monitoring the
validity of our manipulations performed in Appendix \ref{sec:D23Res}
(using sine\textendash Gordon first breather form factors) and in
Appendix \ref{sec:EvaluatingKernel}. From Appendix \ref{sec:CInt},
the calculations in (\ref{subsec:CintTerm1andDescendants}) and (\ref{subsec:CintTerm2andDescendants})
were also cross-checked, whereas the validity of the rest of the Appendix
(i.e. cancellation of the 4th order singularities) is verified by
the match between the predicted time evolution of $D_{23}$ and our
analytical considerations, discussed below.{} 

\subsection{Cancellation of $mL$ terms in $\tilde{D}_{23}$}

\begin{figure}[H]
\subfloat[$\Re e\,e^{imt}\frac{\tilde{D}_{23}}{g/2}$]{\begin{centering}
\includegraphics[width=0.45\columnwidth]{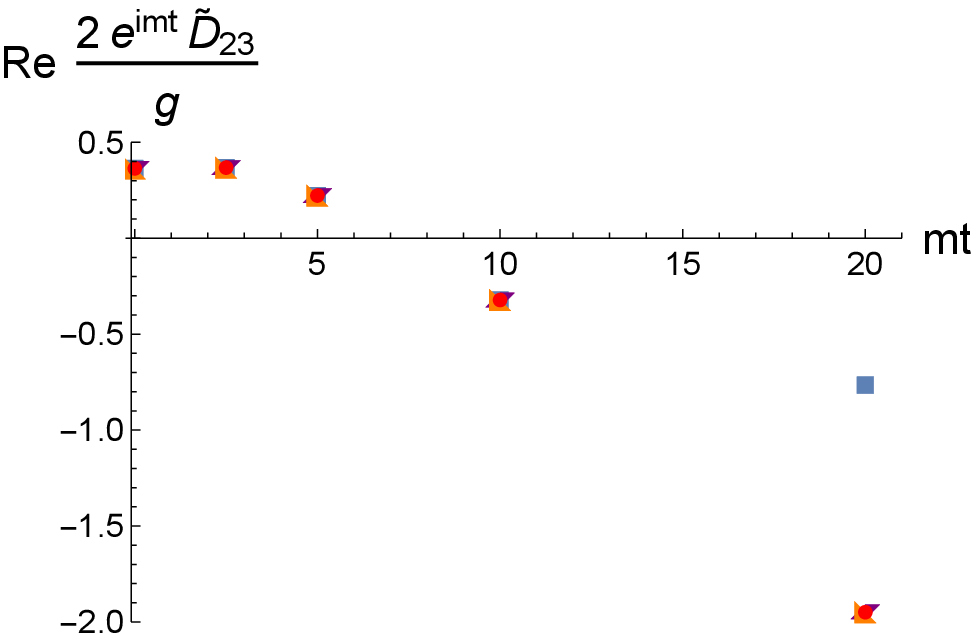}
\par\end{centering}
}\subfloat[\foreignlanguage{english}{$\Im m\,e^{imt}\frac{\tilde{D}_{23}}{g/2}$}]{\begin{centering}
\includegraphics[width=0.45\textwidth]{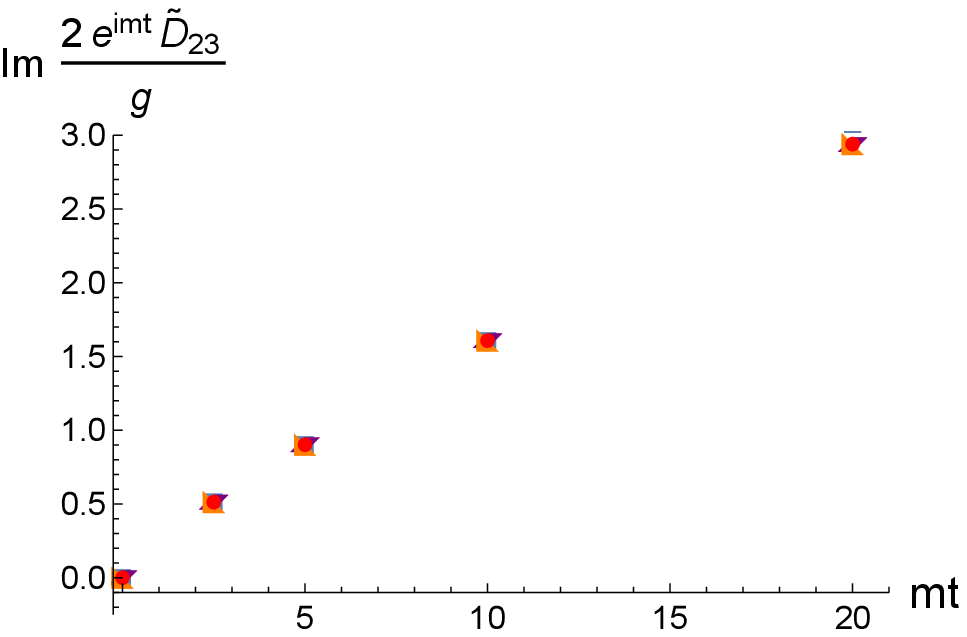}
\par\end{centering}
}\\

\caption{\label{fig:Real-and-imaginary}Real and imaginary parts of $e^{imt}\frac{\tilde{D}_{23}}{g/2}$
for time instants $mt=0,2.5,5,10$ and $20$ for a quench in the Ising
model. Results for system sizes $mL=30,40,50$ and $60$ are shown
with blue, purple, orange and red symbols, but with a single exception
these cannot really be distinguished due to their almost perfect overlap.}
\end{figure}

During the evaluation of $\tilde{D}_{23}$ in Appendix \ref{sec:D23Res}
and \ref{sec:CInt} we ignored discussing and showing the cancellation
of $\mathcal{O}(mL)$ type terms. Here we verify their cancellation
by considering the numerical values for $\tilde{D}_{23}$ for various
time instants and $mL$ system sizes. For simplicity, we use the Ising
case $S=-1$ for our checks with analytic expression for the form
factors from \cite{IsingFF} and with a singular $K$ function

\begin{equation}
K_{Ising}(\vartheta)=\frac{-ig^{2}}{\sinh2\vartheta}\;.
\end{equation}
Note that the only way to obtain a quench with a zero momentum particle
in the Ising model is when quenching from the ferromagnetic to the
paramagnetic phase \cite{CalabreseEsslerFagotti1,CalabreseEsslerFagotti2,CalabreseEsslerFagotti3}.
For such a quench calculations based on a form factor expansion presupposing
a small post-quench density are not expected to give accurate results.
Nevertheless, the cancellation of volume dependent terms is related
to the order-by-order structure of the expansion independent of the
eventual behaviour of the expansion itself. We evaluated

\begin{equation}
\tilde{D}_{23}=C_{23}-Z_{1}C_{12}-(Z_{2}-Z_{1}^{2})C_{01}
\end{equation}
numerically for time instants $mt=0,2.5,5,10,20$ and volume sizes
$mL=30,40,50,60$ performing discrete summation on the quantised rapidities.
To make the summation finite we introduced a rapidity cut-off $\vartheta_{c}=4.834$
for the particles involved. In Fig. (\ref{fig:Real-and-imaginary})
we show the numerical values for $e^{imt}\frac{\tilde{D}_{23}}{g/2}$
with $g=1$.

\begin{figure}[H]
\subfloat[$|\Delta\Re e\,e^{imt}\frac{\tilde{D}_{23}}{g/2}|$]{\begin{centering}
\includegraphics[width=0.45\columnwidth]{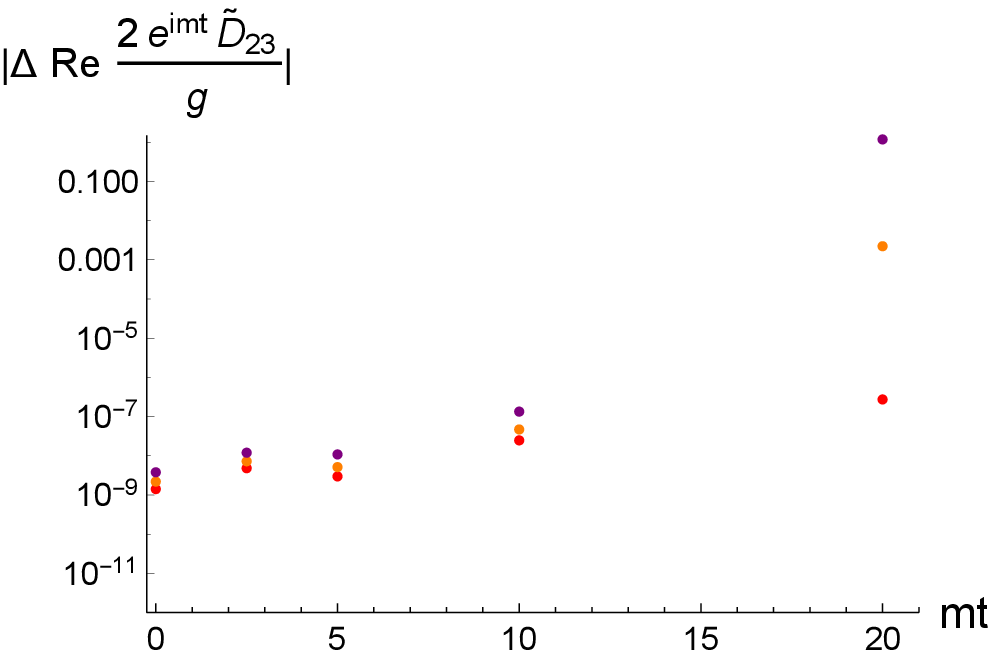}
\par\end{centering}
}\subfloat[$|\Delta\Im m\,e^{imt}\frac{\tilde{D}_{23}}{g/2}|$]{\begin{centering}
\includegraphics[width=0.45\columnwidth]{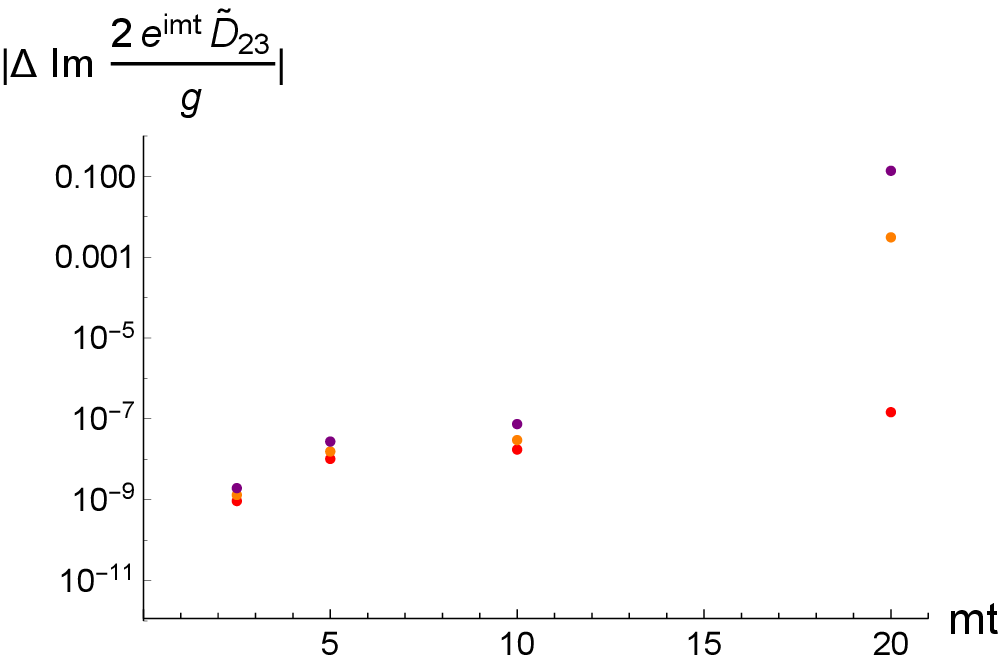}
\par\end{centering}
}

\caption{\label{fig:Modulus-of-differences}Modulus of differences between
real and imaginary parts of $e^{imt}\frac{\tilde{D}_{23}}{g/2}$ for
different system sizes at time instants $mt=0,2.5,5,10$ and $20$
for a quench in the Ising model. The purple, orange, and red dots
correspond to the difference of the value obtained for $30-40$, $40-50$
and $50-60$ in units of $mL$.}
\end{figure}

The source of the only observable deviation (for the case $mL=30$
and $mt=20$ ) is numerical inaccuracy resulting from the oscillatory
nature of the terms of the sum. In Fig. (\ref{fig:Modulus-of-differences})
we plot the difference between the numerical values obtained for various
system sizes at fixed times.

These results demonstrate that terms growing with $mL$ indeed cancel.

\subsection{Time dependence from $C^{intBI-II}$}

\begin{figure}[H]
\subfloat[$\Re e\,e^{imt}\frac{C^{intBI-II}}{F_{1}g/2}-mL\frac{Ker''g^{4}}{32}$]{\begin{centering}
\includegraphics[width=0.45\columnwidth]{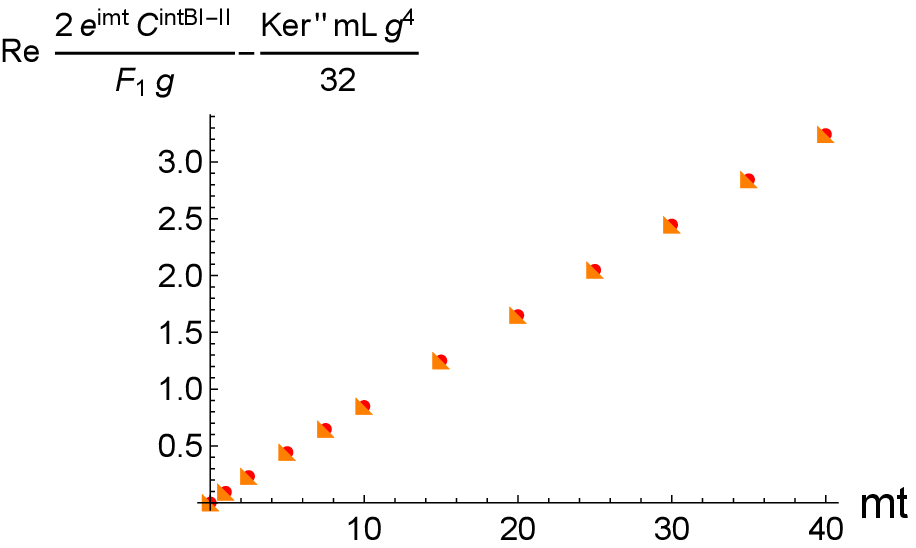}
\par\end{centering}
}\subfloat[\foreignlanguage{english}{$\Im m\,e^{imt}\frac{C^{intBI-II}}{F_{1}g/2}-mL\frac{Ker''g^{4}}{32}$}]{\begin{centering}
\includegraphics[width=0.45\textwidth]{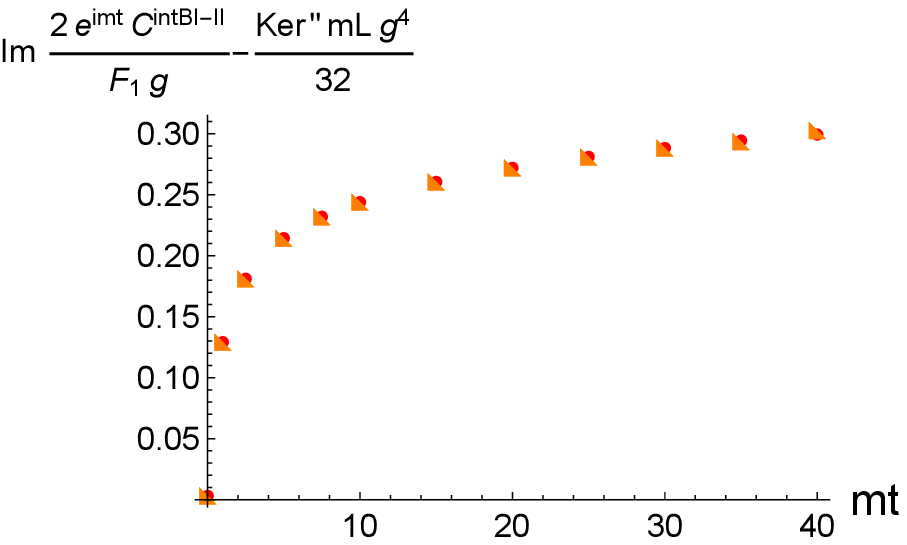}
\par\end{centering}
}\\

\caption{\label{fig:Real-and-imaginaryCintBI-II}Real and imaginary parts of
$e^{imt}\frac{C^{intBI-II}}{F_{1}g/2}-mL\frac{Ker''g^{4}}{32}$ calculated
with a discrete sum for time instants $mt=0.1,1,2.5,5,7.5,10,15,20,25,30,35$
and $40$ for a quench in the Ising model. The orange and red symbols
correspond to system sizes $mL=50$ and $60$. $g=1$.}
\end{figure}

In this subsection we present numerical results to confirm that the
time dependence of

\begin{equation}
\frac{g}{2}F_{1}e^{-imt}\sum_{I\neq0}\frac{|K(\vartheta_{1})|^{2}Ker(\vartheta_{1},t,R)}{\bar{\rho}_{2}(\vartheta_{1})}\frac{\frac{F_{5}^{\varepsilon}(\vartheta_{1})}{F_{1}\Omega(\vartheta)}+\frac{K'(\vartheta_{1})}{K(\vartheta_{1})}-\frac{F_{5}^{\varepsilon}(-\vartheta_{1})}{F_{1}\Omega(\vartheta)}-\frac{K'(-\vartheta_{1})}{K(-\vartheta_{1})}}{4\sinh\vartheta_{1}}\label{CIntBI-IISumSecondAppG}
\end{equation}
in the term $C^{intBI-II}$ defined in (\ref{CIntBI-II}), (\ref{CIntBI-IISecond})
is of the form

\begin{equation}
\frac{g}{2}F_{1}e^{-imt}\left(\Upsilon_{1}mt+i\Upsilon_{2}\log mt\right)\;,
\end{equation}
and that

\begin{equation}
\Upsilon_{1}mt=\int_{-\infty}^{\infty}\frac{d\vartheta}{2\pi}\frac{\left(|K(\vartheta)|^{2}\Re e\,Ker_{stac}(\vartheta,t,R)\,\Re e\,\left(\frac{F_{5}^{\varepsilon}(\vartheta)}{F_{1}\Omega(\vartheta)}+\frac{K'(\vartheta)}{K(\vartheta_{1})}-\frac{F_{5}^{\varepsilon}(-\vartheta)}{F_{1}\Omega(\vartheta)}-\frac{K'(-\vartheta)}{K(-\vartheta)}\right)\tanh\vartheta\right)'}{4\sinh\vartheta}\;.\label{CIntBI-IILinearTimeDepAppG}
\end{equation}
For the numerical analysis we used the Ising model to demonstrate
these statements with

\begin{equation}
K(\vartheta)=-i\frac{g^{2}}{2\sinh\vartheta}\;,
\end{equation}
where $F_{5}^{\varepsilon}(\vartheta)=\frac{8F_{1}}{\sinh\vartheta}$,
yielding

\begin{equation}
\frac{g}{2}F_{1}e^{-imt}\sum_{I\neq0}\frac{|K(\vartheta_{1})|^{2}\left(2-\cosh\vartheta_{1}\right)Ker(\vartheta_{1},t,R)}{\bar{\rho}_{2}(\vartheta_{1})2\sinh^{2}\vartheta_{1}}\label{CIntBI-IISumAppGIsing}
\end{equation}
and

\begin{equation}
\Upsilon_{1}mt=\int_{-\infty}^{\infty}\frac{d\vartheta}{2\pi}\frac{\left(|K(\vartheta)|^{2}\left(2-\cosh\vartheta_{1}\right)/\cosh\vartheta\,\Re e\,Ker_{stac}(\vartheta,t,R)\right)'}{2\sinh\vartheta}\;.\label{CIntBI-IILinearTimeDepAppGIsing}
\end{equation}
Figure \ref{fig:Real-and-imaginaryCintBI-II} shows the numerical
results for (\ref{CIntBI-IISumAppGIsing}) for system sizes $mL=50,60$
and for various time instants. The kernel $Ker$ was determined by
numerical integration and we subtracted the $mL$ residue term

\begin{equation}
\frac{g}{2}F_{1}e^{-imt}\frac{g^{4}}{4}\frac{mLKer(\vartheta,t,0)''}{8}\;,
\end{equation}
from (\ref{CIntBI-IIintegralRep}) to check the resulted $mL$ independence.

\begin{figure}[H]
\subfloat[$\Re e\,e^{imt}\frac{C^{intBI-II}}{F_{1}g/2}-mL\frac{Ker''g^{4}}{32}$]{\begin{centering}
\includegraphics[width=0.45\columnwidth]{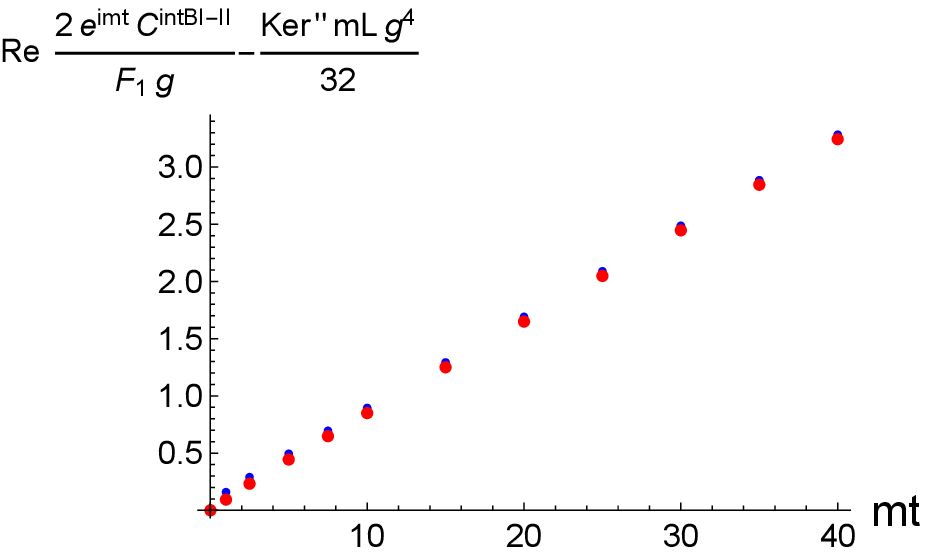}
\par\end{centering}
}\subfloat[\foreignlanguage{english}{$\Im m\,e^{imt}\frac{C^{intBI-II}}{F_{1}g/2}-mL\frac{Ker''g^{4}}{32}$}]{\begin{centering}
\includegraphics[width=0.45\textwidth]{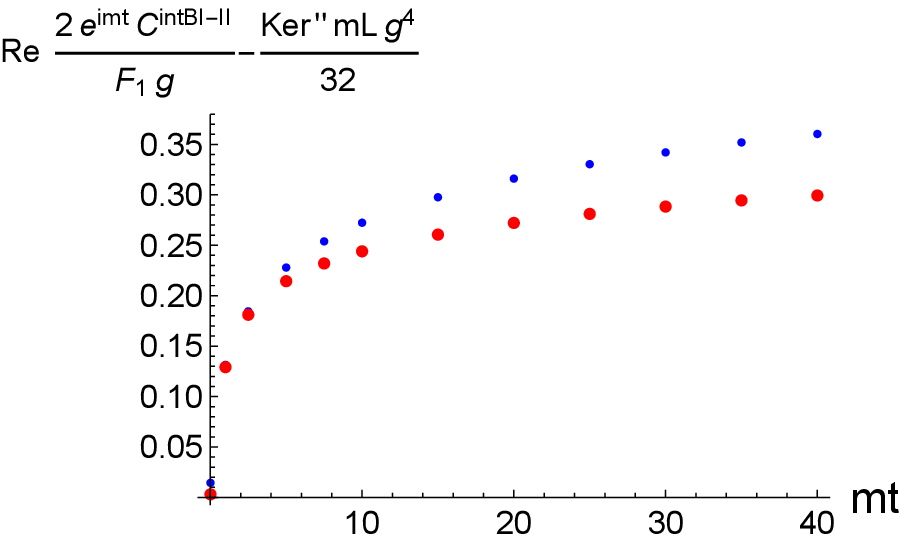}
\par\end{centering}
}\\

\caption{\label{fig:Real-and-imaginaryCintBI-IIAnal}Real and imaginary parts
of $e^{imt}\frac{C^{intBI-II}}{F_{1}g/2}-mL\frac{Ker''g^{4}}{32}$
calculated by a discrete sum and a continuous integral using $Ker_{stac}$
for time instants $mt=0.1,1,2.5,5,7.5,10,15,20,25,30,35$ and $40$
for a quench in the Ising model. The red dots correspond to system
sizes $mL=60$ and the blue dots to the analytic results. $g=1$.}
\end{figure}

Figure \ref{fig:Real-and-imaginaryCintBI-IIAnal} shows the real and
imaginary factor multiplying the term $\frac{g}{2}F_{1}e^{-imt}$
in $C^{intBI-II}$ calculated by a discrete summation (\ref{CIntBI-IISumAppGIsing})
and by the integration (\ref{CIntBI-IILinearTimeDepAppGIsing}). For
the imaginary parts, we also keep the $F_{S}+F_{C}$ part from $Ker_{stac}$
(\ref{KerFresnel}) but drop the $\sqrt{\sinh^{2}\vartheta}$ type
function, $f$.

\begin{figure}[H]
\subfloat[$\Re e\,e^{imt}\frac{C^{intBI-II}}{F_{1}g/2}-mL\frac{Ker''g^{4}}{32}$]{\begin{centering}
\includegraphics[width=0.45\columnwidth]{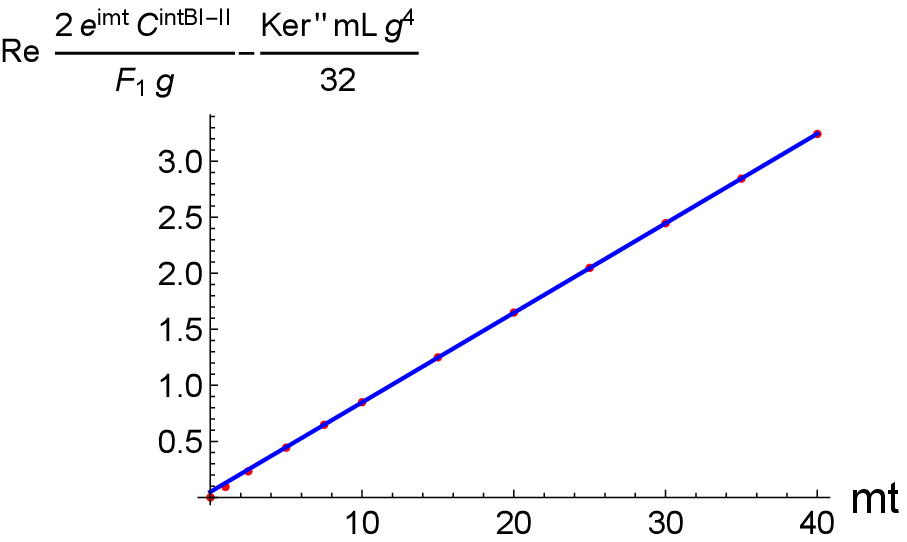}
\par\end{centering}
}\subfloat[\foreignlanguage{english}{$\Im m\,e^{imt}\frac{C^{intBI-II}}{F_{1}g/2}-mL\frac{Ker''g^{4}}{32}$}]{\begin{centering}
\includegraphics[width=0.45\textwidth]{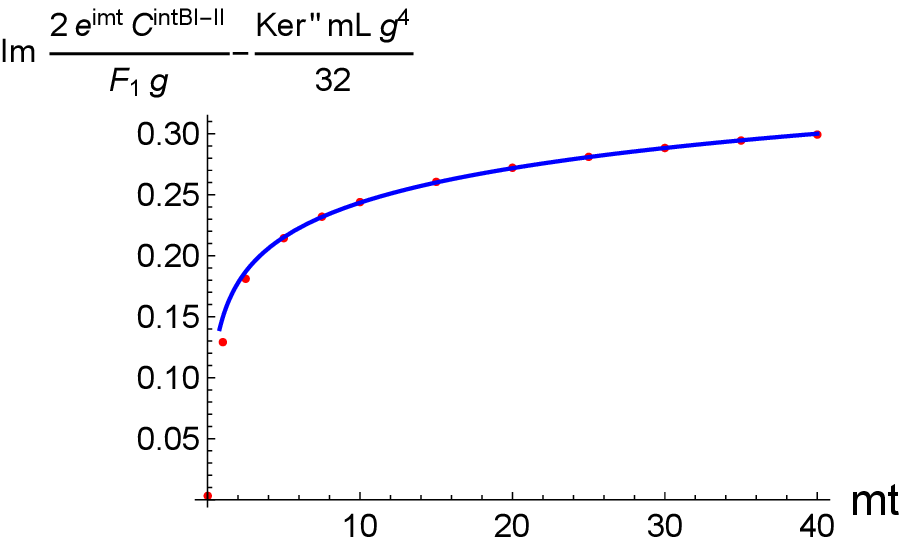}
\par\end{centering}
}\\

\caption{\label{fig:Real-and-imaginaryCintBI-IIFits}Real and imaginary parts
of $e^{imt}\frac{C^{intBI-II}}{F_{1}g/2}-mL\frac{Ker''g^{4}}{32}$
calculated by a discrete sum using $Ker_{stac}$ for various time
instants for a quench in the Ising model. The red dots correspond
to system sizes $mL=60$ and the blue line to the fitted curves of
type $a+b\,mt$ and $a+b\,\ln mt$. }
\end{figure}

We can conclude that for the real part multiplying $\frac{g}{2}F_{1}e^{-imt}$
using the real part of $Ker_{stac}$ in (\ref{CIntBI-IILinearTimeDepAppGIsing})
gives a correct result. For the imaginary part, however, the imaginary
part without the singular $f$ from $Ker_{stac}$ alone is not able
to reproduce the result of the discrete summation. On the other hand,
the time dependence of the imaginary coefficient is only logarithmic,
and therefore the calculation of this sub-leading time dependence
is not addressed in this work. We also checked if the linear time
dependence can be described by 
\[
\frac{g^{4}}{4}\frac{1}{\pi}mt\;,
\]
as predicted in Appendix \ref{sec:EvaluatingKernel} and if the logarithmic
time dependence for the imaginary part is a correct assumption. In
Fig. \ref{fig:Real-and-imaginaryCintBI-IIFits} we therefore fit the
functions $a+b\,mt$ and $a+b\,\ln mt$ to the real and imaginary
parts of 
\[
e^{imt}\frac{C^{intBI-II}}{F_{1}g/2}-mL\frac{Ker''g^{4}}{32}
\]
calculated by the discrete summation with $mL=60$ omitting the first
3 data points corresponding to the shortest times $mt=0.01,1,2.5$.

From the linear regression, the coefficient of the linearly time-dependent
term is $0.0799167$ which is to be compared with $\frac{1}{4\pi}=0.0795775$
as $g=1$. The agreement is excellent, and although the error of the
fitted parameter is $7\times10^{-5}$, the match is convincing.

\subsection{Comparing the time dependence of $\tilde{D}_{23}$ with the analytic
results}

In this subsection we study the time dependence of 
\[
e^{imt}\frac{\tilde{D}_{23}}{g/2}
\]
for the Ising model with $K=-i\frac{g^{2}}{\sinh2\vartheta}$ for
$mL=70$. We compute this quantity using discrete finite volume summation
for various time instants, and fit its real and imaginary parts with
the appropriate function dependences $a+b\,\sqrt{mt}+c\,mt+d\,mt\ln mt$
and $a+b\,\sqrt{mt}+c\,mt$ dictated by our analysis done in Appendix
\ref{sec:EvaluatingKernel}, omitting the first 7 data points corresponding
to short times.

\begin{figure}[H]
\subfloat[$\Re e\,e^{imt}\frac{\tilde{D}_{23}}{g/2}$]{\begin{centering}
\includegraphics[width=0.45\columnwidth]{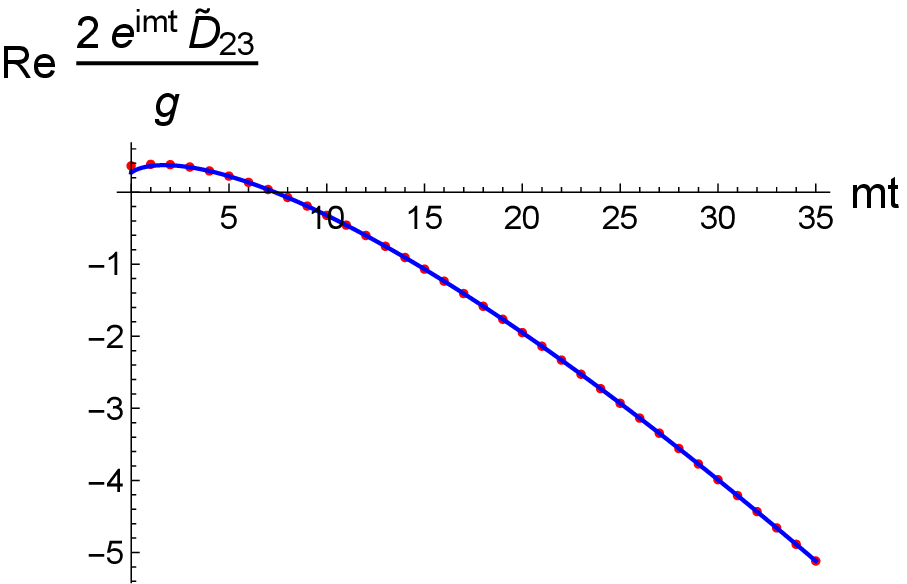}
\par\end{centering}
}\subfloat[\foreignlanguage{english}{$\Im m\,e^{imt}\frac{\tilde{D}_{23}}{g/2}$}]{\begin{centering}
\includegraphics[width=0.45\textwidth]{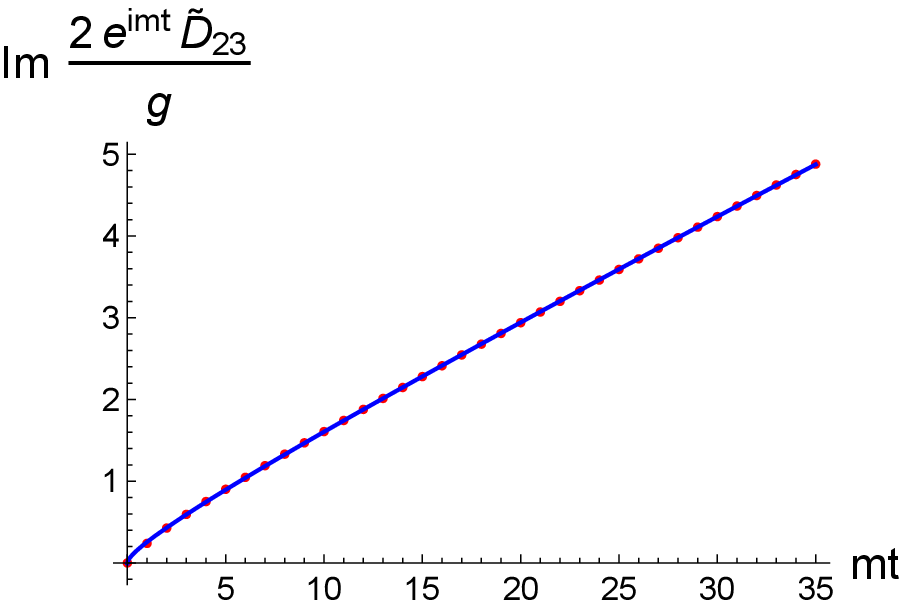}
\par\end{centering}
}\\

\caption{\label{fig:Real-and-imaginaryFits}Real and imaginary parts of $e^{imt}\frac{\tilde{D}_{23}}{g/2}$
for time instants $mt=0,1,2...,35$ for a quench in the Ising model.
The red dots correspond to system sizes $mL=70$ and the blue curve
to to the fitted curves of type $a+b\,\sqrt{mt}+c\,mt+d\,mt\ln mt$
and $a+b\,\sqrt{mt}+c\,mt$. $g=1$.}
\end{figure}

Concerning the real parts, the values obtained from the fit are $d=-0.0825821$
and $c=0.149943$, which must be compared with $-\frac{1}{4\pi}=-0.0795775$
and $\mathcal{K}+\frac{3}{4\pi}=0.131292$ resulting from (\ref{D23(t)})
and (\ref{Gammas}) with $g=1$ and $S=-1$. The accuracy of the fit
itself is around $10^{-3}-10^{-4}$. The difference between the fitted
parameters and the analytic predictions is now a bit larger compared
to the previous subsection. However, neglecting more data points for
short times the fitted values move towards the analytic predictions.
We note however, that omitting too many points leads to a deterioration
of the fit quality, as for an accurate determination of a logarithmic
term data points over several orders of magnitude should be used,
which is not possible to extract due to the inaccuracy of the discrete
sum for large times. As an additional test we also tried the fitting
function $a+b\,\sqrt{mt}+c\,mt$ and noted that the fit residuals
were two orders of magnitude larger than for the case including the
term $mt\ln mt$, which is a confirmation of the presence of the term
$mt\ln mt$ in the time dependence.

For the imaginary part the parameter $c$ was found to be $c=0.114809$
which must be compared with $\frac{1}{8}=0.125$ for $g=1$ according
to (\ref{D23(t)}) and (\ref{Gammas}). The total estimated error
of the fitted value is of the order $10^{-3}.$ In principle we should
have either included $\ln mt$ in the fitting function or subtracted
the contribution of $\Im m\,e^{imt}\frac{C^{intBI-II}}{F_{1}g/2}-mL\frac{Ker''g^{4}}{32}$
discussed in the previous subsection from the data points, but since
this correction is rather small it was simply discarded. 

\begin{thebibliography}{10}
\bibitem{NewtonCradle}T. Kinoshita, T. Wenger, and D. S. Weiss, \emph{Nature}
\textbf{440} (2006) 900\textendash 903.

\bibitem{ExperimentalNoThermalization1}S. Trotzky, Y.-A. Chen, A.
Flesch, I. P. McCulloch, U. Schollw{ö}ck, J. Eisert and I. Bloch,
\emph{Nature Phys.} \textbf{8} (2012) 325-330, arXiv:1101.2659 {[}cond-mat.quant-gas{]}.

\bibitem{ExperimentalNoThermalization3}M. Gring, M. Kuhnert, T. Langen,
T. Kitagawa, B. Rauer, M. Schreitl, I. Mazets, D. A. Smith, E. Demler
and J. Schmiedmayer, \emph{Science} \textbf{337} (2012) 1318-1322,
arXiv:1112.0013.

\bibitem{GGEExperimental}T. Langen, S. Erne, R. Geiger, B. Rauer,
T. Schweigler, M. Kuhnert, W. Rohringer, I. E. Mazets, T. Gasenzer
and J. Schmiedmayer, \emph{Science} \textbf{348} (2015) 207\textendash 211,
arXiv:1411.7185 {[}cond-mat.quant-gas{]}.

\bibitem{ColdAtomSchm1}S. Hofferberth, I. Lesanovsky, B. Fischer,
T. Schumm and J. Schmiedmayer, \textit{Nature} \textbf{449}, 324 (2007),
arXiv:0706.2259 {[}cond-mat.other{]}

\bibitem{ColdAtomSchm2}T. Langen, R. Geiger, M. Kuhnert, B. Rauer
and J. Schmiedmayer, \textit{Nature Physics} \textbf{9}, 640 (2013),
arXiv:1305.3708 {[}cond-mat.quant-gas{]}.

\bibitem{Nagerl}F. Meinert, M. J. Mark, E. Kirilov, K. Lauber, P.
Weinmann, A. J. Daley and H.-C. N{ä}gerl, \emph{Phys. Rev. Lett.}
\textbf{111} (2013) 053003, arXiv:1304.2628 {[}cond-mat.quant-gas{]}.

\bibitem{Fukuhara}T. Fukuhara, P. Schauss, M. Endres, S. Hild, M.
Cheneau, I. Bloch and C. Gross, \textit{Nature} \textbf{502} (2013)
76, arXiv:1305.6598 {[}cond-mat.quant-gas{]}.

\bibitem{Kaufman}A. M. Kaufman, M. E. Tai, A. Lukin, M. Rispoli,
R. Schittko, P. M. Preiss and M. Greiner, \texttt{Quantum thermalization
through entanglement in an isolated many-body system}, ArXiv e-prints
(2016), 1603.04409.

\bibitem{ExperimentalNoThermalization2}M. Cheneau, P. Barmettler,
D. Poletti, M. Endres, P. Schauss, T. Fukuhara, C. Gross, I. Bloch,
C. Kollath and S. Kuhr, \emph{Nature} \textbf{481} (2012) 484-487,
arXiv:1111.0776 {[}cond-mat.quant-gas{]}.

\bibitem{CardyCalabrese}P. Calabrese and J. Cardy, \emph{Phys. Rev.
Lett.} \textbf{96} (2006) 136801, arXiv:cond-mat/0601225; \\
 P. Calabrese and J. Cardy, \emph{J. Stat. Mech.} \textbf{0706} (2007)
P06008, arXiv:0704.1880 {[}cond-mat.stat-mech{]}.

\bibitem{GGEProposal}M. Rigol, V. Dunjko, V. Yurovsky and M. Olshanii,
\emph{Phys. Rev. Lett.} \textbf{98} (2007) 050405, arXiv:cond-mat/0604476
{[}cond-mat.other{]}.

\bibitem{CauxNoGGE}B. Wouters, J. De Nardis, M. Brockmann, D. Fioretto,
M. Rigol and J.-S. Caux\emph{, Phys. Rev. Lett.} \textbf{113} (2014)
117202, arXiv:1405.0172, arXiv:1405.0172 {[}cond-mat.str-el{]}.

\bibitem{PozsgayNoGGE}B. Pozsgay, M. Mesty{á}n, M. A. Werner, M.
Kormos, G. Zar{á}nd and G. Tak{á}cs\emph{, Phys. Rev. Lett.} \textbf{113}
(2014) 117203, arXiv:1405.2843 {[}cond-mat.stat-mech{]}.

\bibitem{Pozsi2}B. Pozsgay, \emph{J. Stat. Mech.} \textbf{9} (2014)
09026, arXiv:1406.4613 {[}cond-mat.stat-mech{]}.

\bibitem{Goldstein}G. Goldstein and N. Andrei, \emph{Phys. Rev. }\textbf{\emph{A}}\textbf{90}
(2014) 043625, arXiv:1405.4224 {[}cond-mat.quant-gas{]}.

\bibitem{EsslerMussardo}F. H. L. Essler, G. Mussardo and M. Panfil,
\emph{Phys. Rev. }\textbf{\emph{A}}\textbf{91} (2015) 051602, arXiv:1411.5352
{[}cond-mat.quant-gas{]}.

\bibitem{ProsenCGGE}E. Ilievski, J. De Nardis, B. Wouters, J.-S.
Caux, F. H. L. Essler and T. Prosen\emph{, Phys. Rev. Lett.} \textbf{115}
(2015) 157201, arXiv:1507.02993 {[}quant-ph{]}.

\bibitem{IlievskiQuasiLocal}E. Ilievski, M. Medenjak, T. Prosen and
L. Zadnik, \emph{J. Stat. Mech.} \textbf{1606} (2016) 064008, arXiv:1603.00440
{[}cond-mat.stat-mech{]}.

\bibitem{Cazalilla}M. A. Cazalilla, \textit{Phys. Rev. Lett.} \textbf{97}(15),
(2006)156403, arXiv:cond-mat/0606236v2 {[}cond-mat.stat-mech{]}.

\bibitem{Silva}A. Silva, \textit{Phys. Rev. Lett.} \textbf{101} (2008)
120603, arXiv:0806.4301 {[}cond-mat.stat-mech{]}.

\bibitem{SotiriadisCalabreseCardy}S. Sotiriadis, P. Calabrese and
J. Cardy, \textit{EPL (Europhysics Letters)} \textbf{87} (2009) 20002,
arXiv:0903.0895 {[}cond-mat.stat-mech{]}

\bibitem{FiorettoMussardo} D. Fioretto and G. Mussardo,\textit{ New
J. Phys.} \textbf{12} (2010) 055015, arXiv:0911.3345 {[}cond-mat.stat-mech{]}.

\bibitem{D=00003D0000F3raZar=00003D0000E1nd}B. D{ó}ra, M. Haque
and G. Zar{á}nd, \textit{Phys. Rev. Lett.} \textbf{106}(15) (2011)
156406, arXiv:1011.6655 {[}cond-mat.str-el{]}.

\bibitem{CalabreseEsslerFagotti1}P. Calabrese, F. H. L. Essler and
M. Fagotti, \textit{Phys. Rev. Lett.} \textbf{106} (2011) 227203,
arXiv:1104.0154 {[}cond-mat.str-el{]}.

\bibitem{CalabreseEsslerFagotti2}P. Calabrese, F. H. L. Essler and
M. Fagotti, \textit{J. Stat. Mech.} \textbf{7} (2012) 07016, arXiv:1204.
3911 {[}cond-mat.quant-gas{]}.

\bibitem{CalabreseEsslerFagotti3}P. Calabrese, F. H. L. Essler and
M. Fagotti, \textit{J. Stat. Mech.} \textbf{7} (2012) 07022, arXiv:1205.2211
{[}cond-mat.stat-mech{]}.

\bibitem{EsslerEvangelisti}F. H. L. Essler, S. Evangelisti and M.
Fagotti, \textit{Phys. Rev. Lett.} \textbf{109}(24) (2012) 247206,
arXiv:1208.1961 {[}cond-mat.stat-mech{]}.

\bibitem{ColluraSotiriadis}M. Collura, S. Sotiriadis and P. Calabrese,
\textit{Phys. Rev. Lett.} \textbf{110}(24) (2013) 245301, arXiv:1303.3795
{[}cond-mat.quant-gas{]}.

\bibitem{HeylPolkovnikov}M. Heyl, A. Polkovnikov and S. Kehrein,
\textit{Phys. Rev. Lett.} \textbf{110} (2013)135704, arXiv:1206.2505
{[}cond-mat.stat-mech{]}.

\bibitem{BucciantiniKormosCalabrese}L. Bucciantini, M. Kormos and
P. Calabrese, \textit{J. Phys. A: Math. Theor.}\textbf{ 47} (2014)175002,
arXiv:1401.7250 {[}cond-mat.stat-mech{]}.

\bibitem{KormosColluraHardCore}M. Kormos, M. Collura and P. Calabrese,
\emph{Phys. Rev. }\textbf{\emph{A}}\textbf{89}(1) (2014) 013609, arXiv:1307.2142
{[}cond-mat.quant-gas{]}.

\bibitem{SpyrosCalabrese}S. Sotiriadis and P. Calabrese, \textit{J.
Stat. Mech.} \textbf{7} (2014) 07024, arXiv:1403.7431 {[}cond-mat.stat-mech{]}.

\bibitem{SpyrosMemory}S. Sotiriadis, \textit{Phys. Rev.} \textbf{A
94}(3) (2016) 031605, arXiv:1507.07915 {[}cond-mat.stat-mech{]}.

\bibitem{Andrei1}D. Iyer and N. Andrei, \textit{Phys. Rev. Lett.}
\textbf{109}, (2012) 115304, arXiv:1206.2410 {[}cond-mat.quant-gas{]}. 

\bibitem{Andrei2}D. Iyer, H. Guan and N. Andrei, \textit{Phys. Rev.}
\textbf{A 87}, (2013) 053628,arXiv:1304.0506 {[}cond-mat.quant-gas{]}.

\bibitem{Andrei3}W. Liu and N. Andrei, \textit{Phys. Rev. Lett. }\textbf{112},
(2014) 257204, arXiv:1311.1118 {[}cond-mat.quant-gas{]}.

\bibitem{Andrei4}H. Guan and N. Andrei: \texttt{Quench Dynamics of
the Gaudin-Yang Model,} arXiv:1803.04846 {[}cond-mat.quant-gas{]}.

\bibitem{NardisPiroliCaux}J. De Nardis, L. Piroli and J.-S. Caux,
\textit{J.Phys. A: Math. Theor.} \textbf{48}, (2015) 43FT01, arXiv:1505.03080
{[}cond-mat.quant-gas{]}. 

\bibitem{PiroliPozsgayVernier1}L. Piroli, B. Pozsgay and E. Vernier,
\textit{J. Stat. Mech. }(2017) 023106, arXiv:1611.06126 {[}cond-mat.stat-mech{]}.

\bibitem{PiroliPozsgayVernier2}L. Piroli, B. Pozsgay and E. Vernier:
\texttt{Non-analytic behavior of the Loschmidt echo in XXZ spin chains: exact
results}, arXiv:1803.04380 {[}cond-mat.stat-mech{]}.

\bibitem{Schmiedmayer}T. Schweigler, V. Kasper, S. Erne, B. Rauer,
T. Langen, T. Gasenzer, J. Berges and J. Schmiedmayer\texttt{,} \textit{Nature}
\textbf{545}, 323 (2017), arXiv:1505.03126 {[}cond-mat.quant-gas{]}.

\bibitem{SchmiedmayerPhase}M. Pigneur, T. Berrada, M. Bonneau, T.
Schumm, E. Demler and J. Schmiedmayer: \texttt{Relaxation to a Phase-locked
Equilibrium State in a One-dimensional Bosonic Josephson Junction},
arXiv:1711.06635 {[}quant-ph{]}.

\bibitem{Igloi1}H. Rieger and F. Igl{ó}i, \textit{Phys. Rev. }\textbf{B
84}(16) (2011) 165117, arXiv:1106.5248 {[}cond-mat.stat-mech{]}.

\bibitem{Igloi2}B. Blass, H. Rieger and F. Igl{ó}i, \textit{EPL
(Europhysics Letters)} \textbf{99} (2012) 30004, arXiv:1205.3303 {[}cond-mat.stat-mech{]}.

\bibitem{Evangelesti}S. Evangelisti, \emph{J. Stat. Mech.} \textbf{4}
(2013) 04003, arXiv:1210.4028 {[}cond-mat.stat-mech{]}.

\bibitem{SineGSemiClassical}M. Kormos and G. Zar{á}nd\textit{\emph{,}}\textit{
Phys. Rev.} \textbf{E93} (2016) 062101, arXiv:1507.02708 {[}cond-mat.stat-mech{]};\\
 C. P. Moca, M. Kormos and G. Zar{á}nd, \textit{Phys. Rev. Lett.
}\textbf{119} (2016) 100603, arXiv:1609.00974 {[}cond-mat.stat-mech{]}.

\bibitem{SchurichtEssler}D. Schuricht and F.H.L. Essler, \emph{J.
Stat. Mech.} \textbf{1204} (2012) P04017, arXiv:1203.5080 {[}cond-mat.str-el{]}.

\bibitem{BertiniSineG} B. Bertini, D. Schuricht and F.H.L. Essler,
\emph{J. Stat. Mech.} \textbf{1410} (2014) P10035, arXiv:1405.4813
{[}cond-mat.stat-mech{]}.

\bibitem{SchurichtCubero}A. C. Cubero, and D. Schuricht, \textit{J.
Stat. Mech.} \textbf{1710} (2017) 103106, arXiv:1707.09218v2 {[}cond-mat.stat-mech{]}.

\bibitem{DelfinoOscOlder}G. Delfino, \emph{J. Phys. A: Math. Theor.}
\textbf{47} (2014) 402001, arXiv:1405.6553 {[}cond-mat.stat-mech{]}.

\bibitem{DelfinoOscNewer}G. Delfino, J. Viti, \emph{J. Phys. A: Math.
Theor.} \textbf{50} (2017) 084004, arXiv:1608.07612 {[}cond-mat.stat-mech{]}.

\bibitem{E8Quench}K. H{ó}ds{á}gi, M. Kormos, and G. Tak{á}cs:
\texttt{Quench dynamics of the Ising field theory in a magnetic field},\textit{
}arXiv:1803.01158 {[}cond-mat.stat-mech{]}.

\bibitem{sGOverlaps}D.X. Horv{á}th and G. Tak{á}cs, \emph{Phys.
Lett. }\textbf{B771} (2017) 539\textendash 545, arXiv:1704.00594 {[}cond-mat.stat-mech{]}.

\bibitem{SotiriadisTakacsMussardo}S. Sotiriadis, G. Tak{á}cs and
G. Mussardo, \emph{Phys. Lett.} \textbf{B734} (2014) 52-57, arXiv:1311.4418
{[}cond-mat.stat-mech{]}.

\bibitem{InitalStateIntEqHierarchcy}D. X. Horv{á}th, S. Sotiriadis
and G. Tak{á}cs, \textit{Nucl. Phys.} \textbf{B902} (2016) 508-547,
arXiv:1510.01735 {[}cond-mat.stat-mech{]}.

\bibitem{PiroliPozsiVernier}L. Piroli, B. Pozsgay and E. Vernier,
\emph{Nucl. Phys.} \textbf{B925} (2017) 362-402, arXiv:1709.04796.

\bibitem{GhoshalZamo}S. Ghoshal and A.B. Zamolodchikov, \emph{Int.
J. Mod. Phys.} \textbf{A9} (1994) 3841-3886, arXiv:hep-th/9306002.

\bibitem{gDorey}P. Dorey, M. Pillin, R. Tateo, and G. M. T. Watts,
\emph{Nucl. Phys.} \textbf{B594} (2001) 625\textendash 659, arXiv:hep-th/0007077.

\bibitem{gPallaBajnokTakacs2}Z. Bajnok, L. Palla, and G. Tak{á}cs,
\emph{Nucl. Phys.} \textbf{B772} (2007) 290\textendash 322, arXiv:hep-th/0611176.

\bibitem{FinVolFF1and2}B. Pozsgay and G. Tak{á}cs, \emph{Nucl.
Phys.} \textbf{B788} (2008) 167-208, arXiv: 0706.1445 {[}hep-th{]}.\\
 B. Pozsgay and G. Tak{á}cs, \emph{Nucl. Phys.} \textbf{B788} (2008)
209-251, arXiv: 0706.3605 {[}hep-th{]}.

\bibitem{OnePointFunctions} M. Kormos and B. Pozsgay,\textit{ JHEP}
\textbf{1004} (2010) 112, arXiv:1002.2783 {[}hep-th{]}.

\bibitem{gPallaBajnokTakacs}Z. Bajnok, L. Palla, and G. Tak{á}cs,
\emph{Nucl. Phys.} \textbf{B716} (2005) 519\textendash 542, arXiv:hep-th/0412192.

\bibitem{bff}Z. Bajnok, L. Palla, and G. Tak{á}cs, \emph{Nucl.
Phys.} \textbf{B750} (2006) 179-212, arXiv:hep-th/0603171.

\bibitem{Tibi}T. Rakovszky, M. Mesty{á}n, M. Collura, M. Kormos
and G. Tak{á}cs, \textit{Nucl. Phys.} \textbf{B} 911 (2016) 805-845,
arXiv:1607.01068v2 {[}cond-mat.stat-mech{]}.

\bibitem{IsingDiag1}E. Lieb, T. Schultz, and D. Mattis, ``Two soluble
models of an antiferromagnetic chain,\textquotedblright{} Annals of
Physics 16 (1961) 407\textendash 466.

\bibitem{IsingDiag2}P. Pfeuty, ``The one-dimensional Ising model
with a transverse field,\textquotedblright{} Annals of Physics 57
(1970) 79\textendash 90.

\bibitem{FiniteTCorr}B. Pozsgay and G. Tak{á}cs, \emph{J. Stat.
Mech. }\textbf{1011} (2010) 11012, arXiv: 1008.3810 {[}hep-th{]}.\\
 I.M. Sz{é}cs{é}nyi and G. Tak{á}cs, \emph{J. Stat. Mech.} \textbf{1212
}(2012) P12002, arXiv: 1210.0331 {[}hep-th{]}.

\bibitem{EsslerKonik}F.H.L. Essler and R.M. Konik, \emph{J. Stat.
Mech.} \textbf{0909} (2009) P09018, arXiv: 0907.0779 {[}cond-mat.str-el{]}.

\bibitem{L=00003D0000FCscher}M. L{ü}scher, \emph{Commun. Math.
Phys. }\textbf{104} (1986) 177.

\bibitem{KlassenM}T. R. Klassen and E. Melzer, \emph{Nucl. Phys.}
\textbf{B362} (1991) 329\textendash 388.

\bibitem{BajnokJanik}Z. Bajnok and R. A. Janik, \emph{Nucl. Phys.}
\textbf{B807} (2009) 625\textendash 650, arXiv: 0807.0399 {[}hep-th{]}.

\bibitem{Hatsuda}Y. Hatsuda and R. Suzuki, \emph{JHEP} \textbf{0809}
(2008) 025, arXiv: 0807.0643 {[}hep-th{]}.

\bibitem{GiuseppeReview}G. Mussardo, \emph{Phys. Rept.} \textbf{218}
(1992) 215-379.

\bibitem{SmirnovBook}F.A. Smirnov: \texttt{Form Factors in Completely
Integrable Models of Quantum Field Theory}, \emph{Adv. Ser. Math.
Phys.} \textbf{14} (1992) 1-208.

\bibitem{Lukyanov}S.L. Lukyanov, \emph{Mod. Phys. Lett. }\textbf{A12}
(1997) 2543\textendash 2550, arXiv: hep-th/9703190.

\bibitem{IsingFF}P. Fonseca and A.B. Zamolodchikov,\textbf{ }\textit{{}
J. Stat. Phys.}{} \textbf{110}{} (2003) 527\textendash 590, arXiv:hep-th/0112167 
\end{thebibliography}
\end{document}